\documentclass[11pt]{article}
\usepackage[margin=1.0in]{geometry}

\usepackage{constants}
\newconstantfamily{c}{symbol=c}
\usepackage{preamble}
\usepackage{braket}
\usepackage{xcolor}
\usepackage[export]{adjustbox}
\newtheorem*{CM3p2}{Theorem 3.2 of~\cite{collins2017weingarten}}
\newtheorem*{CM4p11}{Theorem 4.11 of~\cite{collins2017weingarten}}
\newtheorem*{CM4p10}{Theorem 4.10 of~\cite{collins2017weingarten}}
\newtheorem*{ACQl6}{Lemma 6 of~\cite{aharonov2021quantum}}
\newtheorem*{ACQl8}{Lemma 8 of~\cite{aharonov2021quantum}}
\newtheorem*{ACQl10}{Lemma 10 of~\cite{aharonov2021quantum}}

\title{Exponential separations between learning with and without quantum memory}
\author{Sitan Chen\thanks{Email: \texttt{sitanc@berkeley.edu} This work was supported in part by NSF Award 2103300, NSF CAREER Award CCF-1453261, NSF Large CCF-1565235 and Ankur Moitra's ONR Young Investigator Award. Part of this work was completed while visiting the Simons Institute for the Theory of Computing.} \\
UC Berkeley
 \and 
Jordan Cotler\thanks{Email: \texttt{jcotler@fas.harvard.edu}. This work is supported by
a Junior Fellowship from the Harvard Society of Fellows, as well as in part by the Department of
Energy under grant DE-SC0007870.} \\
Harvard University
\and
Hsin-Yuan Huang\thanks{Email: \texttt{hsinyuan@caltech.edu}. This work is supported by the J. Yang \& Family Foundation and the Google PhD Fellowship.} \\
Caltech
\and
Jerry Li\thanks{Email: \texttt{jerrl@microsoft.com}. Part of this work was completed while visiting the Simons Institute for the Theory of Computing.} \\
Microsoft Research
}

\DeclareMathAlphabet\mathbfcal{OMS}{cmsy}{b}{n}

\DeclareMathOperator{\Expect}{\mathbb{E}}
\DeclareMathOperator{\HS}{HS}

\newcommand{\ketbra}[2]{\lvert #1 \rangle \! \langle #2 \rvert}

\newcommand{\bH}{\mathbb{H}}

\newcommand{\U}{\vec{U}}

\newcommand{\Sig}{\vec{\Sigma}}

\newcommand{\tr}{\mathsf{tr}}

\newcommand{\rhomm}{\rho_{\mathsf{mm}}}

\renewcommand{\epsilon}{\varepsilon}
\newcommand{\eps}{\epsilon}

\DeclareMathOperator{\leaf}{leaf}

\usepackage{multirow}

\begin{document}

\maketitle



\begin{abstract}
    We study the power of quantum memory for learning properties of quantum systems and dynamics, which is of great importance in physics and chemistry.
    Many state-of-the-art learning algorithms require access to an additional external quantum memory.
    While such a quantum memory is not required \emph{a priori}, in many cases, algorithms that do not utilize quantum memory require much more data than those which do.
    We show that this trade-off is inherent in a wide range of learning problems. Our results include the following:
    \begin{itemize}
        \item We show that to perform shadow tomography on an $n$-qubit state $\rho$ with $M$ observables, any algorithm without quantum memory requires $\tilde{\Omega}(\min(M, 2^n))$ samples of $\rho$ in the worst case.
        Up to logarithmic factors, this matches the upper bound of \cite{huang2020predicting}, and completely resolves an open question in~\cite{aaronson2018shadow, aaronson2019gentle}.
        \item We establish exponential separations between algorithms with and without quantum memory for purity testing, distinguishing scrambling and depolarizing evolutions, as well as uncovering symmetry in physical dynamics. 
        Our separations improve and generalize prior work of~\cite{aharonov2021quantum} by allowing for a broader class of algorithms without quantum memory.
        \item We give the first tradeoff between quantum memory and sample complexity.
        More precisely, we prove that to estimate absolute values of all $n$-qubit Pauli observables, algorithms with $k < n$ qubits of quantum memory require at least $\Omega(2^{(n-k)/3})$ samples, but there is an algorithm using $n$-qubit quantum memory which only requires $\mathcal{O}(n)$ samples.
    \end{itemize}
    The separations we show are sufficiently large and could already be evident, for instance, with tens of qubits.
    This provides a concrete path towards demonstrating real-world advantage for learning algorithms with quantum memory.
\end{abstract}

\thispagestyle{empty}
\clearpage
\tableofcontents
\addtocontents{toc}{\protect\thispagestyle{empty}}
\pagenumbering{gobble}

\clearpage
\pagenumbering{arabic} 

\section{Introduction}

Learning properties of the physical world is at the heart of many scientific disciplines, including physics, chemistry, and material science.
The suite of emerging quantum technology, from quantum sensors, quantum computers, to quantum memories for retrieving, processing, and storing quantum information, provides the potential to significantly improve upon what we could achieve with existing technology.
Because the physical world is quantum-mechanical, it is natural to expect quantum information technology to provide an advantage in various tasks.
Yet, the potential remains to be justified both experimentally and theoretically.

In this work, we establish provable advantages in a wide range of tasks for predicting properties of physical systems and dynamics.
We consider \emph{learning algorithms with external quantum memory}, namely those that can retrieve quantum data from each experiment, store the data in quantum memory, and compute on the quantum memory in an entangled fashion (see Definitions~\ref{def:quantum_memory_state}, \ref{def:quantum_memory_channel}, and \ref{defn:sizek}).
This class of algorithms provides an abstract model for the potential experiments that a scientist  assisted by quantum information technology could perform in the future.

We also consider \emph{learning algorithms that do not use external quantum memory}, namely those that can only retrieve classical data by performing measurements after each experiment, store the classical information in classical memory, and process the classical measurement data to learn the properties (see Definitions~\ref{def:classical_memory_state} and \ref{def:classical_memory_channel}).
This class of algorithms is an abstract model that encompasses conventional experiments that one could achieve with existing experimental platforms.

\begin{table}[ht!]
    \centering
    \small
    \begin{tabular}{l|l|l|l}
        Tasks & LB (no q. memory) & UB (no q. memory) & UB (w/ q. memory)\\
        \hline
        Shadow tomogr. & $\widetilde{\Omega}\left( \min\left(M, 2^n\right) \right)$ [Cor~\ref{cor:shadow-random}] & $\widetilde{\mathcal{O}}\left( \min(M, 2^n)\right)$ \cite{huang2020predicting} & $\mathcal{O}(n \log(M)^2)$ \cite{buadescu2020improved} \\
        Shadow tomogr. (P) & $\Omega(2^n)$ [Cor~\ref{cor:shadow-pauli}] & $\mathcal{O}(n 2^n)$ \cite{huang2020predicting} & $\mathcal{O}(n)$ \cite{huang2021information} \\
        Purity testing & $\Omega(2^{n/2})$ [Thm~\ref{thm:purity_lower}] & $\mathcal{O}(2^{n/2})$ [Thm~\ref{thm:purity_upper}] & $\mathcal{O}(1)$ \cite{montanaro2013survey} \\
        Depolar. vs unitary & $\Omega(2^{n/3})$ [Thm~\ref{thm:channelhard1}] & $\mathcal{O}(2^{n/2})$ [Cor~\ref{thm:depo_vs_scram_upper}] & $\mathcal{O}(1)$ \cite{aharonov2021quantum} \\
        Classify symmetry & $\Omega(2^{n/3.5})$ [Thm~\ref{thm:channelhard2}] & $\mathcal{O}(2^{n})$ [Thm~\ref{thm:symmetry_tomo}] & $\mathcal{O}(1)$ \cite{aharonov2021quantum} \\
        \hline
        \hline
        Shadow tomogr. (P) & \multicolumn{3}{c}{Lower bound: $\Omega(2^{(n-k) / 3})$ }\\
         w/ $k$-qubit memory & \multicolumn{3}{c}{See [Thm~\ref{thm:pauli_memory}]}
    \end{tabular}
    \caption{\small A table of the upper and lower bounds on sample/query complexity for learning algorithms with quantum memory (``w/ q. memory'') and without quantum memory (``no q. memory''). For shadow tomography (``Shadow tomogr.''), we consider the sample complexity $T$ needed to achieve a small constant error for estimating expectation values of $M$ observables.
    The $\tilde{\Omega}$ and $\tilde{\mathcal{O}}$ neglects contributions from $\log(M)$ in the upper and lower bound without quantum memory.
    Similarly, we consider the sample complexity $T$ for predicting all $4^n$ Pauli observables to a constant error, denoted by the task: ``Shadow tomogr. (P)''.
    In ``Shadow tomogr. (P) w/ $k$-qubit memory'', we consider a simplified task of predicting absolute value for the expectation of Pauli observables up to a constant error.
    }
    \label{tab:summary}
\end{table}

In this work, we provide a set of mathematical tools for proving exponential separations between these two classes of learning algorithms.
Recently, two independent works \cite{huang2021information, aharonov2021quantum} have established exponential separations for some quantum learning tasks.
However, their techniques are rather cumbersome and work in more restrictive settings.
In this work, we design new, flexible frameworks for proving lower bounds against learning algorithms without quantum memory that strengthen and generalize those presented in \cite{huang2021information, aharonov2021quantum}. 
By using these new ideas, we are able to both significantly tighten their bounds, as well as derive novel bounds for more general and realistic settings that previous techniques could not handle.

\paragraph{Towards quantum advantage with quantum memory.}
These quantitative and qualitative improvements to the prior known separations have an important consequence for the near-term goal of demonstrating a provable quantum advantage.
As we will explain in more detail in the following, our bounds demonstrate that there exist quantum learning problems on $n$-qubit states for which there are copy-efficient and gate-efficient quantum algorithms that only require $\mathcal{O}(n)$ bits of quantum memory, but at the same time, any algorithm without quantum memory unconditionally requires $\Omega (2^n)$ copies.

The lower bounds derived in this work are sufficiently tight that this gap is already noticeable at the scale of tens of qubits.
In addition, our results guarantee that any protocol with quantum memory of insufficient size must require an exponentially increasing sample complexity; see the last row in Table~\ref{tab:summary}.
Together, these results imply that quantum computers with less than a hundred qubits could already assist experimentalists in extracting properties of physical systems and dynamics beyond conventional experiments that do not utilize quantum memory.
We believe the implementation of such protocols is an important open experimental question that may be within reach in the coming years.
To achieve that, we would need a better understanding on how the presence of noise in the quantum computers affect the advantage studied in this work.

\subsection{Our Results}
We now informally describe the problems we will consider throughout this paper, as well as our results; we will include pointers to the appropriate sections for the formal theorem statements.

\subsubsection{Learning physical systems}
We first consider the setting in which there is a physical source that generates copies of an unknown $n$-qubit state $\rho$ one at a time. 
The goal is to learn the physical system (or some properties thereof) from as few copies of it as possible.
A learning algorithm with quantum memory can store many copies of the quantum state, and perform joint quantum processing on the product state, followed by an entangled measurement on the quantum memory. 
On the other hand, a learning algorithm without quantum memory can only perform measurements on each copy of the quantum state, and learn from the resulting classical measurement data. However, note that the choice of measurement applied to the $i$-th copy could depend on all previous measurement outcomes.

The standard approach for learning physical systems is quantum state tomography \cite{gross2010quantum, o2017efficient, haah2017sample, guta2020fast}, where the goal is to learn a full representation of $\rho$ up to $\epsilon$ error in trace distance.
However, it is well known \cite{o2017efficient, haah2017sample} that quantum state tomography requires exponentially many copies $T = 2^{\Omega(n)}$ for learning algorithms with and without quantum memory.
Hence, such an approach is intrinsically inefficient.

\paragraph{Shadow tomography} To circumvent the exponential scaling, beginning with the work of~\cite{aaronson2018shadow}, many researchers \cite{brandao2017quantum, aaronson2019gentle, buadescu2020improved, huang2020predicting, chen2020robust} have considered the task of predicting many properties of the quantum state $\rho$ from very few copies of the unknown state. This task is known as \emph{shadow tomography} or \emph{shadow estimation}.
Here, we are given $M$ observables $O_1, \ldots, O_M$ satisfying $\norm{O_i}_\infty \leq 1$, and the goal is to accurately estimate $\Tr (O_i \rho)$ to error $\epsilon$, for all $i = 1, \ldots, M$.
When $M < 2^{\Omega (n)}$, this can be solved using much fewer copies than is required for full state tomography.
In particular, there is a simple algorithm using $\mathcal{O}(M\log M / \eps^2)$ copies that simply measures each observable independently. 
Notably, this estimator does not require quantum memory.

However, perhaps surprisingly, there are estimators which can solve this task with a sample complexity which is sublinear in $M$.
The most recent progress is due to \cite{buadescu2020improved}, who showed $\mathcal{O}(n \log(M)^2 / \epsilon^4)$ copies suffice.
This is somewhat unexpected, because the observables $O_1, \ldots, O_M$ can be highly incompatible, and due to Heisenberg's uncertainty principle, one should not be able to efficiently measure these observables simultaneously. 
Nevertheless, by performing quantum data analysis on a few copies of $\rho$, we can circumvent this incompatibility.

These estimators crucially rely on fully entangled measurements on all copies of $\rho$, and hence require a large quantum memory.
An important research direction in shadow tomography is understanding how well a learning algorithm without quantum memory can perform.
Indeed, this was posed as an open question in the first paper introducing the problem of shadow tomography~\cite{aaronson2018shadow}, and has remained an important question in the field ever since.
On the upper bounds side, many papers have developed improved algorithms for special classes of observables \cite{paini2019approximate, huggins2019efficient, huang2020predicting}. See related work for a more in-depth review.
However, when no assumptions are made on the class of observables, $\mathcal{O}(M\log M / \eps^2)$ remains the best known sample complexity.
On the lower bounds side, a very recent work of~\cite{huang2021information} shows that any learning algorithm without quantum memory requires sample complexity at least $\Omega(M^{1/6} / \eps^2)$ for this problem.
However, this still left a large gap between the known upper and lower bounds.

In this work, we completely resolve the sample complexity of shadow tomography without quantum memory up to logarithmic factors, by providing a lower bound which (nearly) matches the best known upper bound.
In particular, we show:
\begin{theorem}[General shadow tomography---informal, see Corollary~\ref{cor:shadow-random} and Theorem~\ref{thm:shadowtomo}]\label{thm:randomobservables_informal}
	For all $M > 1$, there exists a set of $M$ observables so that any learning algorithm without quantum memory that solves the shadow tomography problem for this set of observables requires $T = \tilde{\Theta}\left( \min(M, 2^n) / \epsilon^2\right)$ copies.\footnote{Throughout the paper, we say $g = \tilde{\Theta} (f)$ if $g = \Theta (f \log^c f)$ for some fixed $c$. We similarly define $\tilde{\mathcal{O}}$ and $\tilde{\Omega}$.}
\end{theorem}
\noindent
We note that this bound is tight up to logarithmic factors: there is an algorithm with sample complexity $\mathcal{O}\left( \min(M\log M, 2^n\log M) / \epsilon^2\right))$~\cite{huang2020predicting} that uses no quantum memory.
At a high level, this bound says that in general, algorithms without quantum memory cannot do much better than directly measuring each observable.
Up to these logarithmic factors, this fully answers the question posed in~\cite{aaronson2018shadow}.


\paragraph{Shadow Tomography with Pauli observables} An important and well-studied special case of shadow tomography is the case where the observables are the Pauli matrices, or a subset thereof.
This problem naturally arises in the study of quantum chemistry \cite{mcclean2016theory, huggins2019efficient, huang2021efficient}.
Indeed, it is this instance which~\cite{huang2021information} used to prove their $\Omega(M^{1/6} / \eps^2)$ lower bound.
In particular, they show that if the observables are the set of $M = \mathcal{O}(4^n)$ Pauli matrices, then any algorithm which uses no quantum memory and solves the shadow tomography problem must use at least $\Omega (2^{n/3} / \eps^2)$ copies.
In contrast, algorithms with quantum memory can solve this instance using $\mathcal{O}(n)$ copies~\cite{huang2021information}.
As before, the lower bound against algorithms without quantum memory is not tight: in particular, the best known sample complexity upper bound for such algorithms is $\widetilde{\mathcal{O}}(2^n)$ \cite{huang2020predicting}.

In this work, we provide a tight bound on the sample complexity of this problem without quantum memory, up to lower order terms:
\begin{theorem}[Pauli observables -- informal, see Corollary~\ref{cor:shadow-pauli} and Theorem~\ref{thm:shadowtomo}]\label{thm:pauliobservables_informal}
	To predict the expectation values of all Pauli observables in an unknown $n$-qubit state $\rho$, any learning algorithm without quantum memory requires $T = \Omega \left( 2^n \right)$ copies of $\rho$.
\end{theorem}
\noindent
Given the previously developed machinery, the proof of this is almost shockingly simple.
Indeed, the entire calculation fits in roughly half a page.
This serves as a demonstration of the power of the techniques we develop in this paper; in contrast, the proof of the bound in~\cite{huang2021information} is substantially more involved, not to mention much looser.

\paragraph{Purity testing}
We next apply our techniques to the slightly different setting of quantum property testing.
We are given copies of an $n$ qubit state $\rho$, but rather than learn statistics of $\rho$, the goal is to determine whether or not $\rho$ satisfies some condition, or is far from satisfying it.
Some examples include mixedness testing (is $\rho$ the maximally mixed state or far from it?) and its generalization identity testing (is $\rho = \sigma$ for some given state $\sigma$ or far from it?).
In this work, we consider the well-studied problem of \emph{purity testing} \cite{ekert2002direct}, where the goal is to distinguish if $\rho$ is rank-$1$, or if it is far from it.

This problem is natural in its own right, but it is of particular interest for us, because it is a setting where there are exponential separations between algorithms that use $\mathcal{O} (n)$ qubits of memory, and those that do not use quantum memory.
In fact, this shows up even for the special case of distinguishing whether $\rho$ is a pure state or if $\rho$ is the maximally mixed state.
For this case, there is a simple algorithm based on the SWAP test that uses $2$-entangled measurements---that is, measurements on $\rho^{\otimes 2}$---and succeeds with $\mathcal{O} (1)$ samples.
Note that this algorithm only requires $n$ qubits of additional quantum memory to store the single-copy from a previous experiment.
Prior work of~\cite{aharonov2021quantum} implied as a corollary that any such algorithm must use at least $\Omega(2^{2n / 7})$ samples, but as with shadow tomography, this bound was not tight.

In this work, we completely resolve this question, up to constant factors. In Theorem~\ref{thm:purity_upper}, we show that there is an algorithm that uses unentangled measurements with $\mathcal{O}(2^{n / 2})$ samples. And, we give the following matching lower bound:

\begin{theorem}[Purity testing---informal, see Theorem~\ref{thm:purity_lower}]\label{thm:purity_informal}
    Any algorithm which can distinguish between $\rho$ being a pure state and $\rho$ being the maximally mixed state without using external quantum memory requires $\Omega (2^{n / 2})$ samples.
\end{theorem}
\noindent

\subsubsection{Smooth tradeoffs for learning with quantum memory}
\label{sec:intro_bounded}
In Section~\ref{sec:bounded} we revisit the problem of shadow tomography with Pauli observables.
While our previously established bounds demonstrate exponential separations when the algorithm is given no quantum memory, it is natural to ask whether any algorithm that achieves exponentially better sample complexities must be using a \emph{large} amount of quantum memory.
In particular, is it possible that there is an algorithm that only uses a constant number of qubits of external quantum memory, but which still achieves $\mathcal{O} (1)$ sample complexity?
We demonstrate that this is impossible, by demonstrating a lower bound which scales with the number of qubits of quantum memory.
Formally, we show the following lower bound even for the simpler task of predicting the absolute value of the expectations of the Pauli observables:

\begin{restatable}[Shadow tomography with bounded quantum memory]{theorem}{paulimemory}\label{thm:pauli_memory}
	Any learning algorithm with $k$ qubits of quantum memory requires
	$T \ge \Omega\left(2^{(n - k)/3}\right)$ copies of $\rho$ to predict $\abs{\Tr(P\rho)}$ for all $n$-qubit Pauli observables $P$ with at least probability 2/3.
\end{restatable}

This bound is meaningful for $k \leq n$, and we complement this with a simple upper bound which uses $2$-entangled measurements (and thus $n$ qubits of additional quantum memory to store a single-copy of $\rho$ from the previous experiment) that solves this problem with $\mathcal{O} (n)$ copies.
We remark that this still leaves a large part of the landscape open, and we believe it is an interesting open question to fully characterize the sample complexity of this problem as a function of $k$.
However, this result is the first step towards understanding the tradeoffs between sample complexity and quantum memory in a more fine-grained fashion.

\subsubsection{Learning quantum dynamics}

We now turn our attention to the problem of learning the dynamics of a quantum system.
Here, we consider an unknown physical process whose properties we wish to understand.
We may perform experiments by interacting with this physical process, and the goal is to learn the desired property of the process with as few experiments as possible.
More formally, there is an unknown quantum channel $\mathcal{C}$ over $n$ qubit states.
We interact with this channel by passing a quantum state through it and measuring the resulting state.
Algorithms with quantum memory may store some quantum information about the result, and subsequent interactions may be entangled with previous outcomes.
In contrast, algorithms without quantum memory can only store classical information about the output, although as before, interactions can be chosen adaptively based on the outcomes of previous experiments.
In both settings, the goal is to learn some property of $\mathcal{C}$ while minimizing the number of experiments.

A very similar setting was considered in~\cite{aharonov2021quantum}, however, with one key qualitative difference: in~\cite{aharonov2021quantum}, they only consider algorithms without quantum memory that prepare $n$-qubit states $\rho$, pass them through the channel $\mathcal{C}$, and measure the resulting state.
For this restricted class of algorithms, they prove exponential separations between the power of algorithms with and without quantum memory.

In general however, an algorithm (even without quantum memory) could also prepare a state with $m > n$ qubits, pass part of it through $\mathcal{C}$, and perform some arbitrary measurement on the result.
Unfortunately, the proof technique in~\cite{aharonov2021quantum} breaks down even when $m = n + 1$.
In particular, the addition of one extra ancilla causes the bound to grow exponentially in the number of experiments.
In this work, we generalize and strengthen the lower bounds of~\cite{aharonov2021quantum} by removing this restriction on ancilla qubits, as well as quantitatively improving some of their bounds.
As a result, we demonstrate the first exponential separations for these tasks between \emph{general} algorithms with and without quantum memory.

\paragraph{Distinguishing the completely depolarizing channel} The first task we consider is testing whether or not $\mathcal{C}$ is the completely depolarizing channel, that is, the channel which sends every state to the maximally mixed state.
Understanding the depolarization of a given quantum channel is a well-studied problem, and is an important component of quantum error mitigation, see e.g.~\cite{temme2017error} and references within.

We consider the following simple case, previously studied in~\cite{brandao2019models,aharonov2021quantum}: distinguish between the case where $\mathcal{C}$ is the completely depolarizing channel, and the case where $\mathcal{C}$ is a random unitary channel, that is,  $\mathcal{C} [\rho] = U \rho U^\dagger$ for some Haar random unitary matrix $U$.
Even for this very simple case, we demonstrate that algorithms without quantum memory require exponentially more copies of $\mathcal{C}$ than those with quantum memory.
Specifically, we show:
\begin{theorem}[Depolarizing vs. unitary channel---informal, see Theorem~\ref{thm:channelhard1}]
\label{thm:channelhard1-inf}
Any algorithm without quantum memory that can distinguish between $\mathcal{C}$ being the completely depolarizing channel and $\mathcal{C}$ being a Haar random unitary channel requires $\Omega (2^{n / 3})$ experiments.
\end{theorem}
As mentioned earlier, this quantitatively improves the lower bound given in~\cite{aharonov2021quantum}, and also applies against all algorithms without quantum memory; in contrast, the lower bound of~\cite{aharonov2021quantum} only applied to algorithms that did not use ancilla qubits.
Combining this bound with the upper bound in~\cite{aharonov2021quantum}, who demonstrated an algorithm that uses quantum memory that only requires $\mathcal{O} (1)$ interactions with $\mathcal{C}$, yields the claimed exponential separation.

\paragraph{Distinguishing time-reversal symmetries} We next turn our attention to the problem of distinguishing between different types of symmetries in a physical evolution.
Determining symmetries is one of the most fundamental problems in experimental physics.  We will focus on the case of global symmetries, and in particular time-reversal symmetries.  Our question will be: does a unitary manifest time-reversal symmetry, and if so what kind?

More precisely, suppose we are given a quantum channel $\mathcal{C}$, and we are promised it is of the form $\mathcal{C} [\rho] = A \rho A^\dagger$, where $A$ is drawn from the Haar measure of either the class of unitary matrices, orthogonal matrices, or symplectic matrices.  Ordinary unitary matrices do not possess time-reversal symmetry, whereas orthogonal matrices and symplectic matrices realize distinct manifestations of time-reversal symmetry (i.e., orthogonal matrices are equal to their own transpose, and symplectic matrices are equal to their own symplectic transpose).  The goal is to distinguish which case we are in, with as few experiments as possible.
For this problem, there is an algorithm that uses quantum memory that succeeds with $\mathcal{O} (1)$ experiments, however,~\cite{aharonov2021quantum} previously showed that for the restricted class of algorithms without quantum memory they considered, $\Omega (2^{2n / 7})$ experiments were necessary.
We generalize this bound to the class of all algorithms without quantum memory:
\begin{theorem}[Distinguishing time-reversal symmetries---informal, see Theorems~\ref{thm:channelhard1},~\ref{thm:Ochannelhard1}, and~\ref{thm:Spchannelhard1}]
\label{thm:symmetries-inf}
Any algorithm without quantum memory that can distinguish between time-reversal symmetries with high probability requires $\Omega (2^{2n / 7})$ experiments.
\end{theorem}

\section{Technical Overview}
\label{sec:technical_overview}

As is common in the property testing literature, the starting point for our lower bounds is Le Cam's two-point method (see Lemma~\ref{lem:lecam}). We would like to exhibit a simple ``null hypothesis'' and a ``mixture of alternatives'' and argue that distinguishing between these two scenarios with nontrivial advantage requires a large number of quantum measurements. 

As a first running example, consider the task of shadow tomography with Pauli observables, where the goal is to estimate the expectation value of all Pauli observables to $\epsilon$-error.
The corresponding distinguishing task will be to decide whether the unknown state is the maximally mixed state $\rhomm$ or otherwise equal to $\rho_P\triangleq \frac{1}{2^n}(\Id + \epsilon P)$ for a uniformly random signed Pauli matrix $P\neq \Id$. Given a learning algorithm that makes $T$ measurements, let $q_0$ and $q_1 = \E[P]{q^P_1}$ denote the (classical) distributions over measurement outcomes induced by the algorithm in these two situations, where $q^P_1$ denotes the distribution over outcomes when the unknown state is $\rho_P$. To prove the $T = \Omega(2^n/\epsilon^2)$ lower bound in Theorem~\ref{thm:pauliobservables_informal}, it suffices to show that for $T = o(2^n/\epsilon^2)$, $\tvd(q_0,\E[P]{q^P_1}) = o(1)$ (see Section~\ref{sec:intro_distinguish}).

We emphasize that while this general approach of proving statistical lower bounds by showing estimates on the total variation between a single ``simple'' distribution and a mixture is standard, the general technique we use for proving these estimates is a departure from standard approaches. Indeed, whereas the latter typically first pass to chi-squared or KL divergence \cite{paninski2008coincidence,auer2002nonstochastic,bubeck2018entropic,bubeck2020entanglement,ingster2012nonparametric}, we work directly with total variation. We remark that this is in some sense necessary: for our Pauli observables lower bound for instance, passing to chi-squared divergence in the first step will result in a lower bound of at most $T = \Omega(\mathrm{poly}(n, 1/\epsilon))$.

One of the general techniques that we employ to bound total variation is to reduce to showing {\bf one-sided likelihood ratio bounds}, that is, to go through the following elementary fact (see also Lemma~\ref{lem:oneside2pt}).

\begin{fact}\label{fact:oneside}
    Let $0 \le \delta < 1$. Given probability distributions $p, q$ on a finite domain $\Omega$, if $p(x)/q(x) > 1 - \delta$ for all $x\in\Omega$, then $\tvd(p,q) \le \delta$.
\end{fact}

Fact~\ref{fact:oneside} tells us that to bound $\tvd(q_0,\E[P]{q^P_1})$, it suffices to show a \emph{uniform} lower bound on the likelihood ratio \begin{equation} \label{eq:TVleaf-basic}
    \E[P]{q_1^P(x)}/q_0(x).
\end{equation} 
In words, it suffices to show that no sequence of measurement outcomes from applying the learning algorithm to samples from the mixture of alternatives is too unlikely compared to applying it to samples from the null hypothesis.
While such an estimate need not hold in general, e.g. for our channel learning and bounded memory lower bounds which call for more involved techniques, it turns out to hold for all of our lower bounds on learning quantum states.
Furthermore, our lower bounds on learning quantum states all match known upper bounds up to a logarithmic factor.
Indeed, the fact that such an approach can yield tight or nearly tight lower bounds for these problems is quite surprising.




\begin{figure}[t]
    \centering
    \includegraphics[width=0.8\textwidth]{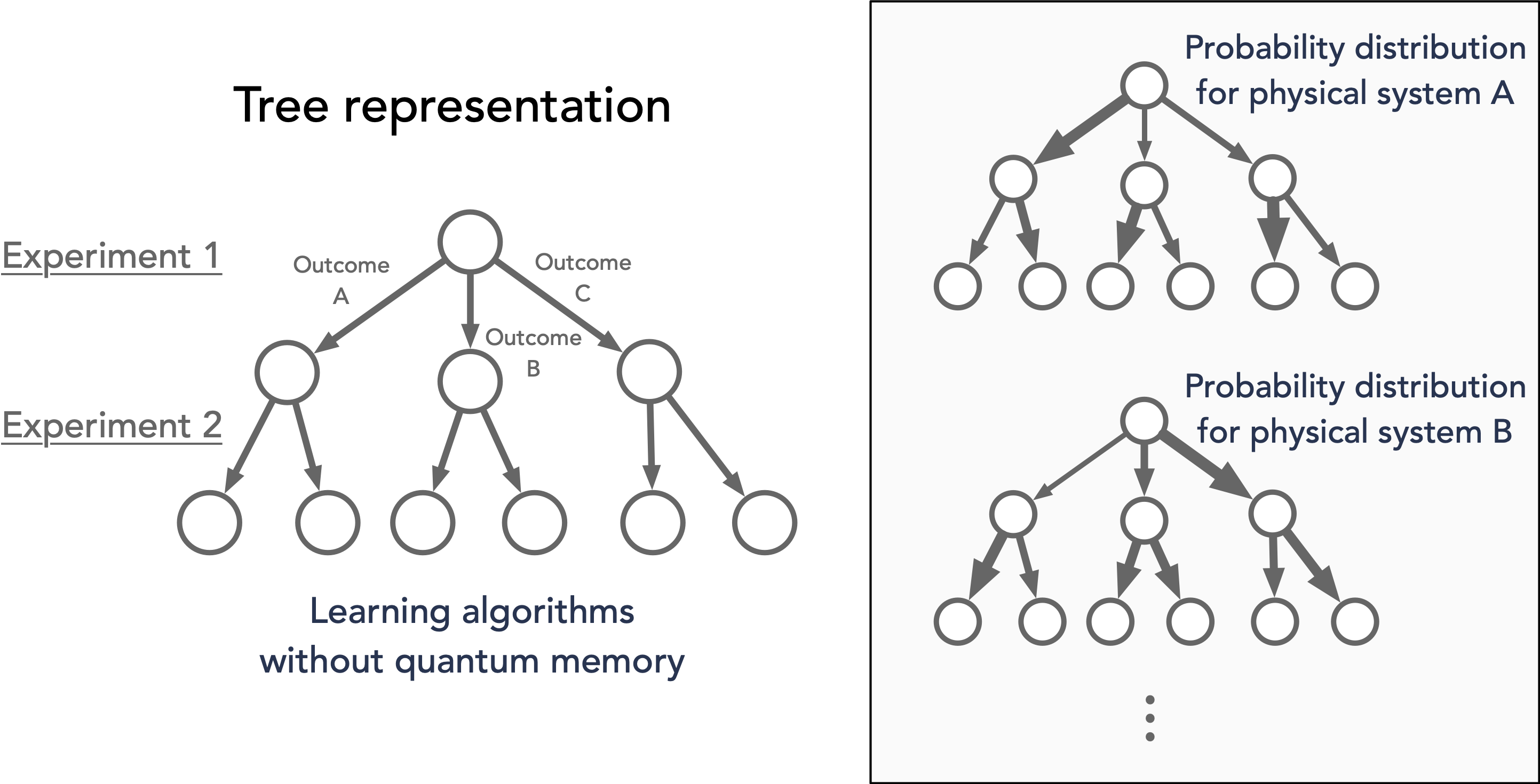}
    \caption{Illustration of the tree representation for learning algorithms without quantum memory. When the unknown physical system (or physical process) is different, the probability to traverse each edge will be different. Learning succeeds if for distinct physical systems, the distributions over leaf nodes are sufficiently different. }
    \label{fig:TreeRepre}
\end{figure}

\subsection{Tree representation} A primary technical hurdle to overcome which does not manifest in analogous classical settings is the fact that the quantum experiments that the learner runs may be chosen \emph{adaptively} based on the outcomes of earlier experiments. 
Such algorithms are naturally represented as a tree~\cite{bubeck2020entanglement,aharonov2021quantum,huang2021information}.
Formally, we associate to any such algorithm a rooted tree $\mathcal{T}$ in the following way. The root $r$ of the tree corresponds to the initial state of the algorithm.
At any node $u$, the algorithm performs an experiment.
Concretely, if the goal is to learn a quantum state, it makes a POVM measurement on a fresh copy of the state; alternatively, if the goal is to learn a quantum channel, it prepares an input quantum state, passes it through the channel, and makes a POVM measurement on the output.
After the experiment, the state of the algorithm moves to a child node of $u$ depending on the experimental outcome.
In this way, each node encodes all experimental outcomes that have been seen so far; correspondingly, the $T$-th layer of the rooted tree corresponds to all possible states of the algorithm after $T$ experiments.


The tree structure naturally encodes the adaptivity of the learning algorithm, as the choice of experiment that the algorithm performs at a given node can depend on all prior experimental outcomes.
Bounding the aforementioned total variation distance thus amounts to bounding the total variation distance between the induced distribution on leaves under the null hypothesis versus under the mixture of alternatives.

With this tree formalism in place, we can now describe our different techniques for bounding total variation, which roughly fall into two categories:

\begin{enumerate}[leftmargin=*]
    \item {\bf Edge-based}: Bound the information the learner can gain from traversing \emph{any} edge in the tree.
    \item {\bf Path-based}: Utilize multi-linear structure in the tree and high-order moments of the unknown states or channels to bound the information gain for traversing an entire path jointly.
\end{enumerate}

\begin{figure}[t]
    \centering
    \includegraphics[width=0.74\textwidth]{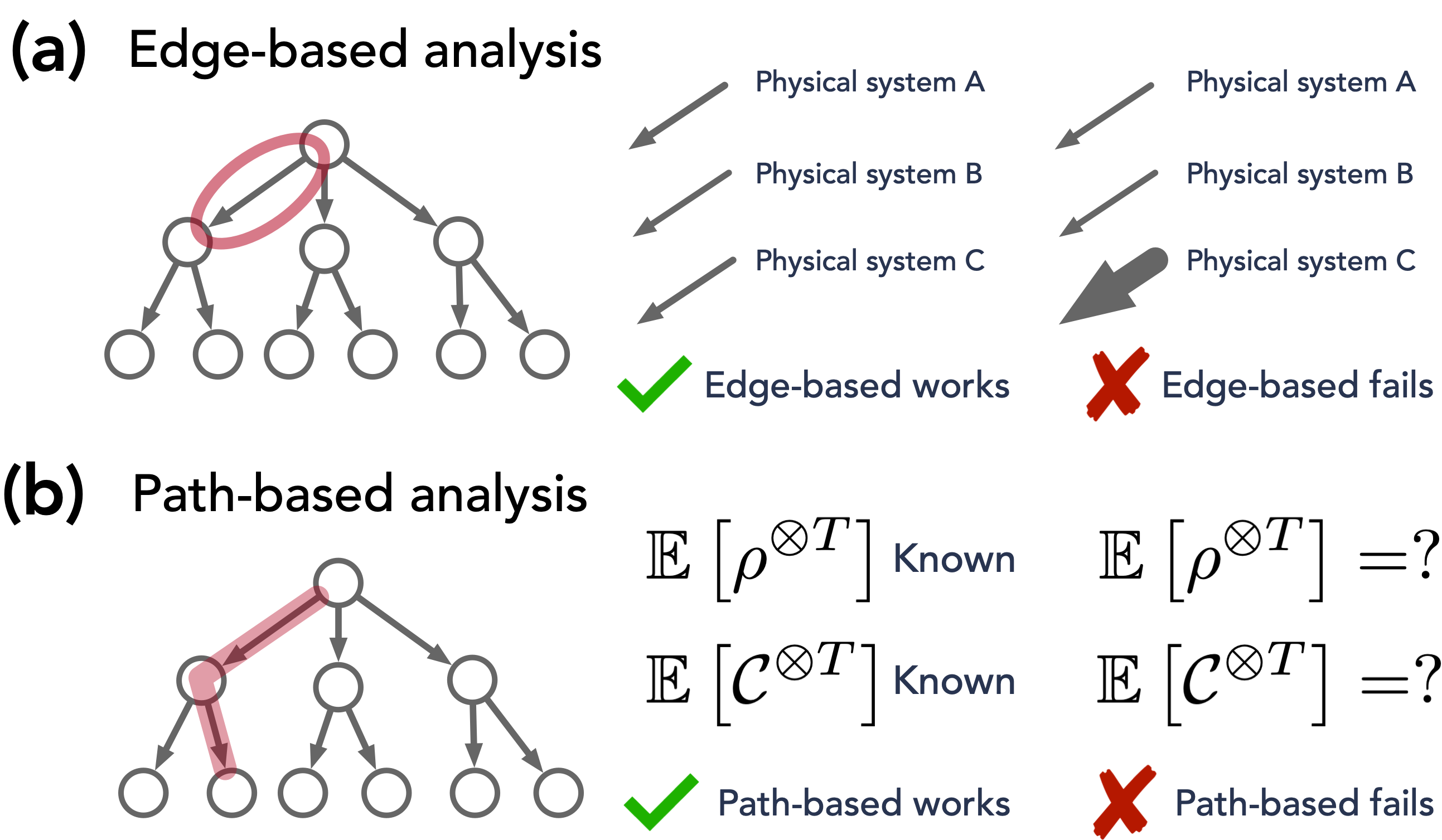}
    \caption{Illustration of edge-based and path-based analysis proving lower bounds for learning algorithms without quantum memory. Edge-based analysis works when the probability to traverse every edge does not differ too much across different unknown physical system (or process). Path-based analysis works when we have an adequate characterization for the higher moments for the distribution over unknown physical system (or process). }
    \label{fig:Edge-Path}
\end{figure}

\subsection{Edge-based analysis}
\label{subsec:edge_overview}

We formalize the notion of ``information'' gained from any given edge as follows. 
As an example, consider again the task of shadow tomography for Pauli observables, and suppose for simplicity of exposition that at every node $u$, the algorithm performs some projective measurement $\brc{\ketbra{\psi^u_s}{\psi^u_s}}_s$, where every $s$ corresponds to a child of $u$.\footnote{As we will show, the proceeding discussion extends to general POVMs.}
For each edge $e_{u, s}$, we can consider the following notion of {\bf average discrepancy} along an edge, namely the squared difference between the probability of traversing $e_{u,s}$ when the underlying state is $\rhomm$ versus when it is $\rho_P$, averaged over the mixture of alternatives $P$:
\begin{equation} \label{eq:avediffmmandP}
    \E[P]{\left( \bra{\psi^u_s} \rho_{\mathrm{mm}} \ket{\psi^u_s} - \bra{\psi^u_s} \rho_P \ket{\psi^u_s} \right)^2},
\end{equation}
where $\bra{\psi^u_s} \rho_{(\cdot)} \ket{\psi^u_s}$ is the probability for traversing edge $e_{u, s}$ when the unknown state is $\rho_{(\cdot)}$.

Given bounds on the average discrepancy in Eq.~\eqref{eq:avediffmmandP}, we can leverage a convexity argument to obtain a one-sided likelihood ratio bound (see the calculation beginning at Eq. \label{eq:convexity_start}).
The one-sided likelihood ratio bound would then give us the lower bound on $T$ through Fact~\ref{fact:oneside}.
A similar convexity argument appeared implicitly in \cite{bubeck2020entanglement}.
We generalize this argument, and in doing so, elucidate a new quantity $\delta(O_1, \ldots, O_M)$ (ses Eq.~\eqref{eq:highincomp}) for a collection of any $M$ observables $O_1, \ldots, O_M$ that, loosely speaking, measures how much information an algorithm without quantum memory can learn from a single measurement.
For instance, in the Pauli setting, we can upper bound $\delta$ by $\mathcal{O}(1/2^n)$ (see Lemma~\ref{lem:deltapauli}), yielding our near-optimal lower bound against algorithms without quantum memory for shadow tomography with Pauli observables. For shadow tomography with general observables, we obtain an analogous result, by applying the bound to a collection of Haar-random observables.


At this juncture, it is instructive to underscore how crucial it is that \emph{one-sided} bounds on the likelihood ratio suffice to bound total variation.
Indeed, even though a bound on \eqref{eq:avediffmmandP} implies that traversing any given edge is roughly equally likely on average under the null hypothesis as it is under the mixture of alternatives, it \emph{does not} imply that traversing any given \emph{path} from root to leaf $\ell$ is roughly equally likely.
That is, the likelihood ratio $\E[P]{p^{\rho_P}(\ell)}/p^{\rho_\mathrm{mm}}(\ell)$ could be quite large; in fact it could be much larger than 1, because the discrepancies introduced by the individual edges along the path could easily compound.
For instance, if in every experiment the learner simply measured in the computational basis, then for the root-to-leaf path corresponding to observing $\ket{0}$ after every measurement, \eqref{eq:TVleaf-basic} would be of order $(1+\epsilon)^T$. 
If we simply upper bounded total variation by the average absolute difference between \eqref{eq:TVleaf-basic} and 1, we would at best obtain a $T = \Omega(n/\epsilon)$ sample complexity lower bound.

\paragraph{Shadow tomography with $k$-qubit memory.} For proving lower bounds against learning algorithms with $k$ qubits of memory, we can employ a similar idea.
Whereas previously, we merely associated to each node the probability of reaching it, we now associate to it a $2^k \times 2^k$ positive-semidefinite (PSD) matrix representing the unnormalized quantum state of the $k$-qubit memory. Each edge can then be regarded as a completely-positive linear map on PSD matrices.
We can define and upper bound a notion of average discrepancy along an edge generalizing \eqref{eq:avediffmmandP} (see Lemma~\ref{fact:secondmoment_sizek}).

The main difference is that the aforementioned convexity argument for passing from an average discrepancy bound to a likelihood ratio bound no longer applies to give a uniform lower bound on the likelihood ratio.
Instead, we utilize a {\bf pruning argument}.
By Markov's, a bound on the average discrepancy along an edge implies that the discrepancy for a random Pauli $P$ along that edge is small with high probability over the choice of $P$ (see Definition~\ref{def:badgoodPauli} and Fact~\ref{fact:fewbad}).
In particular, for any leaf $\ell$ in the tree, the fraction of Paulis for which the discrepancy is small for every edge in the path from root to $\ell$ is also large by a union bound.
So for any leaf $\ell$, if we decompose the likelihood ratio \eqref{eq:TVleaf-basic} into the contribution from the Paulis with small discrepancy and large discrepancy.
Those of the latter type are few enough that we can naively prune them from consideration (Lemma~\ref{lem:goodbad}).
The contribution from those of the former type, averaged across leaves, can be related to the total variation distance at depth $T - 1$.
Proceeding by induction we can obtain the desired sample complexity lower bound.

\subsection{Path-based analysis}
\label{subsec:path_overview}

\paragraph{From average discrepancy to multi-linear structure.} The edge-based analysis fails when there is an extensive number of edges where the average discrepancy is sizable. 
As a warmup example, consider the problem of purity testing, where the null hypothesis is that $\rho$ is maximally mixed, and the mixture of alternatives is that $\rho$ is a random pure state. As above, suppose for simplicity of exposition that the quantum experiments the learner runs are all projective measurements $\brc{\ketbra{\psi^u_s}{\psi^u_s}}_s$. Then for all edges $e_{u, s}$ the average discrepancy
$\E[\ket{\phi}]{\left( \bra{\psi^u_s} \rho_{\mathrm{mm}} \ket{\psi^u_s} - \bra{\psi^u_s} \ketbra{\phi}{\phi} \ket{\psi^u_s} \right)^2}$
between the completely mixed state $\rho_{\mathrm{mm}}$ and a random pure state $\ketbra{\phi}{\phi}$ is of order $\mathcal{O}(1/2^{2n})$.
Using the aforementioned convexity argument, this would only yield a trivial lower bound of $\Omega(1)$.

In order to establish a nontrivial lower bound, we consider an analysis that better utilize structure in the tree representation of the learning algorithm.
The central idea is to exploit the \textbf{multi-linear structure} inherent in the tree representation.
We illustrate this idea using the purity testing problem.
For a given leaf $\ell$, suppose the edges on the path from root to $\ell$ are $e_{u_0 = r, s_0}, e_{u_1, s_1}, \ldots, e_{u_{T-1}, s_{T-1}}$, where $e_{u_{T-1}, s_{T-1}}$ connects $u_{T-1}$ and $\ell$.  The probability associated with the leaf $\ell$ under an unknown quantum state $\rho$ is given by
\begin{equation}
    p^{\rho}(\ell) = \prod_{t=0}^{T-1} (\bra{\psi^{u_t}_{s_t}} \rho \ket{\psi^{u_t}_{s_t}}) = \Tr\left( \left(\bigotimes_{t=0}^{T-1} \ketbra{\psi^{u_t}_{s_t}}{\psi^{u_t}_{s_t}}\right) \rho^{\otimes T} \right). \label{eq:path_prob}
\end{equation}
Under the tree representation, the measurement outcomes along the path yield a tensor product of pure states $\bigotimes_{t=0}^{T-1} \ketbra{\psi^{u_t}_{s_t}}{\psi^{u_t}_{s_t}}$ that depends on the learning algorithm. The key point is that the probability \eqref{eq:path_prob} of traversing this path is linear in $\ketbra{\psi^u_s}{\psi^u_s}$ for every edge $e_{u,s}$ on the path, and also linear in $\rho^{\otimes T}$.

How does $p^{\rho}(\ell)$ behave under averaging of $\rho$ across a mixture of alternatives? In our purity testing and channel learning lower bounds, the mixture of alternatives is an average over the Haar measure of some Lie group, so we can readily compute $\E[\rho]{p^{\rho}(\ell)}$ by computing the high-order moment $\E[\rho]{\rho^{\otimes T}}$.

\paragraph{Purity testing.} We first instantiate this for purity testing. When $\rho$ is a random pure state $\ketbra{\phi}{\phi}$, then it is well-known \cite{christandl2006structure,harrow2013church} that $\E{\rho^{\otimes T}}$ is proportional to the projector $\Pi$ onto the symmetric subspace in $(\mathbb{C}^{2^n\times 2^n})^{\otimes T}$ (see Lemma~\ref{lem:churchsymsubspace}). Specifically, the probability for a leaf $\ell$ is exactly $\prod^{T-1}_{t=0}\left(1 - t/2^n\right)^{-1}\cdot\Tr\left(\Pi \bigotimes_{t=0}^{T-1} \ketbra{\psi^{u_t}_{s_t}}{\psi^{u_t}_{s_t}}\right)$. The key inequality we show is that \begin{equation}
    \Tr\left(\Pi \bigotimes_{t=0}^{T-1} \ketbra{\psi^{u_t}_{s_t}}{\psi^{u_t}_{s_t}}\right) \ge 1, \label{eq:maincombo_preview}
\end{equation}
i.e. the squared norm of the projection of $\bigotimes_{t=0}^{T-1} \ketbra{\psi^{u_t}_{s_t}}{\psi^{u_t}_{s_t}}$ onto the symmetric subspace, is always at least 1 (see Lemma~\ref{lem:maincombo}). The upshot is that we obtain a uniform one-sided lower bound on the likelihood ratio of $\prod^{T-1}_{t=0}\left(1 - t/2^n\right)^{-1}$, and for $T = o(2^{n/2})$ this quantity is $1 - o(1)$ as desired.

To get intuition for this one-sided lower bound, imagine the projective measurements $\brc{\ketbra{\psi^u_s}{\psi^u_s}}$ were all in the same basis. Then the sequence of outcomes which would be most unlikely for the mixture of alternatives (random pure state) relative to the null hypothesis (maximally mixed state) would be one in which all $\ket{\psi^u_s}$ along the path were mutually orthogonal. 
Note that in such a case, we actually have $\Tr\left(\Pi \bigotimes_{t=0}^{T-1} \ketbra{\psi^{u_t}_{s_t}}{\psi^{u_t}_{s_t}}\right) = 1$.
Moreover, the quantity $\prod^{T-1}_{t=0}\left(1 - t/2^n\right)$ is the probability that all $T$ outcomes from measuring $\rho = \rhomm$ are distinct. Equivalently, this is the probability of seeing a collision among $T$ draws from the uniform distribution $[2^n]$, and this is exactly why we obtain an $\Omega(2^{n/2})$ lower bound for purity testing. We remark that extending this intuition to a formal proof of Eq.~\eqref{eq:maincombo_preview} is nontrivial and requires a careful tensor analysis that crucially exploits the multi-linear structure afforded by the tree representation.


\paragraph{Extending to channel learning.} The multi-linear structure also plays a crucial role for establishing lower bounds for channel learning tasks.
Consider the task of distinguishing between whether the unknown channel $\mathcal{C}$ is a completely depolarizing channel $\mathcal{D}$ or a random channel from some distribution.
Each edge $e_{u, s}$ in the tree representation of an algorithm (without quantum memory) for learning quantum channels is characterized by an $(n+n')$-qubit input state $\ket{\phi^u}$ and an $(n+n')$-qubit measurement outcome $\ket{\psi^u_s}$.
Again for simplicity, we only consider projective measurements in the technical overview. 

Similar to purity testing, we consider a sequence of edges $\brc{e_{u_t,s_t}}$ on a path from root to leaf~$\ell$.
The probability that we end up at $\ell$ when we employ the learning algorithm on an unknown quantum channel $\mathcal{C}$ can be shown to be
\begin{equation}
    p^{\mathcal{C}}(\ell) = \prod_{t=0}^{T-1} \bra{\psi^{u_t}_{s_t}} (\mathcal{C} \otimes \Id)(\ketbra{\phi^{u_t}}{\phi^{u_t}}) \ket{\psi^{u_t}_{s_t}} = \Tr\left( \left(\bigotimes_{t=0}^{T-1} \ketbra{\psi^{u_t}_{s_t}}{\psi^{u_t}_{s_t}} \right) \left(\bigotimes_{t=0}^{T-1} (\mathcal{C} \otimes \Id)(\ketbra{\phi^{u_t}}{\phi^{u_t}}) \right) \right).
\end{equation}
Hence, the average probability $\E[\mathcal{C}]{p^{\mathcal{C}}(\ell)}$ would be a multi-linear function in $\E[\mathcal{C}]{\bigotimes_{t=0}^{T-1} (\mathcal{C} \otimes \Id)}$, the $T$-th moment of the unknown quantum channel $\mathcal{C}$, and elements $\ket{\phi^u}, \ket{\psi^u_s}$ in the algorithm.

The main complication in channel learning tasks is that it becomes much more challenging to obtain a uniform one-sided likelihood ratio bound.
Instead, we explicitly expand the $T$-th moment of the unknown quantum channel $\mathcal{C}$ into a linear combination of basic tensors, e.g. permutation operators, that arise from a suitable Weingarten expansion of the integral over $\mathcal{C}$.
The total variation distance can be upper bounded by a sum over absolute values of multi-linear functions,
\begin{equation}
    \sum_{\ell: \mathrm{leaves}} |p^{\mathcal{D}}(\ell) - \E[\mathcal{C}]{p^{\mathcal{C}}(\ell)}| \leq \sum_{\{u_t, s_t\}_{t=0}^{T-1}} \sum_A \Bigg| \underbrace{\Tr\left(A \bigotimes_{t=0}^{T-1} (\ketbra{\phi^{u_t}}{\phi^{u_t}} \otimes \ketbra{\psi^{u_t}_{s_t}}{\psi^{u_t}_{s_t}}) \right)}_{\text{Multi-linear function}} \Bigg|.
\end{equation}
The sum over $A$ depends on the $T$-th moment of the unknown channel $\E[\mathcal{C}]{(\mathcal{C} \otimes \Id)^{\otimes T}}$.
For each multi-linear function, the interactions between the tensor components $\ket{\phi^{u_t}}, \ket{\psi^{u_t}_{s_t}}$ can be rather complex. Hence a central tool in our analysis is to utilize tensor network diagrams \cite{orus2019tensor} to bound and disentangle the relations between each components.

The goal of the tensor network manipulation is to upper bound the total variation distance by
a sum over multi-linear functions without the absolute values,
\begin{equation}
    \sum_{\{u_t, s_t\}_{t=0}^{T-1}} \sum_B \Tr\left(B \bigotimes_{t=0}^{T-1} (\ketbra{\phi^{u_t}}{\phi^{u_t}} \otimes \ketbra{\psi^{u_t}_{s_t}}{\psi^{u_t}_{s_t}}) \right). \label{eq:noabs}
\end{equation}
Then we can utilize the resolution of identity $\sum_{s} \ketbra{\psi^{u}_{s}}{\psi^{u}_{s}} = \Id$ at the bottom layer of the tree to collapse the summation over all leaves (nodes at depth $T$) to summation over nodes at depth $T-1$. In other words, we will upper bound $\Tr(B \left( X \otimes \ketbra{\phi^{u_{T-1}}}{\phi^{u_{T-1}}} \otimes \Id \right)) \leq \Tr(B' X)$ for some tensor $B'$, which lets us upper bound \eqref{eq:noabs} by
\begin{equation}
    \sum_{\{u_t, s_t\}_{t=0}^{T-2}} \sum_{B'} \Tr\left(B' \bigotimes_{t=0}^{T-2} (\ketbra{\phi^{u_t}}{\phi^{u_t}} \otimes \ketbra{\psi^{u_t}_{s_t}}{\psi^{u_t}_{s_t}}) \right).
\end{equation}
We inductively perform this collapsing procedure to reduce the summation over all paths on the tree to a single value and conclude the lower bound proof. 


\section{Related Work}

The field of learning and testing properties of quantum systems is broad and a full review of the literature is out of scope for this paper.
For conciseness, here we only review the most relevant lines of work.
See e.g.~\cite{montanaro2013survey} for a more complete survey. 

\paragraph{Learning properties of quantum states} When it comes to learning properties of quantum states, arguably the most canonical problem is quantum tomography, which is the problem of learning the unknown quantum state $\rho$ to high accuracy, usually in trace norm or fidelity.
See e.g.~\cite{hradil1997quantum,gross2010quantum,blume2010optimal,banaszek2013focus} and references within.
It was not until recently that the sample complexity of this problem was fully characterized~\cite{haah2017sample,o2016efficient}.
These papers demonstrated that $\Theta( 2^{2n} )$ samples are necessary and sufficient for this problem.
In particular, the sample complexity is exponential in $n$.

As a way around this barrier,~\cite{aaronson2018shadow} proposed the shadow tomography problem.
Ever since, there has been an active line of work improving the sample complexity upper bounds for this problem~\cite{brandao2017quantum,aaronson2019gentle,huang2020predicting,buadescu2020improved,huang2021information}. However, the algorithms presented in these papers largely require heavily entangled measurements, and thus a great deal of quantum memory.
Some notable exceptions are~\cite{huang2020predicting, huang2021information}, who gave algorithms for shadow tomography that do not use quantum memory that match the lower bound we achieve in the general case, and improved algorithms for shadow tomography in special cases.
The special case that we consider in this paper where the observables are Pauli matrices is also of independent interest, as it is a useful subroutine in many quantum computing applications, see discussions in \cite{mcclean2016theory, kandala2017hardware, paini2019approximate, yen2019measuring, gokhale2019minimizing, huggins2019efficient, cotler2020quantum, crawford2020efficient, huang2021efficient, huang2021provably}.

In quantum property testing, besides purity testing, there have been a number of well-studied problems, including quantum state discrimination~\cite{chefles2000quantum,audenaert2008asymptotic,barnett2009quantum}, mixedness testing~\cite{o2015quantum,bubeck2020entanglement} and more generally spectrum testing~\cite{o2015quantum} and state certification~\cite{buadescu2019quantum,chen2021toward}.
However, for these problems, while in some settings there are separations between the power of algorithms with and without quantum memory, these settings inherently limit the separations to be at most polynomial, rather than exponential.

\paragraph{Quantum memory tradeoffs for learning} Understanding the power of algorithms without quantum memory has been posed as an open question in a number of learning theoretic settings, such as shadow tomography~\cite{aaronson2018shadow}, as well as spectrum testing and general state tomography~\cite{wright2016learn}.
However, until recently, lower bounds against the power of algorithms without quantum memory usually only applied to the non-adaptive setting, where the algorithm had to commit to the set of measurements ahead of time, see e.g.~\cite{haah2017sample,chen2021toward}.
The first work which demonstrated a separation against general algorithms was~\cite{bubeck2020entanglement}, who showed a polynomial gap for mixedness testing and quantum identity testing.
Subsequently, the aforementioned works of~\cite{huang2020predicting,aharonov2021quantum} demonstrated exponential separations for shadow tomography and channel learning.
In many ways this work can be thought of as a synthesis of the ideas in~\cite{bubeck2020entanglement} with the ones developed in~\cite{huang2020predicting,aharonov2021quantum} to both tighten and generalize the bounds obtained by the latter.
We also note that memory tradeoffs have also been demonstrated for classical learning problems, e.g.~\cite{steinhardt2016memory,raz2018fast,raz2017time,kol2017time,moshkovitz2017mixing,moshkovitz2018entropy,sharan2019memory}.
It would be interesting to see if there are any technical connections between these two lines of inquiry.

\paragraph{Learning quantum dynamics} Similar to learning quantum states, the task of learning a full description of a quantum process/channel is known as quantum process tomography \cite{mohseni2008quantum}. Due to the exceedingly high sample complexity, most of the recent works have focused on learning restricted classes of dynamics. A large class of algorithms assume the dynamics to be governed by a time-independent Hamiltonian and the goal is to reconstruct the Hamiltonian \cite{wiebe2014hamiltonian, wang2017experimental, evans2019scalable}. Another class of quantum channels that have been subject to active study is Pauli channels \cite{harper2020efficient, flammia2020efficient, flammia2021pauli}, which are useful for characterizing noise in a quantum device.
More recently, \cite{huang2021information, aharonov2021quantum, rossi2021quantum, chen2021quantum} studied the task of learning properties of quantum channels using information-theoretic bounds.



\section{Preliminaries}


Given a (possibly unnormalized) state $\sigma$ on $m \ge k$ qubits, let $\Tr_{>k}(\sigma)$ denote the state on $k$ qubits obtained by tracing out the last $m - k$ qubits. We denote the maximally mixed state on $n$ qubits by $\rhomm\in\co^{2^n\times 2^n}$. Given signed Pauli matrix $P\in \co^{2^n\times 2^n}$, define $\rho_P \triangleq (\Id + P)/2^n$.
We will also use the small-o notation $f(n) = o(g(n))$ if for all $c > 0$ there exists a $k > 0$, such that $0 \leq f(n) < c g(n), \forall n \geq k$.
A fact we will often use in the proof is the following: if $g(n) = o(h(n))$ implies $f(n) = o(1)$, then $f(n) = \Omega(1)$ implies $g(n) = \Omega(h(n))$.
Given a matrix $M$, we will use $M^{\dagger}$ to denote conjugate transpose and $M^t$ to denote transpose.
Given probability distributions $p,q$ over some domain $\Omega$, define the \emph{total variation distance} $\tvd(p,q)\triangleq \sup_{S\subseteq\Omega} p(S) - q(S)$.

\subsection{Tail bounds}

We also recall the following properties for subexponential random variables. A mean-zero random variable $Z$ is $\lambda$-subexponential if it satisfies the tail bound $\Pr{|Z| > t} \le 2e^{-2t/\lambda}$.

\begin{fact}[See e.g. Lemma 2.7.6 in \cite{vershynin2018high}]\label{fact:square_subgauss}
    If $X$ is sub-Gaussian with variance proxy $\sigma^2$, then $Z\triangleq X^2 - \E{X^2}$ is $O(\sigma^2)$-subexponential.
\end{fact}

\begin{fact}[Bernstein's, see e.g. Corollary 2.8.3 in \cite{vershynin2018high}]\label{fact:sum_subexp}
    If $Z_1,\ldots,Z_m$ are independent, mean-zero, $\lambda$-subexponential random variables, then for $Z\triangleq \frac{1}{m}\sum_i Z_i$ and any $t > 0$, \begin{equation}
        \Pr{\abs{Z} > t} \le \exp\left(-\frac{m}{2}\left(\min\left(\frac{t^2}{\lambda^2}, \frac{t}{\lambda} \right)\right)\right).
    \end{equation}
\end{fact}

We will also use concentration of measure for the Haar measures on the orthogonal and unitary groups:

\begin{lemma}[See e.g. \cite{anderson2010introduction}, Corollary 4.4.28]\label{lem:levy}
    For $G = O(d), U(d)$, let $f: G\to\R$ be $L$-Lipschitz with respect to the Frobenius norm. There is an absolute constant $c > 0$ such that for $x$ sampled from the Haar measure on $G$, $\Pr{|f(x) - \E{f(x)}| > c \cdot L\sqrt{\log(1/\delta)/d}} \le \delta$.
\end{lemma}

\subsection{Basic results in quantum information theory}

In the following, we present several standard definitions in quantum information theory. We also provide the basic lemmas that illustrate properties of these quantum objects.

\begin{definition}[Quantum channel]
    A quantum channel $\mathcal{C}$ from $n$-qubit to $m$-qubit is a linear operator from $\mathbb{H}^{2^n \times 2^n}$ to $\mathbb{H}^{2^m \times 2^m}$ given by
    \begin{equation}
        \mathcal{C}(X) = \sum_{i} K_i X K_i^\dagger,
    \end{equation}
    where $K_i \in \mathbb{H}^{2^m \times 2^n}$ satisfies $\sum_i K_i^\dagger K_i = \mathds{1}$.
    When $n = m$, we let $\mathcal{I}$ denote the \emph{identity channel}, that is, the channel that acts trivially on $\mathbb{H}^{2^n\times 2^n}$.
\end{definition}

\begin{definition}[POVMs]
\label{def:povm_basic}
    A positive operator-valued measure (POVM) on $n$-qubit states is given by positive-semidefinite matrices $\brc{F_s}_s$, where the set of matrices $F_s\in\mathbb{C}^{2^n\times 2^n}$ satisfies $\sum_s F_s = \mathds{1}$.
    The probability for obtaining classical outcome $s$ when we measure the density matrix $\rho\in\mathbb{H}^{2^n\times 2^n}$ is $\Tr(F_s \rho)$.
\end{definition}

In Section~\ref{sec:bounded}, we will need to reason about how states transform upon being measured; we summarize the formalism for this in the following definition:

\begin{definition}[Post-measurement states] \label{eq:POVMpostmeasurement}
    Given a POVM on $n$-qubit states $\brc{F_s}_s$ where each POVM element has some Cholesky decomposition $F_s = M_s^{\dagger} M_s$ for $M_s\in\mathbb{C}^{2^{n+m} \times 2^n}$, the post-measurement quantum state upon measuring $\rho$ with this POVM and observing outcome $s$ is given by
    \begin{equation}
        \frac{M_s \rho M_s^\dagger}{\Tr(M_s^\dagger M_s \rho)} \in \mathbb{H}^{2^{n+m} \times 2^{n+m}}.
    \end{equation} 
\end{definition}

We also define a restricted version of POVM where each POVM element is a rank-$1$ matrix.

\begin{definition}[Rank-$1$ POVM]
    A rank-$1$ positive operator-valued measure (POVM) on $n$-qubit state is given by $\{\sqrt{w_s 2^n} \ketbra{\psi_s}{\psi_s}\}_s$, where $w_s \geq 0, w_s \in \mathbb{R}$ and pure states $\{\ket{\psi_s}\}_s$ satisfy
    \begin{equation} \label{eq:rank1POVMnorm}
        \sum_s w_s 2^n \ketbra{\psi_s}{\psi_s} = \mathds{1}.
    \end{equation}
    The above normalization condition implies $\sum_s w_s = 1$.
    The probability for obtaining the classical outcome $s$ when we measure the density matrix $\rho \in \mathbb{H}^{2^{n} \times 2^n}$ is $w_s 2^n \bra{\psi_s} \rho \ket{\psi_s}$.
    And the post-measurement quantum state is given by $\ketbra{\psi_s}{\psi_s}.$
\end{definition}

A nice property of rank-$1$ POVM is that it can be used to simulate arbitrary POVM. Hence, rank-$1$ POVM is as powerful as any POVM from an information-theoretic perspective. The following folklore lemma illustrates this fact. We provide a proof of this lemma for completeness.

\begin{lemma}[Simulating POVM with Rank-$1$ POVM] \label{lem:simPOVMwithr1POVm}
When we only consider the classical outcome of the POVM measurement and neglect the post-measurement quantum state, then an POVM $\{F_s\}_s$ on $n$-qubit states can be simulated by a rank-$1$ POVM on $n$-qubit states with some post-processing.
\end{lemma}
\begin{proof}
Let $F_s$ have Cholesky decomposition $M_s^{\dagger}M_s$. 
Consider the diagonalization
\begin{equation}
M_s^\dagger M_s = \sum_{p} \tilde{w}_{sp} \ketbra{\psi_{sp}}{\psi_{sp}} = \sum_{p} w_{sp} 2^n \ketbra{\psi_{sp}}{\psi_{sp}},
\end{equation}
where $w_{sp} = \tilde{w}_{sp} / 2^n$.
Because $M_s^\dagger M_s$ is positive-semidefinite, we have $w_{sp} \geq 0$.
$\{ \sqrt{w_{sp} 2^n} \ketbra{\psi_{sp}}{\psi_{sp}} \}_{sp}$ forms a rank-$1$ POVM because $\sum_{sp} w_{sp} 2^n \ketbra{\psi_{sp}}{\psi_{sp}} = \sum_{s} M_s^\dagger M_s = \mathds{1}$.
To simulate the POVM $\{F_s\}_s$, we measure the rank-$1$ POVM on $\rho$, which gives the classical outcome $sp$ with probability
\begin{equation}
    w_{sp} 2^n \bra{\psi_{sp}} \rho \ket{\psi_{sp}} = \Tr( w_{sp} 2^n \ketbra{\psi_{sp}}{\psi_{sp}} \rho ),
\end{equation}
then output $s$. Hence, we output classical outcome $s$ with probability
\begin{equation}
    \sum_{p} \Tr( w_{sp} 2^n \ketbra{\psi_{sp}}{\psi_{sp}} \rho ) = \Tr(M_s^\dagger M_s \rho).
\end{equation}
Thus the rank-$1$ POVM with some post-processing simulates the POVM $\{M_s\}_s$.
\end{proof}

\begin{definition}[Pauli observables] \label{def:pauli}
    We define an $n$-qubit Pauli observable $P$ to be a tensor product of $n$ operators from the set $\left\{ I = \begin{pmatrix}
    1 & 0 \\
    0 & 1
    \end{pmatrix}, X = \begin{pmatrix}
    0 & 1 \\
    1 & 0
    \end{pmatrix}, Y =
    \begin{pmatrix}
    0 & -\mathrm{i} \\
    \mathrm{i} & 0
    \end{pmatrix}, Z =
    \begin{pmatrix}
    1 & 0 \\
    0 & -1
    \end{pmatrix}\right\}$.
\end{definition}

\begin{lemma}\label{lem:2design}
The sum of the tensor product of two Pauli observables is given by
\begin{equation}
    \sum_{P \in \{I, X, Y, Z\}^{\otimes n}} P \otimes P = 2^n \mathrm{SWAP}_n,
\end{equation}
where $\mathrm{SWAP}_n$ is the swap operator on two copies of $n$-qubit state.
\end{lemma}
\begin{proof}
It is easy to check the following equality,
\begin{equation}
    I\otimes I + X \otimes X + Y \otimes Y + Z \otimes Z = \begin{pmatrix}
    2 & 0 & 0 & 0 \\
    0 & 0 & 2 & 0 \\
    0 & 2 & 0 & 0 \\
    0 & 0 & 0 & 2
    \end{pmatrix} = 2 \mathrm{SWAP}_1.
\end{equation}
Then note that $\sum_{P \in \{I, X, Y, Z\}^{\otimes n}} P \otimes P = \left(\sum_{\sigma \in \{I, X, Y, Z\}} \sigma \otimes \sigma\right)^{\otimes n} = 2^n \mathrm{SWAP}_1^{\otimes n} = 2^n \mathrm{SWAP}_n$.
\end{proof}

In the following, we give basic definition for permutation operators over a tensor product space. In particular, we will use the same notation $\pi$ for an element in the symmetric group as well as the permutation operator over the tensor product space.

\begin{definition}[Permutation operators]
For $T > 0$, we consider $S_T$ to be the symmetric group of degree $T$. For any permutation $\pi \in S_T$, we consider the action of $\pi$ on a tensor product of $T$ states to be given by the following
\begin{equation}
    \pi (\ket{\psi_1} \otimes \ldots \otimes \ket{\psi_T}) = \ket{\psi_{\pi^{-1}(1)}} \otimes \ldots \otimes \ket{\psi_{\pi^{-1}(T)}}, \forall \psi_1, \ldots, \psi_T.
\end{equation}
We linearly extend the action of $\pi$ to any element in the tensor product space.
\end{definition}

\begin{lemma}[Haar integration over states, see e.g. \cite{harrow2013church}] \label{lem:churchsymsubspace}
Consider the uniform (Haar) measure over $n$-qubit pure states $\ket{\psi}$, then
\begin{equation}
    \Expect_{\ket{\psi}}[ \ketbra{\psi}{\psi}^{\otimes T} ] = \binom{2^n + T - 1}{T}^{-1} \sum_{\pi \in S_T} \pi,
\end{equation}
where $\pi$ is a permutation operator on a tensor product space of $T$ $n$-qubit pure states and $S_T$ is the symmetric group of degree $T$.
\end{lemma}

\subsection{Primer on information-theoretic lower bounds}
\label{sec:intro_distinguish}

In this section, we give a brief overview for how to prove information-theoretic lower bounds like the ones in this work.

As discussed in Section~\ref{sec:technical_overview}, we are interested in \emph{distinguishing tasks}. Informally, we have the ability to make some number $N$ of measurements of an unknown object $z$ which is promised to lie in one of two possible sets $\calS_0$ and $\calS_1$. Based on the sequence of measurement outcomes we obtain, we must decide whether the object lies in $\calS_0$ or $\calS_1$. \emph{In this work we will focus on tasks where $\calS_0$ is a singleton set.}

We will not formally define what it means to follow a particular ``strategy'' for measuring $z$, as this will be clear in the body of this paper (see e.g. Definitions~\ref{def:treelearnstate}, \ref{def:treechannel}, and \ref{def:treebounded}). In any case, this is not important for the formalism in the present subsection; the only thing one needs to know here is that any $z$ and measurement strategy induce a probability distribution over \emph{transcripts}, namely sequences of measurement outcomes. And as we are interested in showing lower bounds against arbitrary measurement strategies, in the rest of this subsection we may as well fix an arbitrary such strategy.

Under this strategy, to any $z$ is associated a distribution $p^{\le N}_z$ over transcripts\--- when $z$ is the unique element of $\calS_0$, we will denote this by $p^{\le N}_0$. To show a lower bound for the distinguishing task, formally we would like to argue that there exists no algorithm $\calA$ mapping transcripts $T$ to $\brc{0,1}$ for which $\Pr[T\sim p^{\le N}_z]{\calA(T) = i} \ge 2/3$ for all $z\in\calS_i$ and $i = 0,1$ (the constant 2/3 is arbitrary and we could consider any value greater than 1/2).

To show this, it suffices to consider the following \emph{average-case} version of the original distinguishing task. Let $\calD$ be some distribution over $\calS_1$ that we are free to choose, and consider the \emph{hypothesis testing} problem of distinguishing whether a given transcript came from $p^{\le N}_0$ or $\E[z\sim\calD]*{p^{\le N}_z}$. The following fact is an elementary but fundamental result in binary hypothesis testing:
\begin{fact}\label{fact:hypo_test}
    Given distributions $q_0, q_1$ over a domain $\calS$, if $\tvd(q_0,q_1) < 1/3$, there is no algorithm $\calA: \calS\to\brc{0,1}$ for which $\Pr[x\sim q_i]{\calA(x) = i} \ge 2/3$ for both $i = 0,1$.
\end{fact}

\begin{proof}
    Let $\calS'\subseteq\calS$ denote the set of elements $x$ for which $\calA(x) = 0$. Then observe that
    \begin{align}
        \Pr[x\sim q_0]{\calA(x) = 1} + \Pr[x\sim q_1]{\calA(x) = 0} &= 1 - q_0(\calS') + q_1(\calS') \\
        &\ge 1 - \sup_{\calS''\subseteq\calS}\abs*{q_0(\calS'') - q_1(\calS'')} \\
        &= 1 - \tvd(q_0,q_1) \ge 2/3,
    \end{align} so at least one of the terms on the left-hand side is at least $1/3$.
\end{proof}

We can use this to show that, in order to prove a lower bound for the original distinguishing task, it suffices to bound $\tvd(p^{\le N}_0, \E[z\sim\calD]{p^{\le N}_z})$:
\begin{lemma}[Le Cam's two-point method]
    If there exists a distribution $\calD$ over $\calS_1$ for which $\tvd(p^{\le N}_0, \E[z\sim\calD]{p^{\le N}_z}) < 1/3$, there is no algorithm $\calA$ which maps transcripts of $N$ measurement outcomes to $\brc{0,1}$ for which $\Pr[T\sim p^{\le N}_z]{\calA(T) = i} \ge 2/3$ for any $z\in\calS_i$ and $i = 0,1$.
\end{lemma}

\begin{proof}
    Suppose to the contrary that there existed such an algorithm. Let $p^{\le N}_1 \triangleq \E[z\sim\calD]{p^{\le N}_z}$. Then 
    \begin{equation}
        2/3 \le \E[z\sim\calD]*{\Pr[T\sim p^{\le N}_z]{\calA(T) = i}} = \Pr[T\sim p^{\le N}_1]{\calA(T) = i}.
    \end{equation} But by Fact~\ref{fact:hypo_test}, this would contradict the fact that $\tvd(p^{\le N}_0, \E[z\sim\calD]{p^{\le N}_z}) < 1/3$.
\end{proof}

In the distinguishing tasks we consider, the choice of $\calD$ will be fairly clear, so the primary technical difficulty for us will be upper bounding the total variation distance between $p^{\le N}_0$ and $\E[z\sim\calD]{p^{\le N}_z}$.

\subsection{Learning with and without quantum memory}

We begin with the simpler task of learning to predict properties in an unknown quantum system $\rho$. We consider the setting where an algorithm has access to a physical source (an oracle) that generates an unknown quantum state $\rho$. The goal is to learn to predict properties of $\rho$ from as few accesses to the oracle as possible. We denote the number of oracle accesses to be $T$.
We give definitions for learning algorithms that can utilize quantum memory and those that can not utilize quantum memory throughout the experiments.

\begin{definition}[Learning states without quantum memory]\label{def:classical_memory_state}
    The algorithm can obtain classical data from the oracle by performing arbitrary POVM measurements on $\rho$.
    For each access to the oracle, the algorithm can select a POVM $\{F_s\}_s$ and obtain the classical outcome $s$ with probability $\Tr(F_s \rho)$.
    The selection of the POVM can depend on all previous measurement outcomes.
    After $T$ oracle accesses, the algorithm predicts the properties of $\rho$.
\end{definition}

\begin{definition}[Learning states with quantum memory]\label{def:quantum_memory_state}
    The algorithm can obtain quantum state $\rho$ from the oracle and store in the quantum memory.
    After $T$ oracle accesses, the algorithm performs a POVM $\{F_s\}_s$ on $\rho^{\otimes T}$ to predict properties of $\rho$.
\end{definition}

We now describe the general task of learning to predict properties in an unknown quantum channel $\mathcal{C}$. We consider the setting where an algorithm can access an unknown quantum-mechanical process, given by the quantum channel $\mathcal{C}$.
The goal is to learn to predict properties of $\mathcal{C}$ from minimum number of accesses to $\mathcal{C}$.
Similarly, we denote the number of oracle accesses by $T$.
We also give definitions for learning algorithms with and without quantum memory.

\begin{definition}[Learning channels without quantum memory]\label{def:classical_memory_channel}
    The algorithm can obtain classical data from the oracle by preparing an arbitrary initial state $\ketbra{\psi}{\psi}$, evolve under $\mathcal{C}$ to yield the output state $(\mathcal{C}\otimes \mathcal{I})(\ketbra{\psi}{\psi})$, and perform an arbitrary POVM measurement on the output state $(\mathcal{C}\otimes \mathcal{I})(\ketbra{\psi}{\psi})$.
    For each access to the oracle, the algorithm can select a POVM $\{F_s\}_s$ and obtain the classical outcome $s$ with probability $\Tr(F_s (\mathcal{C}\otimes \mathcal{I})(\ketbra{\psi}{\psi}))$.
    The selection of the initial state and the POVM can depend on all previous measurement outcomes.
    After $T$ oracle accesses, the algorithm predicts the properties of $\mathcal{C}$.
\end{definition}

\begin{definition}[Learning channels with quantum memory]\label{def:quantum_memory_channel}
    The algorithm can access the oracle $\mathcal{C}$ as a quantum channel during the quantum computation.
    In particular, we consider a mixed state quantum computation.
    The resulting quantum memory after $T$ oracle accesses can be written by
    \begin{equation}
        \rho^{\mathcal{C}}_T = \mathcal{C}_T (\mathcal{C}\otimes \mathcal{I}) \ldots \mathcal{C}_2 (\mathcal{C}\otimes \mathcal{I}) \mathcal{C}_1 (\mathcal{C}\otimes \mathcal{I})(\rho_0),
    \end{equation}
    where $\rho_0$ is the initial state of the quantum memory with arbitrary number of qubits, $\mathcal{C}_t$ is a quantum channel for all $t = 1, \ldots, T$.
    After $T$ oracle accesses, the algorithm performs a POVM $\{F_s\}_s$ on the quantum memory $\rho^{\mathcal{C}}_T$ to predict properties of $\mathcal{C}$.
\end{definition}

Learning channels is more general than learning states. We can reduce to the definitions for learning states by considering quantum channels $\mathcal{C}$ with $0$-qubit input state and $n$-qubit output state.
Furthermore, for learning states or channels with/without quantum memory, we can restrict the POVM to be rank-$1$ POVM because we never consider post-measurement quantum states in the definitions.

\begin{remark} \label{rem:rank1POVM}
In the definitions of learning states/channels with/without quantum memory, we can replace each POVM $\{F_s\}_s$ by a rank-$1$ POVM $\{\sqrt{w_s 2^n} \ketbra{\psi_s}{\psi_s}\}_s$ using Lemma~\ref{lem:simPOVMwithr1POVm}.
\end{remark}

Finally, we describe an intermediate regime where the algorithm has bounded quantum memory. We only consider definitions for learning states in the main text. The definition for learning channels with bounded quantum memory can be defined similarly.
Intuitively, we consider the algorithms to have a $k$-qubit quantum memory $\sigma$.
Every time the algorithm a new copy of quantum state $\rho$, it performs a partial measurement on the combined state $\sigma \otimes \rho$ that results in a $k$-qubit post-measurement state.
Because we need to consider post-measurement quantum state in this case, we \emph{can not} replace the general POVM $\{F_s\}_s$ by a rank-$1$ POVM using Lemma~\ref{lem:simPOVMwithr1POVm}.
The formal definition is given below.

\begin{definition}[Learning states with size-$k$ quantum memory]\label{defn:sizek}
    The algorithm maintains a $k$-qubit quantum memory $\sigma$. For each access to the oracle, the algorithm can select a POVM measurement $\{F_s\}_s$ on $(\rho \otimes \sigma)$, where each POVM element has some Cholesky decomposition $F_s = M_s^{\dagger} M_s$, and obtain a classical outcome $s$ as well as the first $k$ qubits of the post-measurement state,
    \begin{equation}
        \Tr_{> k}\left(\frac{\mathcal{M}_s (\rho \otimes \sigma) \mathcal{M}_s^\dagger}{\Tr(F_s (\rho \otimes \sigma))}\right),
    \end{equation}
    with probability $\Tr(F_s (\rho \otimes \sigma))$.
    For each access to the oracle, the algorithm can select a POVM $\{F_s\}_s$ that depends on all previous measurement outcomes.
    After $T$ oracle accesses, the algorithm predicts the properties of $\rho$.
\end{definition}

\begin{remark}
One can verify that the definition for learning states with quantum memory is equivalent to the definition for learning states with size-$k$ quantum memory when $k\rightarrow \infty$.
\end{remark}

\subsection{Review of tensor network diagrams}
\label{sec:diagram}

It will be convenient to review the diagrammatic notation for tensor contraction, which we will leverage in several proofs.  These so-called `tensor networks' will render the index contraction of higher-rank tensors more transparent than standard notations. We also refer the interested reader to \cite{landsberg2012tensors,bridgeman2017hand} for a more comprehensive overview of tensor networks.

\subsubsection*{Diagrams for individual tensors}

For our purposes, a rank $(m,n)$ tensor is a multilinear map $T : \mathcal{H}^{*\,\otimes m} \otimes \mathcal{H}^{\otimes n} \to \mathbb{C}$.  If $\{|i\rangle\}$ is an orthonormal basis for $\mathcal{H}$, then in bra-ket notation $T$ can be expressed as
\begin{align}
\label{E:Ttensor1}
T = \sum_{\substack{i_1,...,i_m \\
j_1,...,j_n}} T^{i_1 \cdots i_{m}}_{j_1 \cdots j_n} \,\big(|i_1\rangle \otimes \cdots \otimes |i_m\rangle\big) \big(\langle j_{1}| \otimes \cdots \otimes \langle j_{n}| \big)\,.
\end{align}
for some $T^{i_1 \cdots i_{m}}_{j_1 \cdots j_n} \in \mathbb{C}$.  It is clear that a quantum state $|\Psi\rangle$ on $\mathcal{H}$ is a rank $(1,0)$ tensor, being a map from $\mathcal{H}^* \to \mathbb{C}$.  Accordingly, its dual $\langle \Psi|$ is a $(0,1)$ tensor.  Moreover a matrix $M = \sum_{ij} M_{j}^i \,|i\rangle \langle j|$ is a $(1,1)$ tensor.  We elect to represent $T$ diagrammatically by
\begin{align}
\label{E:Ttensor2}
 \includegraphics[scale=.32, valign = c]{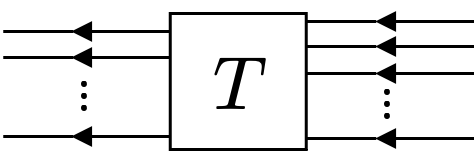}
\end{align}
which has $m$ outgoing legs on the left and $n$ incoming legs on the right.  Each leg in the diagram may be associated with an index of the coefficients $T^{i_1 \cdots i_{m}}_{j_1 \cdots j_n}$\,.  We set the convention that outgoing legs are ordered counter-clockwise and incoming legs are ordered clockwise.  For instance, in~\eqref{E:Ttensor2} the top-left outgoing leg corresponds to $i_1$, the leg below to $i_2$, and so on.  Likewise the top-right incoming leg corresponds to $j_1$, the leg below to $j_2$, and so on.

\subsubsection*{Tensor contraction}
We next explain how to depict tensor network contractions diagrammatically.  For sake of illustration, suppose we have a rank $(2,1)$ tensor
\begin{align}
\label{Atensor1}
 \includegraphics[scale=.32, valign = c]{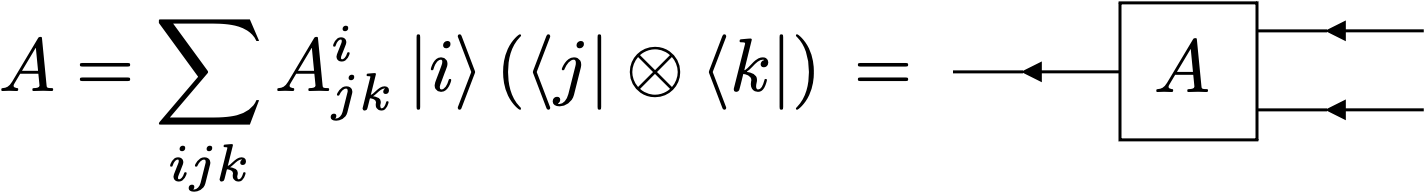}
\end{align}
and a rank $(1,2)$ tensor
\begin{align}
 \includegraphics[scale=.32, valign = c]{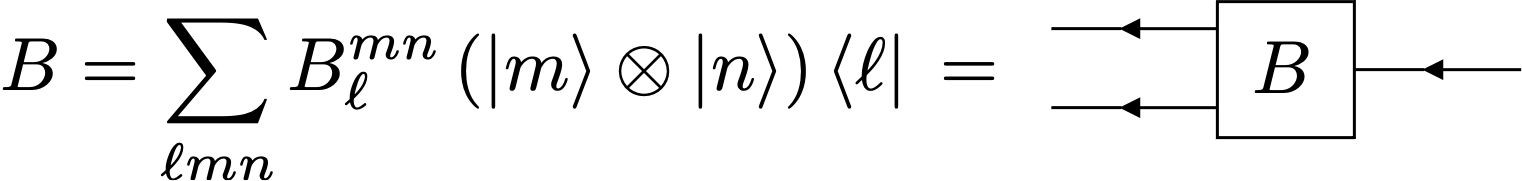}
\end{align}
Now suppose we want to compute the tensor network contraction corresponding to
\begin{equation}
\label{E:ABcontract1}
\sum_{ijk} A^i_{jk} B^{jk}_i\,.
\end{equation}
Here lower indices are contracted with upper indices because this represents contracting vectors with covectors.  The contraction in~\eqref{E:ABcontract1} is depicted diagrammatically as
\begin{align}
\label{E:ABcontract2}
 \includegraphics[scale=.32, valign = c]{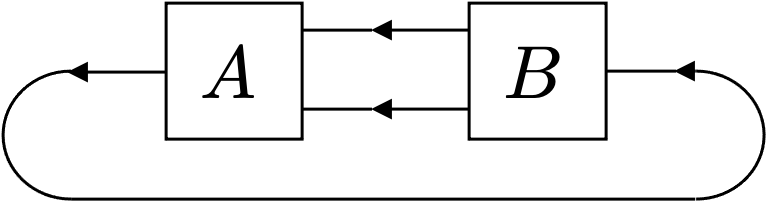}
\end{align}
Comparing the diagram with~\eqref{E:ABcontract1}, we see that contracted indices corresponding to outgoing and incoming lines which are glued together.  The fact that vectors are to be contracted with covectors is reflected in the fact that we are only allowed to glue together lines in a manner consistent with their orientations.

As another example, given a matrix $M = \sum_{ij} M_j^i \,|i\rangle \langle j|$, the trace can be written as
\begin{align}
     \includegraphics[scale=.32, valign = c]{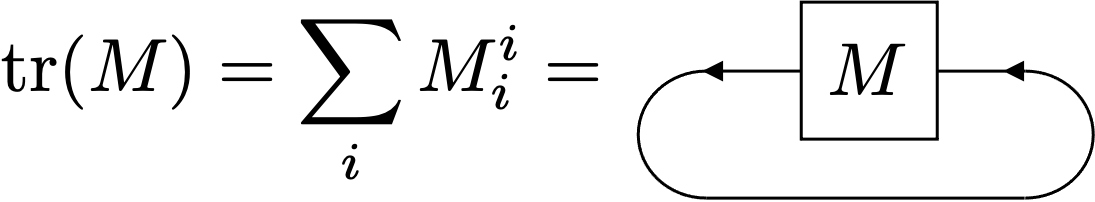}
\end{align}
If $M_1, M_2,...,M_k$ are matrices, then the product $M_1 M_2 \cdots M_k$ is depicted by
\begin{align}
     \includegraphics[scale=.32, valign = c]{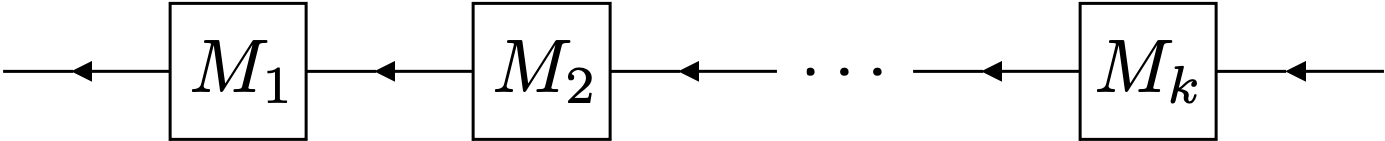}
\end{align}

\subsubsection*{Multiplication by a scalar}

Given a tensor $T$, multiplication by a scalar $\alpha$ is often denoted by $\alpha \, T$.  In our diagrammatic notation, we will simply write
\begin{equation}
 \includegraphics[scale=.32, valign = c]{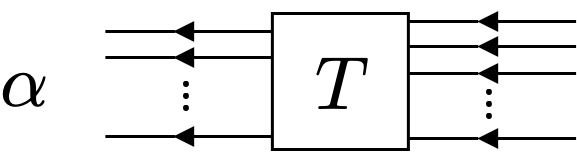}
\end{equation}

\subsubsection*{Tensor products}

Given two tensors $T_1, T_2$, we can form the tensor product $T_1 \otimes T_2$.  We will denote this diagrammatically as
\begin{equation}
 \includegraphics[scale=.32, valign = c]{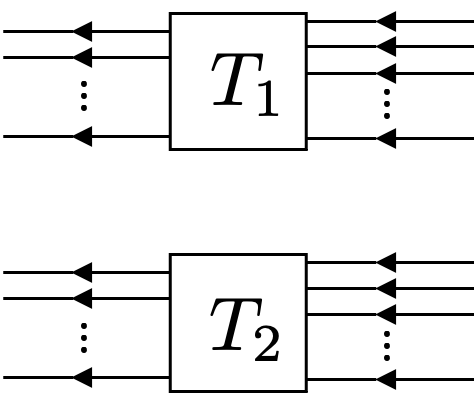}
\end{equation}
or also
\begin{equation}
 \includegraphics[scale=.32, valign = c]{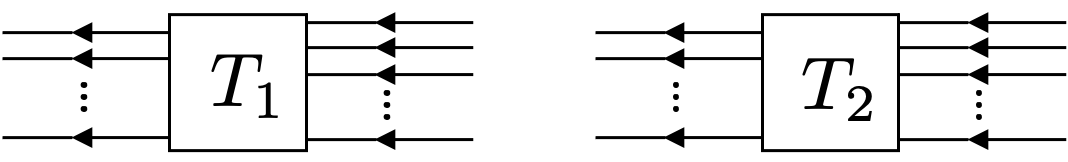}
\end{equation}
More generally, how to read off the order of a tensor product (e.g.~$T_1 \otimes T_2$ or $T_2 \otimes T_1$) from a diagram will be clear in context.

\subsubsection*{Taking norms}

Often it will be convenient to compute the norm of a matrix in tensor notation.  For instance, if $M$ is a matrix, then its 1-norm $\|M\|_1$ can be expressed diagrammatically as
\begin{equation}
 \includegraphics[scale=.32, valign = c]{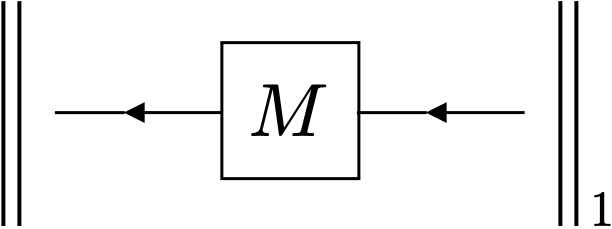}
\end{equation}
Here we are simply taking the diagrammatic notation for $M$ as a stand-in within the expression $\|M\|_1$.  This is particularly convenient in circumstances where $M$ is given by a tensor network contraction whose structure we wish to emphasize; for instance, the 1-norm of $M = \sum_{ijk\ell} A_{k\ell}^i B_j^{k\ell} \,|i\rangle \langle j|$ is conveniently depicted by
\begin{equation}
 \includegraphics[scale=.32, valign = c]{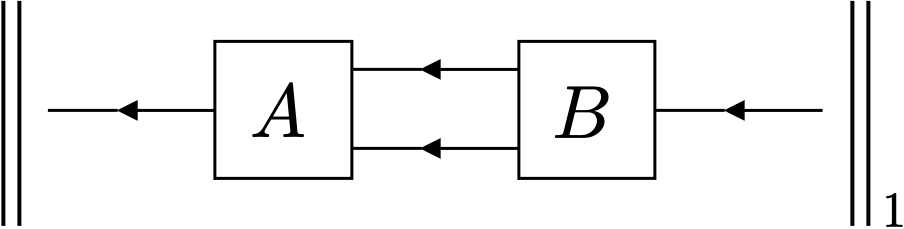}
\end{equation}

\subsubsection*{Tensors with legs of different dimensions}

So far we have considered rank $(m,n)$ tensors as maps $T : \mathcal{H}^{* \otimes m} \otimes \mathcal{H}^{\otimes n} \to \mathbb{C}$.  More generally we can consider tensors $T : \left(\mathcal{H}_1^* \otimes \cdots \otimes \mathcal{H}_m^*\right)  \otimes \left( \mathcal{H}_{m+1} \otimes \cdots \otimes \mathcal{H}_{m + n} \right) \to \mathbb{C}$ where the tensored Hilbert spaces in the domain need not be isomorphic.  We can use the same diagrammatic notation as above, with the additional restriction that tensor legs can be contracted if they both carry the same dimension (i.e., correspond to a Hilbert space and a dual Hilbert space of the same dimension).

As an example, we can consider the state $|\Psi\rangle$ in $\mathbb{C}^2 \otimes \mathbb{C}^3$, and form its density matrix $|\Psi\rangle \langle \Psi|$.  In our tensor diagram corresponding to this state, the $\mathbb{C}^2$ (qubit) legs will be solid lines and the $\mathbb{C}^3$ (qutrit) legs will be dotted lines.  Performing a partial over the qutrit legs is expressed diagrammatically as
\begin{equation}
\label{E:Psipartial1}
 \includegraphics[scale=.32, valign = c]{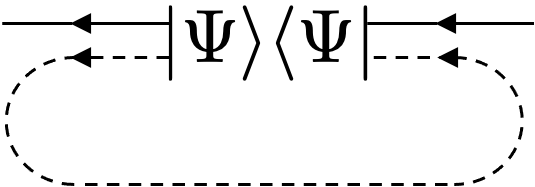}
\end{equation}
We will discuss the diagrammatic notation of partial traces in more detail below.

\subsubsection*{Identity operator}

The identity operator on a Hilbert space $\mathcal{H}$ can be expressed diagrammatically as an oriented line
\begin{equation}
 \includegraphics[scale=.32, valign = c]{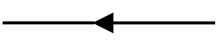}
\end{equation}
We can clearly see that given a state in the Hilbert space
\begin{equation}
 \includegraphics[scale=.32, valign = c]{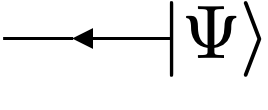}
\end{equation}
if we left-multiply by the identity diagram we will get the same tensor diagram and thus the same state.  Likewise for the dual state
\begin{equation}
 \includegraphics[scale=.32, valign = c]{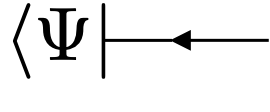}
\end{equation}
if we right-multiply by the identity diagram then we return the same tensor diagram.

Likewise, the identity operator on $k$ copies of the Hilbert space $\mathcal{H}^{\otimes k}$ is just
\begin{equation}
 \includegraphics[scale=.32, valign = c]{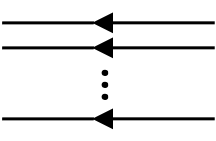}
\end{equation}
In the setting that the Hilbert space under consideration is $\mathcal{H} \otimes \mathcal{H}'$ where each tensor factor has a different dimension, it is convenient to represent tensor legs in $\mathcal{H}$ by solid lines and tensor legs in $\mathcal{H}'$ be dotted line; in this setting the identity operator is
\begin{equation}
 \includegraphics[scale=.32, valign = c]{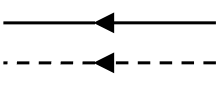}
 \end{equation}
which readily generalizes if there are more than two Hilbert spaces with differing dimensions.

\subsubsection*{Resolutions of the identity}

Suppose $\{|\Psi_i\rangle\}_i$ is an orthonormal basis for $\mathcal{H}$.  Then the resolution of the identity $\sum_i |\Psi_i\rangle \langle \Psi_i| = \mathds{1}$ can be expressed diagrammatically as
\begin{equation}
 \includegraphics[scale=.32, valign = c]{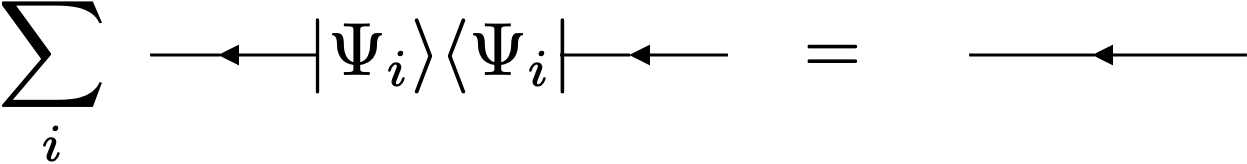}
 \end{equation}
If instead $\{|\Psi_i\rangle\}_i$ is a resolution of the identity for $\mathcal{H} \otimes \mathcal{H}'$ where the two Hilbert spaces have different dimensions, we may analogously denote this diagrammatically by
\begin{equation}
 \includegraphics[scale=.32, valign = c]{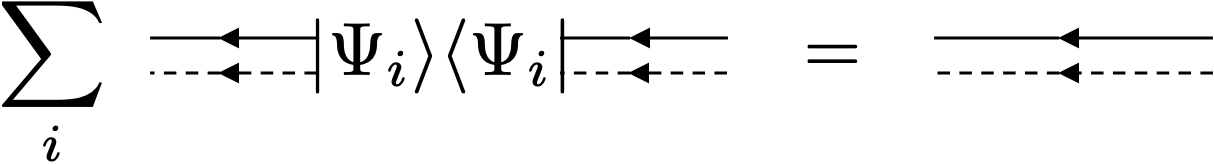}
 \end{equation}

Similarly, if $\{M_s^{\dagger}M_s\}_s$ is a POVM on $\mathcal{H}$ then the resolution of the identity $\sum_s M_s^{\dagger}M_s = \mathds{1}$ can be written as
\begin{equation}
 \includegraphics[scale=.32, valign = c]{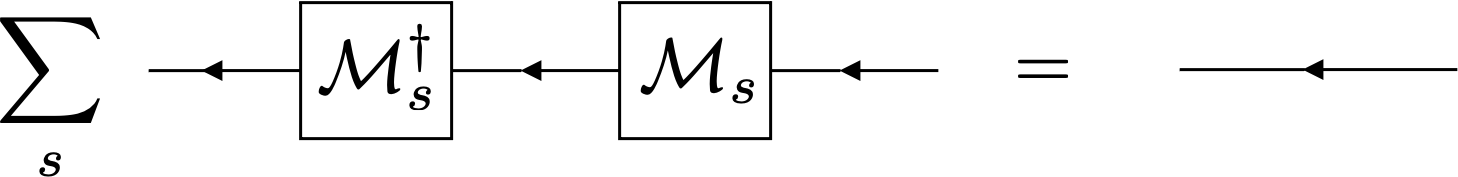}
 \end{equation}
and analogously if the Hilbert space is $\mathcal{H} \otimes \mathcal{H}'$ or has even more tensor factors.

\subsubsection*{Taking traces and partial traces}

Suppose we have a rank $(n,n)$ tensor $T : \mathcal{H}^{* \otimes n} \otimes \mathcal{H}^{\otimes n}$.  Then its trace is given by $\text{tr}(T) = \sum_{i_1,...,i_n} T_{i_1 \cdots i_n}^{i_1 \cdots i_n}$, or diagrammatically
\begin{equation}
\includegraphics[scale=.32, valign = c]{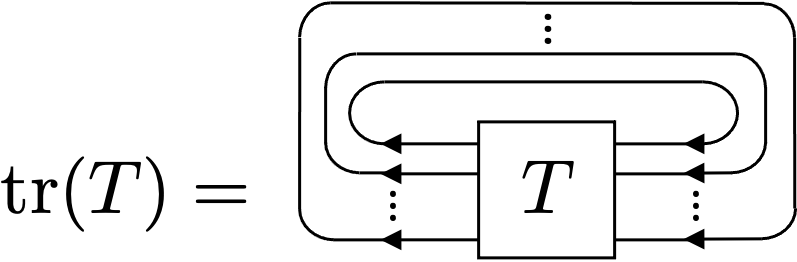}
\end{equation}
A very useful diagrammatic identity is the trace of the identity matrix, which can be regarded as a rank $(1,1)$ tensor $\mathds{1} = \sum_i |i\rangle \langle i|$.  We have
\begin{equation}
\includegraphics[scale=.32, valign = c]{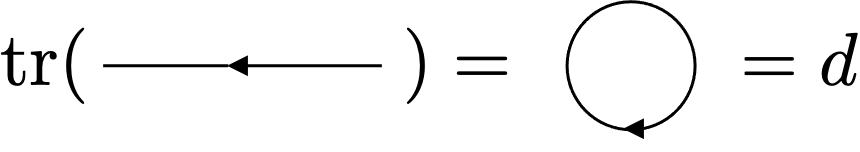}
\end{equation}
and so we see that a closed loop in tensor diagrams equals the dimension of the Hilbert space associated that curve.  As another example, if we have the identity $\mathds{1}_{d \times d} \otimes \mathds{1}_{d' \times d'}$ on $\mathcal{H} \otimes \mathcal{H}'$, where $\dim(\mathcal{H}) = d$, $\dim(\mathcal{H}') = d'$ and we have used subscripts on the identity matrices for emphasis, we have 
\begin{equation}
\includegraphics[scale=.32, valign = c]{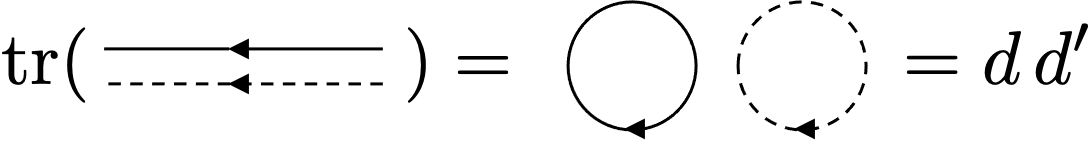}
\end{equation}
where the solid line corresponds to the $\mathcal{H}$ Hilbert space and the dotted line corresponds to the $\mathcal{H}'$ Hilbert space.

We can also take partial traces in similar fashion. We define the partial trace over the `$k$th subsystem' by
\begin{equation}
\text{tr}_k(T) = \sum_{\substack{i_1,...,i_{k-1},i_{k+1},...,i_n \\ j_1,...,j_{k-1},j_{k+1},...,j_n}}\left(\sum_{i_k} T_{j_1 \cdots j_n}^{i_1 \cdots i_n}\right) |i_1\rangle \langle j_1| \otimes \cdots \otimes |i_{k-1}\rangle \langle j_{k-1}| \otimes |i_{k+1}\rangle \langle j_{k+1}| \otimes \cdots \otimes |i_n \rangle \langle j_n|\,.
\end{equation}
Note that $\text{tr}_\ell(\text{tr}_k(T)) = \text{tr}_k(\text{tr}_\ell(T)))$.  Since the operation of taking partial traces is commutative we can use the notation $\text{tr}_{k,\ell}(T)$.  Notice that $\text{tr}_{1,...,n}(T) = \text{tr}(T)$.  That is, taking the partial trace over all subsystems in the tensor is the same as taking the trace of the entire tensor.

Diagrammatically, the partial trace over the first subsystem is given by
\begin{equation}
\includegraphics[scale=.32, valign = c]{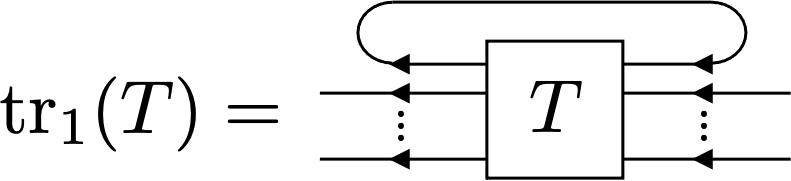}
\end{equation}
The partial trace over the second subsystem is
\begin{equation}
\includegraphics[scale=.32, valign = c]{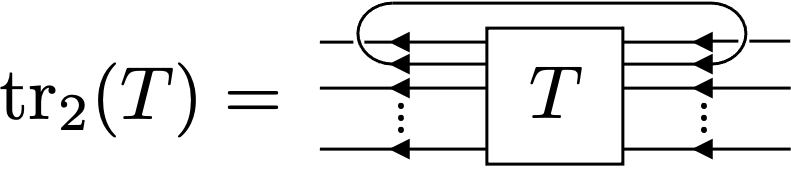}
\end{equation}
and so on.

If we have an tensor with legs corresponding to Hilbert spaces of different dimensions, we can still in some cases take traces or partial traces.  In particular, if $T : (\mathcal{H}_1^* \otimes \cdots \otimes \mathcal{H}_n^*) \otimes (\mathcal{H}_{1}' \otimes \cdots \otimes \mathcal{H}_m') \to \mathbb{C}$, then if $\mathcal{H}_k = \mathcal{H}_k'$ we can still compute the partial trace $\text{tr}_k(T)$.  As a simple example consider the state $|\Psi\rangle$ living on $\mathcal{H} \otimes \mathcal{H}'$.  Then its density matrix $|\Psi\rangle \langle \Psi|$ can be regarded as a $(2,2)$ tensor taking $(\mathcal{H}^* \otimes \mathcal{H}^{'*}) \otimes (\mathcal{H} \otimes \mathcal{H}') \to \mathbb{C}$.  Then we have
\begin{equation}
\includegraphics[scale=.32, valign = c]{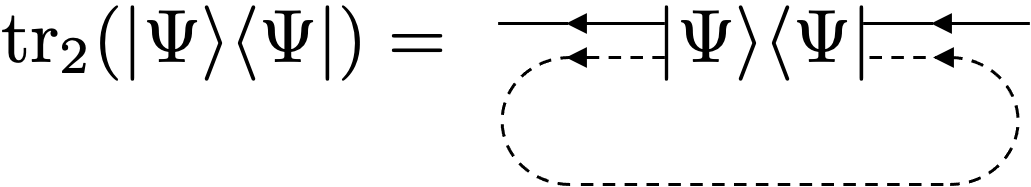}
\end{equation}
which is the same example as~\eqref{E:Psipartial1}; a similar diagram expresses $\text{tr}_1(|\Psi\rangle \langle \Psi|)$.

\subsubsection*{Isotopies}
We remark that tensor network diagrams are to be understood up to isotopy of the tensor legs; that is, deforming or bending the tensor legs does not change the interpretation of the diagram.  For instance, for a product of matrices $M_1 M_2$ we have equivalences like
\begin{equation}
 \includegraphics[scale=.32, valign = c]{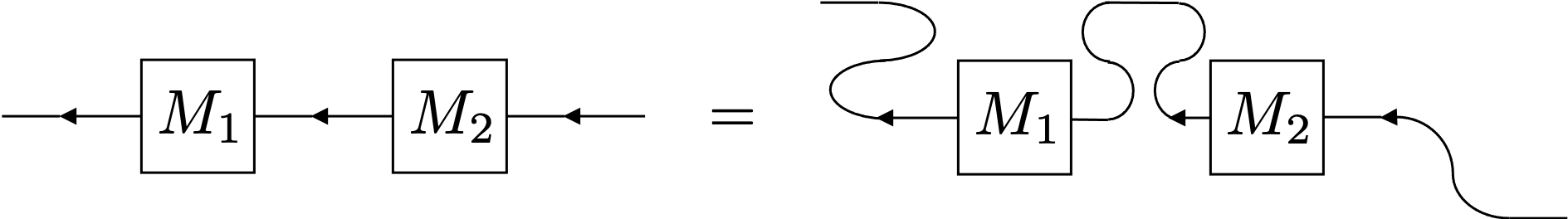}
\end{equation}
and similarly for all other kinds of tensors.

The isotopies are not required to be planar; for instance
\begin{equation}
 \includegraphics[scale=.32, valign = c]{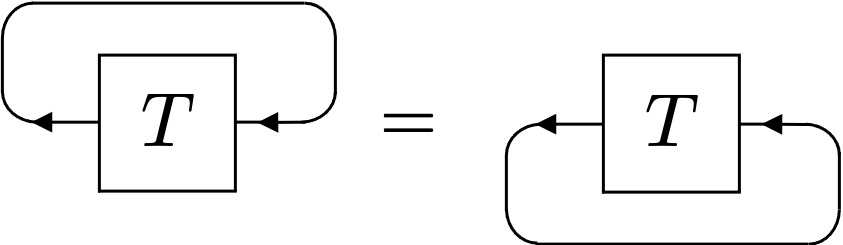}
\end{equation}
We also can allow legs to cross, for instance
\begin{equation}
 \includegraphics[scale=.32, valign = c]{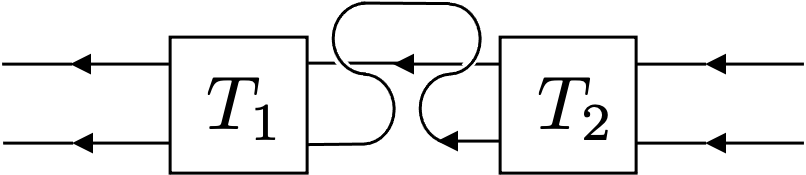}
\end{equation}
We will disregard whether such crossings are overcrossings or undercrossings.

However, we set the convention that we do not change the relative order of the endpoints of the outgoing or incoming legs.  The reason is that permuting the order of the endpoints corresponds would correspond to permuting the tensor factors on which the tensor is defined.  As a transparent example, let $T : (\mathcal{H}_1^* \otimes \mathcal{H}_2^{*}) \otimes (\mathcal{H}_1 \otimes \mathcal{H}_2) \to \mathbb{C}$ be denoted by
\begin{equation}
 \includegraphics[scale=.32, valign = c]{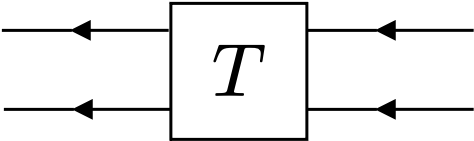}
\end{equation}
Then the diagram
\begin{equation}
 \includegraphics[scale=.32, valign = c]{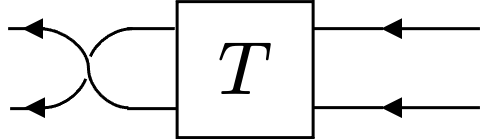}
\end{equation}
corresponds to a tensor $(\mathcal{H}_2^* \otimes \mathcal{H}_1^{*}) \otimes (\mathcal{H}_1 \otimes \mathcal{H}_2) \to \mathbb{C}$ where we note that $\mathcal{H}_1^*$ and $\mathcal{H}_2^*$ have been permuted.  See also the discussion of permutation operators below.

\subsubsection*{Permutation operators}

Consider the permutation group on $k$ elements, $S_k$, and let $\tau$ be an element of the group.  We define a representation of $\tau$, namely $\text{Perm}(\tau)$, which acts on a $k$-copy Hilbert space $\mathcal{H}^{\otimes k}$ as follows.  Letting $|\psi_1\rangle \otimes |\psi_2\rangle \otimes \cdots \otimes |\psi_n\rangle$ be a product state on $\mathcal{H}^{\otimes k}$, we define
\begin{equation}
\label{E:Permdefn1}
\text{Perm}(\tau) |\psi_1\rangle \otimes |\psi_2\rangle \otimes \cdots \otimes |\psi_n\rangle = |\psi_{\tau^{-1}(1)}\rangle \otimes |\psi_{\tau^{-1}(2)}\rangle \otimes \cdots \otimes |\psi_{\tau^{-1}(n)}\rangle 
\end{equation}
which extends to the entire Hilbert space $\mathcal{H}^{\otimes k}$ by linearity.  With these conventions, the representations $\text{Perm}(\tau)$ enjoy the property
\begin{equation}
\text{Perm}(\tau) \cdot \text{Perm}(\sigma) = \text{Perm}(\tau \sigma)
\end{equation}
where $\tau \sigma$ is shorthand for the group product, i.e.~the composition $\tau \circ \sigma$.

These representations of $S_k$ admit a very intuitive tensor diagrams.  Consider, for instance, $S_3$ and $\tau = (123)$.  Then the corresponding tensor diagram is
\begin{equation}
 \includegraphics[scale=.32, valign = c]{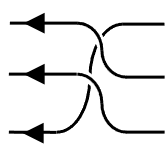}
\end{equation}
This is made very clear by labeling the endpoints of the diagram by
\begin{equation}
 \includegraphics[scale=.32, valign = c]{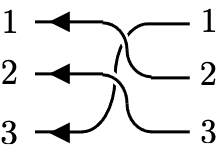}
\end{equation}
This notation generalized accordingly for other permutation representations.  The group product structure is also transparent; for instance $\text{Perm}((123)) \cdot \text{Perm}((12))$ is depicted diagrammatically by
\begin{equation}
 \includegraphics[scale=.32, valign = c]{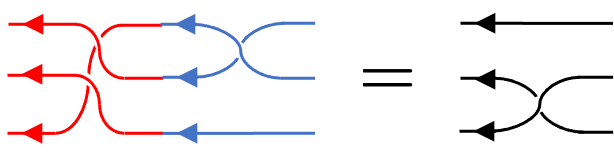}
\end{equation}
where $\text{Perm}((123))$ is given in red and $\text{Perm}((12))$ is given in blue for clarity; the allowed diagrammatic manipulations of performing isotopies without rearranging the endpoints of the tensor legs show that the result of the product is $\text{Perm}((23))$.  A nice feature of the diagrams is that the diagram for $\text{Perm}(\tau^{-1})$ can be obtain from the diagram for $\text{Perm}(\tau)$ by flipping the latter horizontally.

As another example, if we multiply $\text{Perm}((123))$ by a state $|\Psi\rangle$ in $\mathcal{H}^{\otimes 3}$, then we get
\begin{equation}
 \includegraphics[scale=.32, valign = c]{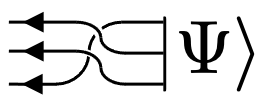}
\end{equation}
from which it is clear that $\text{Perm}((123))$ permutes the tensor factors of the state according to $(123)^{-1} = (132)$.

In some later proofs where there is no ambiguity, we will denote $\text{Perm}(\tau)$ simply by $\tau$.

\subsubsection*{Transposes and partial transposes}

Suppose we have a matrix $M = \sum_{i,j} M_j^i \,|i\rangle \langle j|$ viewed as a rank $(1,1)$ tensor.  We can represent its transpose $M^t = \sum_{i,j} M_j^i \,|j\rangle \langle i|$ diagrammatically by

Here we are dualizing each leg by changing the direction of each arrow, and then reorganizing the legs via isotopy so that the in-arrow comes in from the right and the out-arrow comes out to the left; this isotopy is done in order to match the arrow configuration in the diagram on the left.
\begin{equation}
 \includegraphics[scale=.32, valign = c]{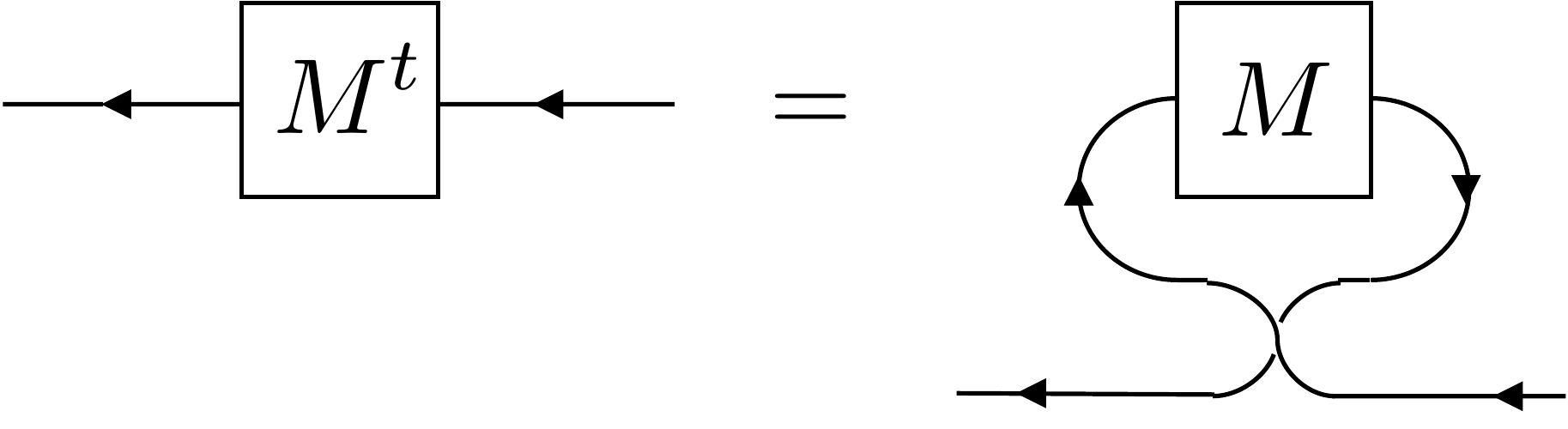}
\end{equation}
If we have a higher-rank tensor, such as a rank $(2,2)$ tensor $T = \sum_{ijk\ell} T_{k\ell}^{ij} \,|i\rangle \langle k| \otimes |j\rangle \langle \ell|$, then we can also perform a partial transposition on a subsystem; for instance, the partial transposition on the second subsystem $\sum_{ijk\ell} T_{k\ell}^{ij} \,|i\rangle \langle k| \otimes |\ell\rangle \langle j|$ is given by
\begin{equation}
 \includegraphics[scale=.32, valign = c]{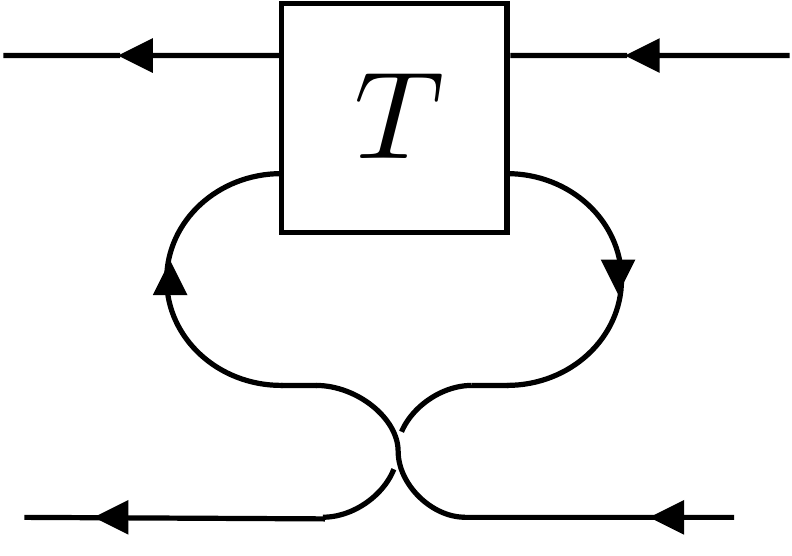}
\end{equation}
This notation extends to higher rank tensors in an analogous fashion.

\subsubsection*{Maximally entangled state}

The maximally entangled state is given by $|\Omega\rangle = \sum_i |i\rangle |i\rangle$ where $\{|i\rangle\}$ is the computational basis.  We treat $|\Omega\rangle$ as unnormalized, and it and its Hermitian conjugate are denoted by
\begin{equation}
 \includegraphics[scale=.32, valign = c]{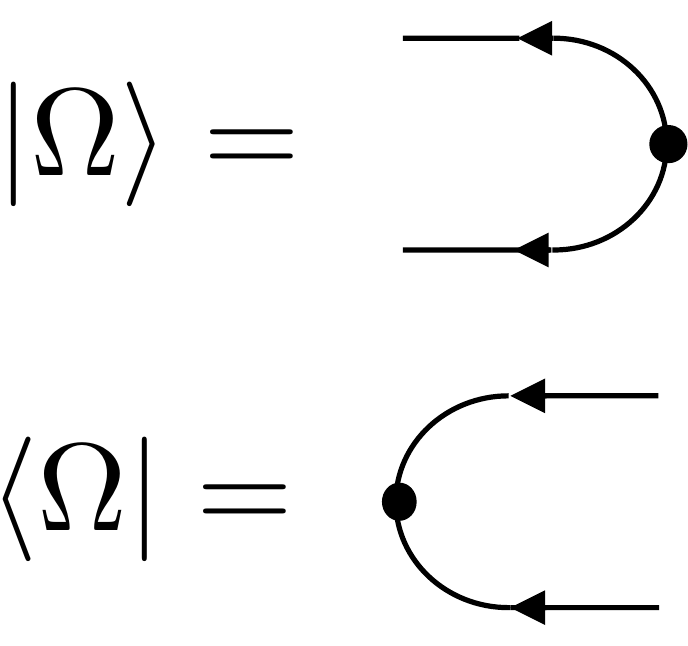}
\end{equation}
Letting $\mathcal{H}_A \simeq \mathcal{H}_B \simeq \mathcal{H}_C$, we have the identities
\begin{align}
\left(\mathds{1}_A \otimes \langle \Omega|_{BC}\right)\left(|\Omega\rangle_{AB} \otimes \mathds{1}_C\right) &= \sum_i |i\rangle_A \langle i|_C \\
\left(\langle \Omega|_{AB} \otimes \mathds{1}_C\right)\left(\mathds{1}_A \otimes |\Omega\rangle_{BC} \right) &= \sum_i |i\rangle_C \langle i|_A
\end{align}
which can be expressed diagrammatically as
\begin{equation}
 \includegraphics[scale=.32, valign = c]{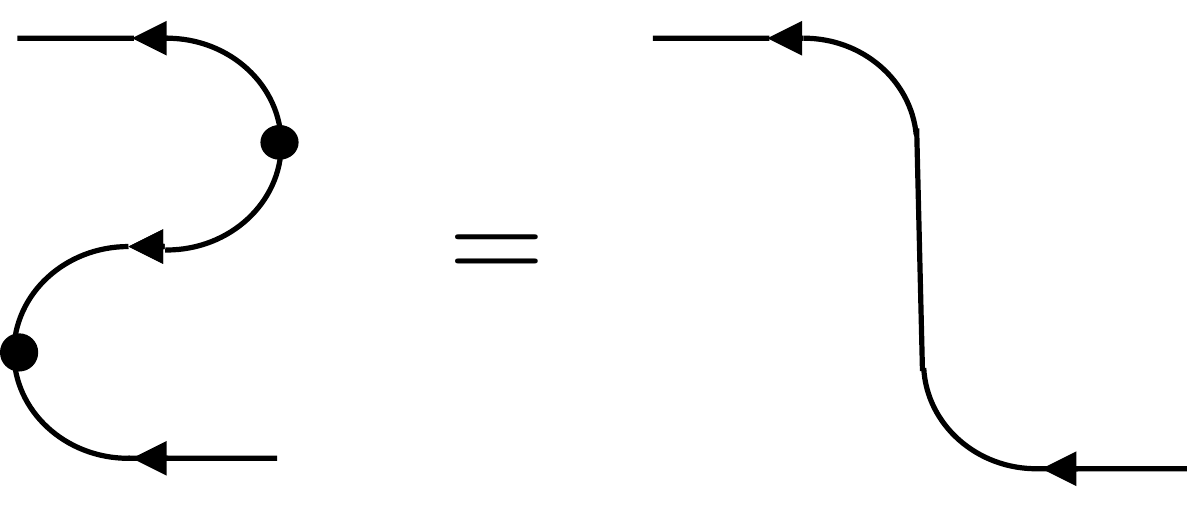}
\end{equation}
\begin{equation}
 \includegraphics[scale=.32, valign = c]{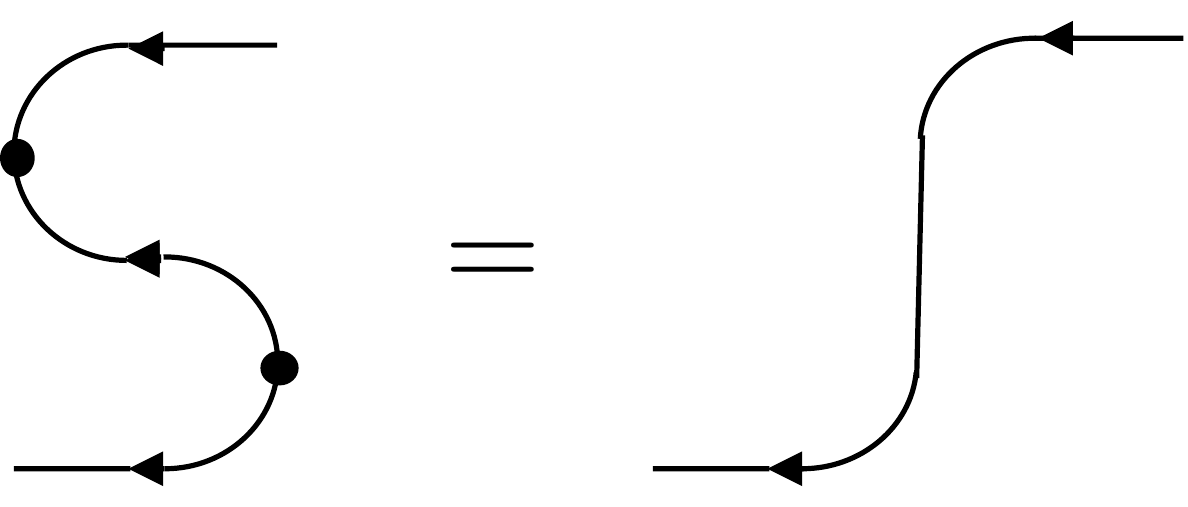}
\end{equation}
We can think of the black dot as being a transpose operation since it changes the orientation of the tensor leg; moreover, two black dots annihilate one another since taking two transposes is the identity operation.


\section{Exponential separations in learning quantum states}
\label{sec:learnstate}

The main contribution of this work is in establishing information-theoretic lower bounds in learning algorithms without quantum memory.
For learning properties of quantum states, a quantum-memoryless algorithm performs a POVM measurement on the unknown quantum state $\rho$ and store the classical measurement outcome in a classical memory.
The state of the classical memory is equivalent to the collection of all previous measurement outcomes.
Because the POVM measurements performed at each time step depends on the previous outcomes, it is naturally to consider a tree representation for the learning algorithm.
Each node on the tree represents the state of the classical memory.
The classical memory state begins from the root of the tree.
Every time we perform a POVM measurement on $\rho$, we move to a child node along the edge that encodes the measurement outcome.
The POVM measurement we perform at each node will be different.
Based on Remark~\ref{rem:rank1POVM}, we only need to consider rank-$1$ POVM.
Hence, we consider each node $u$ to have an associated rank-$1$ POVM $\{\sqrt{w^u_s 2^n} \ketbra{\psi^u_s}{\psi^u_s}\}_s$.
We also associate a probability to each node to represent the probability of arriving at the memory state in the learning process.
Because we consider $T$ measurements, all the leaf nodes are at depth $T$.

\begin{definition}[Tree representation for learning states] \label{def:treelearnstate}
    Fix an unknown $n$-qubit quantum state $\rho$. A learning algorithm without quantum memory can be expressed as a rooted tree $\mathcal{T}$ of depth $T$, where each node on the tree encodes all the measurement outcomes the algorithm has seen so far. The tree satisfies the following properties.
    \begin{itemize}
        \item Each node $u$ is associated with a probability $p^{\rho}(u)$.
        \item For the root $r$ of the tree, $p^{\rho}(r) = 1$.
        \item At each non-leaf node $u$, we measure a POVM $\{\mathcal{M}^u_s\}_{s}$ on $\rho$ to obtain a classical outcome $s$. Each child node $v$ of the node $u$ is connected through the edge $e_{u, s}$.
        \item If $v$ is the child node of $u$ connected through the edge $e_{u, s}$, then
                \begin{equation}
                    p^{\rho}(v) = p^{\rho}(u) w^u_s 2^n \bra{\psi^u_s} \rho \ket{\psi^u_s},
                \end{equation}
                where $\{w^u_s 2^n \ket{\psi^u_s}{\psi^u_s}\}_s$ is a rank-$1$ POVM that depends on the node $u$.
        \item Every root-to-leaf path is of length $T$. Note that for a leaf node $\ell$, $p^{\rho}(\ell)$ is the probability that the classical memory is in state $\ell$ after the learning procedure.
    \end{itemize}
\end{definition}

The general strategy we utilize is a reduction from the prediction task we care about to a two-hypothesis distinguishing problem.
In particular, we consider distinguishing a set of unknown quantum state $\{\rho\}$ from the completely mixed state $\Id / 2^n$.

\begin{definition}[Many-versus-one distinguishing task]
We consider the following two events to happen with equal probability.
\begin{itemize}
\item The unknown state $\rho$ is the $n$-qubit completely mixed state $\Id / 2^n$.
\item The unknown state $\rho$ is sampled uniformly from a set of $n$-qubit states $\{\rho_x\}_x$.
\end{itemize}
The goal of the learning algorithm is to predict which event has happened.
\end{definition}

The basic strategy for proving lower bounds in a two-hypothesis distinguishing problem is the two-point method.
After $T$ measurements on the unknown quantum state $\rho$, the classical memory state will be represented by a leaf $\ell$ of the rooted tree $\mathcal{T}$.
We will then use the information stored in the classical memory to distinguish between the two events.
In order to successfully do so, intuitively, the distribution over the classical memory state in the two events must be sufficiently distinct.
This is made rigorous by the famous two-point method in information-theoretic analysis.
In the following lemma, we describe the two-point method in terms of the tree representation for a learning algorithm without quantum memory.

\begin{lemma}[Le Cam's two-point method, see e.g. Lemma 1 in \cite{yu1997assouad}]\label{lem:lecam}
Consider a learning algorithm without quantum memory that is described by a rooted tree $\mathcal{T}$.
The probability that the learning algorithm solves the many-versus-one distinguishing task correctly is upper bounded by
\begin{equation}
    \frac{1}{2} \sum_{\ell\in \mathrm{leaf}(\mathcal{T})} \left|\Expect_x p^{\rho_x}(\ell) - p^{\Id / 2^n}(\ell)\right|.
\end{equation}
\end{lemma}

It is tempting to try to apply Lemma~\ref{lem:lecam} by uniformly upper bounding the quantity $\left|\Expect_x p^{\rho_x}(\ell) - p^{\Id / 2^n}(\ell)\right|$ for all leaves $\ell$. Unfortunately, it turns out that for some leaves, this quantity can be very large. The good news however is that we do have a uniform \emph{one-sided} bound: as we will show, $p^{\Id / 2^n}(\ell) - \Expect_x p^{\rho_x}(\ell)$ (without the absolute values) can always be upper bounded by a very small value.
It turns out that such a one-sided bound already suffices for applying Le Cam's two-point method.
\begin{lemma}[One-sided bound suffices for Le Cam] \label{lem:oneside2pt}
Consider a learning algorithm without quantum memory that is described by a rooted tree $\mathcal{T}$.
If we have
\begin{equation}\label{eq:onesidebound}
\frac{ \E[x]*{p^{\rho_x}(\ell)} }{p^{\Id / 2^n}(\ell)} \geq 1 - \delta, \quad \forall \ell\in \mathrm{leaf}(\mathcal{T}).
\end{equation}
then the probability that the learning algorithm solves the many-versus-one distinguishing task correctly is upper bounded by $\delta$.
\end{lemma}
\begin{proof}
We utilize the basic fact that $\frac{1}{2}\sum_{i} |p(i) - q(i)| = \sum_{i: p(i) \geq q(i)} p(i) - q(i)$, hence
\begin{align}
        & \frac{1}{2} \sum_{\ell\in \mathrm{leaf}(\mathcal{T})} \left|\Expect_x p^{\rho_x}(\ell) - p^{\Id / 2^n}(\ell)\right|
        = \sum_{\substack{\ell\in \mathrm{leaf}(\mathcal{T}) \\ p^{\Id / 2^n}(\ell) \geq \Expect_x p^{\rho_x}(\ell)}} p^{\Id / 2^n}(\ell) - \Expect_x p^{\rho_x}(\ell)\\
        & \leq \sum_{\substack{\ell\in \mathrm{leaf}(\mathcal{T}) \\ p^{\Id / 2^n}(\ell) \geq \Expect_x p^{\rho_x}(\ell)}} p^{\Id / 2^n}(\ell) \delta \quad \leq \quad \delta.
\end{align}
We can then apply Lemma~\ref{lem:lecam} and conclude the proof of the lemma.
\end{proof}

\subsection{Shadow tomography}

Given $M$ observables $O_1, \ldots, O_M$ with $\norm{O_i}_\infty = 1, \forall i$, the goal of shadow tomography is to predict the expectation values $\Tr(O_i \rho)$ for all the observables up to $\epsilon$-error.
If we consider a class of traceless observables $O_1, \ldots, O_M$ such that
\begin{equation}\label{eq:negativeness}
    O_i = - O_{i + M/2}, \,\, \mbox{and} \,\, \mbox{all eigenvalues of } O_i \mbox{ are } \pm 1, \quad \forall i = 1, \ldots, M/2,
\end{equation}
then the hardness of shadow tomography in learning algorithms without quantum memory can be characterized by the following quantity,
\begin{equation} \label{eq:highincomp}
    \delta(O_1, \ldots, O_M) = \sup_{\ket{\phi}} \frac{2}{M} \sum_{i=1}^{M/2} \bra{\phi} O_i \ket{\phi}^2,
\end{equation}
where $\sup_{\ket{\phi}}$ is taken over the entire space of $n$-qubit pure states.
We establish this hardness relation by the following theorem.

\begin{theorem}[General shadow tomography lower bound] \label{thm:shadowtomo}
Consider $M$ traceless observables $O_1, \ldots, O_M$ satisfying Eq.~\eqref{eq:negativeness}, any learning algorithm without quantum memory requires
\begin{equation}
T \geq \Omega\left( \frac{1}{\epsilon^2 \delta(O_1, \ldots, O_M)} \right)
\end{equation}
copies of $\rho$ to predict expectation values of $\Tr(O_x \rho)$ to at most $\epsilon$-error for all $x = 1, \ldots, M$ with at least a probability of $2/3$.
\end{theorem}
\begin{proof}
We consider a many-versus-one distinguishing task where we want to distinguish between maximally mixed state versus the set of $n$-qubit states $\{\rho_x\}_x$ given by
\begin{equation}
    \rho_x = \frac{\Id + 3 \epsilon O_x}{2^n}.
\end{equation}
Using Eq.~\eqref{eq:negativeness}, we have $\Tr(O_x^2) = 2^n$.
Using $\Tr(O_x^2) = 2^n$ and $\Tr(O_x) = 0$, we have $\Tr(O_x \rho_x) = 3 \epsilon$ and $\Tr(O_x (\Id / 2^n)) = 0$ for all $x = 1, \ldots, M$.
Therefore for any $\rho_x$, there exists an $O_x$ with $\Tr(O_x \rho_x)$ substantially greater than zero.
On the other hand, for $\Id / 2^n$, we have $\Tr(O_x (\Id / 2^n)) = 0$ for all $O_x$.
Together, if a learning algorithm without quantum memory can predict the expectation values of $\Tr(O_x \rho), \forall x = 1, \ldots, M$ up to error $\epsilon$ with probability at least $2/3$, then we can use the learning algorithm to perform the many-versus-one distinguishing task.
Hence, a lower bound for this task also gives a lower bound for shadow tomography task.
Given this result, we will now use Lemma~\ref{lem:oneside2pt} to establish a lower bound on $T$.

To utilize Lemma~\ref{lem:oneside2pt}, we need to establish Eq.~\eqref{eq:onesidebound}.
Consider the tree representation $\mathcal{T}$ of the learning algorithm without quantum memory given in Definition~\ref{def:treelearnstate}.
For any leaf $\ell$ of $\mathcal{T}$, we consider the collection of edges on the path from root $r$ to leaf $\ell$ to be $\{e_{u_t, s_t}\}_{t=1}^T$ where $u_1 = r$.
From the recursive definition for the probability $p^{\rho}(u)$ in Definition~\ref{def:treelearnstate}, we have
\begin{equation}
    p^{\rho}(\ell) = \prod_{t=1}^T w^{u_t}_{s_t} 2^n \bra{\psi^{u_t}_{s_t}} \rho \ket{\psi^{u_t}_{s_t}}.
\end{equation}
We then have the following calculation:
\begin{align}
    \frac{\left(\Expect_x p^{\rho_x}(\ell) \right)}{p^{\Id / 2^n}(\ell)} &= \Expect_x \prod_{t=1}^T \left( \frac{ w^u_s 2^n  + 3\epsilon  w^u_s 2^n \bra{\psi^{u_t}_{s_t}} O_x \ket{\psi^{u_t}_{s_t}} }{w^u_s 2^n} \right) \label{eq:convexity_start}\\
    &= \Expect_x \exp\left( \sum_{t=1}^T \log\left( 1 + 3\epsilon \bra{\psi^{u_t}_{s_t}} O_x \ket{\psi^{u_t}_{s_t}} \right)  \right)\\
    &\geq \exp\left( \sum_{t=1}^T  \Expect_x \log\left( 1 + 3\epsilon \bra{\psi^{u_t}_{s_t}} O_x \ket{\psi^{u_t}_{s_t}} \right)   \right) \label{eq:jensen}\\
    &\geq \exp\left( \sum_{t=1}^T  \frac{1}{M} \sum_{x=1}^{M/2} \log\left( 1 - 9\epsilon^2 \bra{\psi^{u_t}_{s_t}} O_x \ket{\psi^{u_t}_{s_t}}^2 \right)  \right) \label{eq:conditionO} \\
    &\geq \exp\left( - \sum_{t=1}^T  \frac{18}{M} \sum_{x=1}^{M/2}\epsilon^2 \bra{\psi^{u_t}_{s_t}} O_x \ket{\psi^{u_t}_{s_t}}^2  \right) \label{eq:log(1-x)}\\
    &\geq \exp\left( - \sum_{t=1}^T  9 \epsilon^2 \delta(O_1, \ldots, O_M) \right) \label{eq:deltaOcond}
\end{align}
Equation~\eqref{eq:jensen} uses Jensen's inequality.
Equation~\eqref{eq:conditionO} uses the condition on the observables; see Equation~\eqref{eq:negativeness}.
Equation~\eqref{eq:log(1-x)} uses $\log(1-x) \geq -2x, \forall x \in [0, 0.79]$ which is satisfied given $\epsilon < 0.29$.
Equation~\eqref{eq:deltaOcond} uses Equation~\eqref{eq:highincomp}.
Plugging the above analysis to Eq.~\eqref{eq:log(1-x)}, we get
\begin{equation}
    \frac{\left(\Expect_x p^{\rho_x}(\ell) \right)}{p^{\Id / 2^n}(\ell)} \geq \exp\left( -  9 T \epsilon^2 \delta(O_1, \ldots, O_M) \right) \geq 1 - 9 T \epsilon^2 \delta(O_1, \ldots, O_M).
\end{equation}
Utilizing Lemma~\ref{lem:oneside2pt} on a one-sided two-point method, we have that the learning algorithm without quantum memory solves the many-versus-one distinguishing task correctly with probability at most $9 T \epsilon^2 \delta(O_1, \ldots, O_M)$.
Combining with the fact that the learning algorithm without quantum memory solves the many-versus-one distinguishing task correctly with probability at least $2/3$, we have
\begin{equation}
    T \geq \Omega\left(\frac{1}{\epsilon^2 \delta(O_1, \ldots, O_M)} \right).
\end{equation}
This concludes the proof.
\end{proof}

\subsubsection{Lower bound for random observables}

We consider a random collection of observables $O_1, \ldots O_M$ on $n$-qubit systems given by
\begin{align}
&O_1 = U_1 Z_N U_1^\dagger, \,\,\ldots \,\,, O_{M/2} = U_{M/2} Z_N U_{M/2}^\dagger, \label{eq:randomobs1}\\
&O_{M/2 + 1} = -U_{1} Z_N U_{1}^\dagger, \,\, \ldots \,\,, O_{M} = -U_{M/2} Z_N U_{M/2}^\dagger, \label{eq:randomobs2} 
\end{align}
where $U_1, \ldots, U_{M/2}$ are sampled from the Haar measure over the unitary group $\mathbb{U}(2^n)$ and $Z_n = \Id \otimes \ldots \otimes Z$ is the Pauli-Z operator on the $n$-th qubit.
It is easy to check that this collection of observables satisfies Eq.~\eqref{eq:negativeness}.
This random collection of observables is very hard to predict.

\begin{lemma}[$\delta$ for random observables] \label{lem:deltarandomobs}
With a constant probability over the random collection of the observations $\{O_1, \ldots, O_M\}$, we have
\begin{equation} \label{eq:highincomp-random}
    \delta(O_1, \ldots, O_M) = \sup_{\ket{\phi}} \frac{2}{M} \sum_{i=1}^{M/2} \bra{\phi} O_i \ket{\phi}^2 \leq \frac{2304  \pi^3 \log(1 + 2M)}{M} + \frac{4}{M} + \frac{1}{2^n + 1},
\end{equation}
where $\sup_{\ket{\phi}}$ is taken over the entire space of $n$-qubit pure states.
\end{lemma}
\begin{proof}
We prove this lemma by the following strategy: take the covering net over all pure states, obtain exponential concentration inequality, and apply union bound.
We first note that because $U_1, \ldots, U_{M/2}$ are independent random matrices, for a fixed $\ket{\phi}$, we can consider $M/2$ independent and identically distributed pure states $\ket{\psi_1}, \ldots, \ket{\psi_{M/2}}$ sampled from the Haar measure.
Thus we can write 
\begin{equation}
    \frac{\epsilon^2}{M/2} \sum_{i=1}^{M/2} \bra{\psi} U_i Z_n U_i^\dagger \ket{\psi}^2 = \frac{\epsilon^2}{M/2} \sum_{i=1}^{M/2} \bra{\psi_i} Z_n \ket{\psi_i}^2.
\end{equation}
Furthermore we have the following bound on the Lipschitz continuity of the function $f(\ket{\psi}) = \bra{\psi} Z_n \ket{\psi}$,
\begin{align}
    f(\ket{\psi}) - f(\ket{\xi}) &= \bra{\psi} Z_n \ket{\psi} - \bra{\xi} Z_n \ket{\xi}\\
    &\leq (\bra{\psi} - \bra{\xi}) Z_n \ket{\psi} + \bra{\xi} Z_n (\ket{\psi} - \ket{\xi})\\
    &\leq 2 \norm{\ket{\psi} - \ket{\xi}}_2.
\end{align}
Using Levy's lemma and $\E f(\ket{\psi}) = 0$, we have the following concentration inequality
\begin{equation}
    \Pr{ |f(\ket{\psi})| \geq t } \leq 2 \exp\left( - \frac{2^n t^2}{36 \pi^3 } \right).
\end{equation}
This means $f(\ket{\psi})$ is a sub-Gaussian random variable with variance proxy $18 \pi^3 / 2^n$. A well-known fact in high-dimensional probability is that $X^2 - \E X^2$ is a sub-exponential random variable with variance proxy $16 \sigma^2$ if $X$ is a sub-Gaussian random variable with variance proxy $\sigma^2$.
Thus $f(\ket{\psi_i})^2 - \E f(\ket{\psi_i})^2$ is a sub-exponential random variable with variance proxy $\lambda = 288  \pi^3 / 2^n$.
Using Bernstein's inequality, we have
\begin{equation}
    \Pr{ \left|\frac{1}{M/2} \sum_{i=1}^{M/2} f(\ket{\psi_i})^2 - \E f(\ket{\psi_i})^2\right| \geq t } \leq 2 \exp\left( - \frac{M}{2} \min\left( \frac{t^2}{\lambda^2}, \frac{t}{\lambda} \right) \right).
\end{equation}
Using the second moment of Haar-random vectors, we have $\E f(\ket{\psi_i})^2 = 1 / (2^n +1)$.
An $\eta$-covering net $\mathcal{N}_{\eta}$ for $n$-qubit pure state is upper bounded by $(1 + 2 / \eta)^{2 \times 2^n}$.
We can apply union bound to control the fluctuations in the covering net, which gives rise to the following inequality.
\begin{align}
    &\Pr*{\exists \ket{\phi} \in \mathcal{N}_{\eta}, \left|\frac{1}{M/2} \sum_{i=1}^{M/2} \bra{\phi} U_i Z_n U_i^\dagger \ket{\phi}^2 - \frac{1}{2^n + 1} \right| \geq t}\\
    &\leq 2 (1 + 2 / \eta)^{2 \times 2^n} \exp\left( - \frac{M}{2} \min\left( \frac{t^2}{\lambda^2}, \frac{t}{\lambda} \right) \right).
\end{align}
Now, we can extend from the covering net to the entire space of pure states by noting the following Lipschitz continuity property
\begin{align}
    &\frac{1}{M/2} \sum_{i=1}^{M/2} \bra{\phi} U_i Z_n U_i^\dagger \ket{\phi}^2 - \frac{1}{M/2} \sum_{i=1}^{M/2} \bra{\psi} U_i Z_n U_i^\dagger \ket{\psi}^2\\
    &= \frac{1}{M/2} \sum_{i=1}^{M/2} (\bra{\phi} U_i Z_n U_i^\dagger \ket{\phi} - \bra{\psi} U_i Z_n U_i^\dagger \ket{\psi})(\bra{\phi} U_i Z_n U_i^\dagger \ket{\phi} + \bra{\psi} U_i Z_n U_i^\dagger \ket{\psi})\\
    &\leq \frac{1}{M/2} \sum_{i=1}^{M/2} 2 \left|\bra{\phi} U_i Z_n U_i^\dagger \ket{\phi} - \bra{\psi} U_i Z_n U_i^\dagger \ket{\psi}\right| \leq 4 \norm{\ket{\phi} - \ket{\psi}}_2.
\end{align}
This gives rise to the following concentration inequality for any pure states
\begin{align}
    &\Pr*{\exists \ket{\phi}, \left|\frac{1}{M/2} \sum_{i=1}^{M/2} \bra{\phi} U_i Z_n U_i^\dagger \ket{\phi}^2 - \frac{1}{2^n + 1} \right| \geq t + 4 \eta}\\
    &\leq 2 (1 + 2 / \eta)^{2 \times 2^n} \exp\left( - \frac{M}{2} \min\left( \frac{t^2}{\lambda^2}, \frac{t}{\lambda} \right) \right).
\end{align}
We will assume that $M < 2^n$. When $M \geq 2^n$, we can start using classical shadow (an approach for shadow tomography using single-copy incoherent measurement that only requires $\log(M)$ copies), so a lower bound of $M / \log(M)$ would not hold in that regime.
Under the assumption, we choose
\begin{equation}
    \eta = \frac{1}{M}, t = \lambda \times \frac{8 \cdot 2^n \log(1 + 2M)}{M} = \frac{2304  \pi^3 \log(1 + 2M)}{M},
\end{equation}
to yield the following probability bound
\begin{equation}
    \Pr*{\exists \ket{\phi}, \left|\frac{1}{M/2} \sum_{i=1}^{M/2} \bra{\phi} U_i Z_n U_i^\dagger \ket{\phi}^2 - \frac{1}{2^n + 1} \right| \geq t + 4\eta} \leq 2 \exp\left( -2 \right).
\end{equation}
Therefore with constant probability, we have
\begin{equation}
    \sup_{\ket{\phi}} \frac{2}{M} \sum_{i=1}^{M/2} \bra{\phi} U_i Z_n U_i^\dagger \ket{\phi}^2 \leq t + 4\eta + \frac{1}{2^n + 1} = \frac{2304  \pi^3 \log(1 + 2M)}{M} + \frac{4}{M} + \frac{1}{2^n + 1}.
\end{equation}
This concludes the proof of the lemma.
\end{proof}

We can combine Lemma~\ref{lem:deltarandomobs} that upper bounds $\delta(O_1, \ldots, O_M)$ for the random collection of observables and Theorem~\ref{thm:shadowtomo} to obtain the following corollary.
This corollary also implies the existence of a collection of observables $O_1, \ldots, O_M$ such that any learning algorithm without quantum memory requires $\Omega(\min(M / \log(M), 2^n) / \epsilon^2)$ copies of $\rho$ to perform shadow tomography.

\begin{corollary}[Shadow tomography lower bound for random observables] \label{cor:shadow-random}
With a constant probability over the random collection of the observables $\{O_1, \ldots, O_M\}$, any learning algorithm without quantum memory requires
\begin{equation}
T \geq \Omega\left( \frac{\min(M / \log(M), 2^n)}{\epsilon^2} \right)
\end{equation}
copies of $\rho$ to predict expectation values of $\Tr(O_x \rho)$ to at most $\epsilon$-error for all $x = 1, \ldots, M$ with at least a probability of $2/3$.
\end{corollary}

\subsubsection{Lower bound for Pauli observables}

Recall that we define an $n$-qubit Pauli observable to be the tensor product of $n$ operators from the set $\{\Id, X, Y, Z\}$; see Definition~\ref{def:pauli}.
To construct a collection of observables that are traceless and satisfies Equation~\eqref{eq:conditionO}, we define the following collection of $2(4^n-1)$ Pauli observables with plus and minus signs,
\begin{align}
    P_1, \ldots, P_{4^n - 1} &\in \{ +\sigma_1 \otimes \ldots \otimes \sigma_n, \,\, \sigma_i \in \{\Id, X, Y, Z\}, \forall i \} \setminus \{\Id \otimes \ldots \otimes \Id\},\\
    P_{4^n}, \ldots, P_{2(4^n - 1)} &\in \{ -\sigma_1 \otimes \ldots \otimes \sigma_n, \,\, \sigma_i \in \{\Id, X, Y, Z\}, \forall i \} \setminus \{-\Id \otimes \ldots \otimes \Id\},
\end{align}
This collection of observable yields an exponentially small $\delta\left(P_1, \ldots, P_{2 \times (4^n-1)}\right)$.

\begin{lemma}[$\delta$ for Pauli observables] \label{lem:deltapauli}
\begin{equation} \label{eq:highincomp-pauli}
    \delta\left(P_1, \ldots, P_{2 \times (4^n-1)}\right) = \sup_{\ket{\phi}} \frac{1}{4^n-1} \sum_{i=1}^{4^n-1} \bra{\phi} P_i \ket{\phi}^2 = \frac{1}{2^n + 1}.
\end{equation}
\end{lemma}
\begin{proof}
For any $n$-qubit pure state $\ket{\phi}$, we have
\begin{align}
    \frac{1}{4^n-1} \sum_{i=1}^{4^n-1} \bra{\phi} P_i \ket{\phi}^2 &= \frac{1}{4^n-1} \Tr\left(\left( \sum_{i=1}^{4^n-1} P_i \otimes P_i\right) \ketbra{\phi}{\phi}^{\otimes 2} \right)\\
    &= \frac{1}{4^n-1} \Tr\left(\left( 2^n \mathrm{SWAP}_n - \Id^{\otimes 2n} \right) \ketbra{\phi}{\phi}^{\otimes 2} \right)\\
    &= \frac{2^n - 1}{4^n-1} = \frac{1}{2^n + 1}.
\end{align}
The second equality follows from Lemma~\ref{lem:2design}.
\end{proof}

We can combine Lemma~\ref{lem:deltapauli} that characterizes $\delta(P_1, \ldots, P_{2 \times (4^n-1)})$ for the collection of Pauli observables and Theorem~\ref{thm:shadowtomo} to obtain the following corollary.

\begin{corollary}[Shadow tomography lower bound for Pauli observables]\label{cor:shadow-pauli}
Any learning algorithm without quantum memory requires
\begin{equation}
T \geq \Omega\left( 2^n / \epsilon^2 \right)
\end{equation}
copies of $\rho$ to predict expectation values of $\Tr(P_i \rho)$ to at most $\epsilon$-error for all $i = 1, \ldots, 2(4^n - 1)$ with at least a probability of $2/3$.
\end{corollary}

\subsubsection{Upper bound}

We give a upper bound for performing shadow tomography using a learning algorithm with only classical memory. There are two regimes. When $M \leq 2^n$, we apply the naive strategy of measuring each observable sequentially. When $M > 2^n$, we apply the classical shadow protocol \cite{huang2020predicting}.

\begin{theorem}[General shadow tomography upper bound]\label{thm:shadowtomo_upper}
For any collection of observables $\{O_1, \ldots, O_M\}$ on $n$-qubit state with $\norm{O_i}_\infty \leq 1$, there is a learning algorithm without quantum memory that uses
\begin{equation}
T \leq \mathcal{O}\left( \frac{\min(M \log(M), 2^n \log(M))}{\epsilon^2} \right)
\end{equation}
copies of $n$-qubit state $\rho$ to predict expectation values of $\Tr(O_x \rho)$ to at most $\epsilon$-error for all $x = 1, \ldots, M$ with high probability.
\end{theorem}
\begin{proof}
    When $M \leq 2^n$, we measure each observable $O_x$ on the unknown state $\rho$ sequentially.
    For each observable $O_x$, we measure $O_x$ on $\rho$ for $\mathcal{O}(\log(M) / \epsilon^2)$ times, such that the empirical average is close to $\Tr(O_x \rho)$ up to $\epsilon$ error with probability at least $1 - 0.01/M$.
    This gives a total of $T = \mathcal{O}\left(M \log(M) / \epsilon\right)$ measurements on $T$ copies of $\rho$.
    A union bound over all $O_x$ shows that with probability at least $0.99$, we can estimate all $M$ expectation values $\Tr(O_1 \rho), \ldots, \Tr(O_M \rho)$ to $\epsilon$ error.

    When $M > 2^n$, we apply the classical shadow protocol under random Clifford measurement given in \cite{huang2020predicting}.
    We perform a random Clifford measurement on each copy of $\rho$.
    Random Clifford measurement is equivalent to apply a random Clifford unitary sampled uniformly from the Clifford group to $\rho$, then measure in the computational basis.
    We record both the random Clifford unitary and the $n$-bit outcome from computational basis measurement.
    We perform the random Clifford measurement for $T = \mathcal{O}\left( 2^n \log(M) / \epsilon^2 \right)$ times to gather a set of classical measurement data, also referred to as the classical shadow of $\rho$.
    
    By combining Theorem 1 and Equation~(S16) in \cite{huang2020predicting}, we find that,
    for any collection of observables $\{O_1, \ldots, O_M\}$ with $\Tr(O_x^2) \leq 2^n$, with high probability, we can use the classical measurement data (classical shadow of $\rho$) to estimate $\Tr(O_x \rho)$ to $\epsilon$ additive error for all $x = 1, \ldots, M$.
    Because $\norm{O_i}_\infty \leq 1$, we have $\Tr(O_x^2) \leq 2^n \norm{O_i}_\infty^2 \leq 2^n$.
    Hence, we can apply classical shadow protocol with random Clifford measurement to achieve the stated complexity when $M > 2^n$. The stated result follows from combining the complexity from both $M \leq 2^n$ and $M > 2^n$.
\end{proof}

We can see that the above upper bound matches with the lower bound for random observables stated in Corollary~\ref{cor:shadow-random} up to poly-logarithmic scaling in $M$.
When $O_1, \ldots, O_M$ are Pauli observables, we have $M = 2(4^n - 1)$ and the above upper bound becomes $\mathcal{O}(n 2^n / \epsilon^2)$, which matches the lower bound stated in Corollary~\ref{cor:shadow-pauli} up to a factor of $n$.

\subsection{Testing purity of the quantum state}

\subsubsection{Lower bound}

In this subsection, we provide exponential lower bound for testing if a quantum state is pure or maximally mixed. In this section, we will denote $d = 2^n$ to be the Hilbert space dimension.

\begin{theorem}[Purity testing lower bound]\label{thm:purity_lower}
    Any learning algorithm without quantum memory requires \begin{equation}
        T \ge \Omega \left( 2^{n/2} \right)
    \end{equation} copies of $\rho\in\bH^{2^n\times 2^n}$ to distinguish between whether $\rho$ is a pure state or a maximally mixed state with probability at least $2/3$.
\end{theorem}

\begin{proof}
    Let $\calT$ be the tree corresponding to any given learning algorithm for this distinguishing task. By Lemma~\ref{lem:oneside2pt} it suffices to lower bound $\E[v]*{p^{\ketbra{v}{v}}(\ell)/{p^{\rhomm}(\ell)}}$ for all leaves $\ell$.
    If $\brc{e_{u_t,s_t}}_{t=1}^T$ are the edges on the path from root to the leaf $\ell$, then \begin{equation}
        \E[v]*{\frac{p^{\ketbra{v}{v}}(\ell)}{p^{\rhomm}(\ell)}} = \E[v]*{\prod^T_{t = 1} d \braket{\psi^{u_t}_{s_t}|v}^2} = \frac{d^T}{d(d+1)\cdots(d+T-1)}\cdot \sum_{\pi\in\calS_T}\Tr\left(\pi \bigotimes^T_{t=1}\ketbra{\psi^{u_t}_{s_t}}{\psi^{u_t}_{s_t}}\right). \label{eq:integrate_v}
    \end{equation}
    The second equality follows from Lemma~\ref{lem:churchsymsubspace} where $\pi$ is the permutation operator.
    By Lemma~\ref{lem:maincombo} below, the sum in \eqref{eq:integrate_v} is lower bounded by 1, so
    \begin{equation}
        \E[v]*{\frac{p^{\ketbra{v}{v}}(\ell)}{p^{\rhomm}(\ell)}} \ge \frac{d^T}{d(d+1)\cdots(d+T-1)} \ge \prod^{T-1}_{t=0}\left(1 - \frac{t}{d}\right)^{-1}
        \ge \left(1 - \frac{T}{d}\right)^T.
    \end{equation}
    Using Lemma~\ref{lem:oneside2pt}, we have the probability that the given learning algorithm successfully distinguishes the two settings is upper bounded by
    $1 - \left(1 - \frac{T}{d}\right)^T$.
    Therefore, $2/3 \leq 1 - \left(1 - \frac{T}{d}\right)^T$ implying that $T \geq \Omega(\sqrt{d})$.
\end{proof}

The key step in the above proof is the following technical lemma that lower bounds the norm of the projection of any tensor product of pure states to the symmetric subspace.

\begin{lemma}\label{lem:maincombo}
    For any collection of pure states $\ket{\psi_1},\ldots,\ket{\psi_T}\in\bH^d$, \begin{equation}
        \sum_{\pi\in\calS_T}\Tr\left(\pi\bigotimes^T_{t=1}\ketbra{\psi_t}{\psi_t}\right) \ge 1. \label{eq:maincombo}
    \end{equation}
\end{lemma}

\begin{proof}
    Let $\Pi$ denote the projector to the symmetric subspace in $(\mathbb{C}^{2^n})^{\otimes T}$. Note that \eqref{eq:maincombo} is equivalent to the statement that $\Tr\left(\Pi\bigotimes^T_{t=1}\ketbra{\psi_t}{\psi_t}\right) \ge 1/T!$. This is clearly true for $T = 1$; we proceed by induction on $T$. Let $\wt{\Pi}$ denote the projector to the symmetric subspace in $(\mathbb{C}^{2^n})^{\otimes T-1}$, and define the (unnormalized) state $\ket{\wt{\psi}} \triangleq \wt{\Pi}\bigotimes^{T}_{t=2}\ket{\psi_t}$.
    
    As $\wt{\Pi}$ is a projector, we have \begin{equation}
        \braket{\wt{\psi}|\wt{\psi}} = \Braket{\bigotimes^T_{t=2}\psi_t \wt{\Pi}|\wt{\Pi}\bigotimes^T_{t=2}\psi_t} = \Tr\left(\wt{\Pi}\bigotimes^T_{t=2}\ketbra{\psi_t}{\psi_t}\right) \ge \frac{1}{(T-1)!},
    \end{equation} where the last step follows by the inductive hypothesis.
    
    We can rewrite the left-hand side of \eqref{eq:maincombo} as $\sum_{\pi\in\calS_T}\Tr\left(\pi \ketbra{\psi_1}{\psi_1} \otimes \ketbra{\wt{\psi}}{\wt{\psi}}\right)$ and decompose this sum into $\pi$ for which $\pi(1) = 1$ and all other $\pi$. Note that \begin{equation}
        \sum_{\pi\in\calS_T: \pi(1) = 1}\Tr\left(\pi\ketbra{\psi_1}{\psi_1} \otimes \ketbra{\wt{\psi}}{\wt{\psi}}\right) = \sum_{\wt{\pi}\in\calS_{T-1}} \Tr\left(\wt{\pi}\ketbra{\wt{\psi}}{\wt{\psi}}\right) = (T-1)!\braket{\wt{\psi}|\wt{\psi}} \ge 1.
    \end{equation}
    It remains to argue that $\sum_{\pi\in \calS_T: \pi(1)\neq 1} \Tr\left(\pi\ketbra{\psi_1}{\psi_1} \otimes \ketbra{\wt{\psi}}{\wt{\psi}}\right) \ge 0$.
    
    \begin{figure}
        \centering
\includegraphics[scale=.32, valign = c]{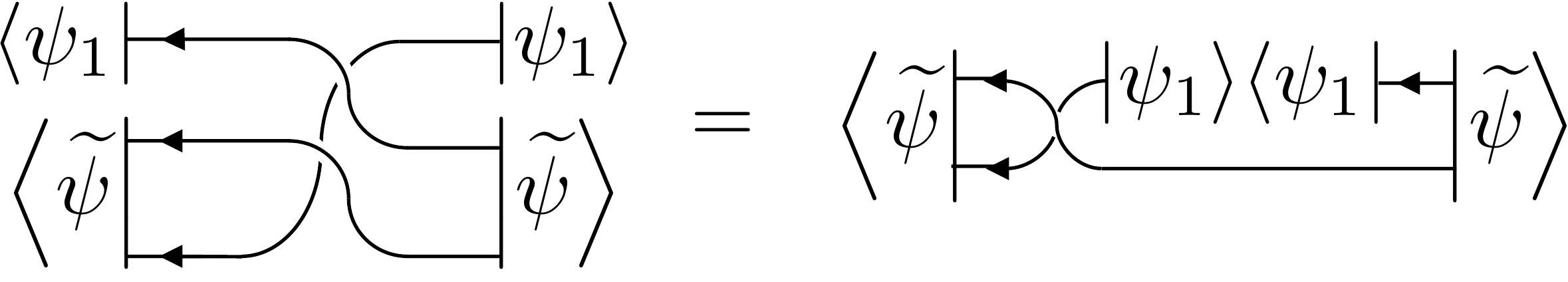}
        \caption{Illustration of the equality \eqref{eq:rearrange} for $T = 3$, $\pi = (123)$. The corresponding permutation $\check{\pi} = (12)$ can be seen on the right-hand side.}
        \label{fig:purity_rearrange}
    \end{figure}
    
    Consider the map which sends any $\pi\in\calS_T$ for which $\pi(1)\neq 1$ to $\check{\pi}\in\calS_{T-1}$ defined as follows. For any $2\le i\le T$, for which $\pi(i)\neq 1$, $\check{\pi}(i - 1)\triangleq \pi(i) - 1$ and for $i = \pi^{-1}(1) > 1$, $\check{\pi}(i - 1)\triangleq \pi(1)$. Then for any $\pi\in\calS_T$, \begin{align}
        \Tr\left(\pi\ketbra{\psi_1}{\psi_1} \otimes \ketbra{\wt{\psi}}{\wt{\psi}}\right) &= \Tr\left(\check{\pi}(\underbrace{\Id\otimes\cdots \otimes\Id}_{\pi(1) - 2}\otimes \ketbra{\psi_1}{\psi_1}\otimes\underbrace{\Id\otimes\cdots\otimes \Id}_{T - \pi(1)}) \ketbra{\wt{\psi}}{\wt{\psi}}\right) \label{eq:rearrange}\\
        &= \Tr\left((\Id\otimes\cdots \otimes\Id\otimes \ketbra{\psi_1}{\psi_1}\otimes\Id\otimes\cdots\otimes \Id) \ketbra{\wt{\psi}}{\wt{\psi}} P^{\dagger}_{\check{\pi}}\right) \\
        &= \Tr\left((\Id\otimes\cdots \otimes\Id\otimes \ketbra{\psi_1}{\psi_1}\otimes\Id\otimes\cdots\otimes \Id) \ketbra{\wt{\psi}}{\wt{\psi}}\right) \ge 0,
    \end{align} as claimed, where the first step \eqref{eq:rearrange} is illustrated in Figure~\ref{fig:purity_rearrange}.
\end{proof}

\subsubsection{Upper bound}

Here we give a simple algorithm for the above distinguishing task that matches the lower bound in Theorem~\ref{thm:purity_lower} up to constant factors.

\begin{theorem}\label{thm:purity_upper}
    There is a learning algorithm without quantum memory which takes $T = O(2^{n/2})$ copies of $\rho$ to distinguish between whether $\rho$ is a pure state or maximally mixed.
\end{theorem}

To prove this, we will use the following well-known result from classical distribution testing:

\begin{theorem}[\cite{chan2014optimal,diakonikolas2014testing,canonne2018testing}]\label{thm:ilias}
    Given $0 < \epsilon<1$ and sample access to a distribution $q$ over $[d]$, there is an algorithm {\sc TestUniformityL2}($q,d,\epsilon)$ that uses $T = O(\sqrt{d}/\epsilon^2)$ samples from $q$ and with probability $9/10$ distinguishes whether $q$ is the uniform distribution over $[d]$ or $\epsilon/\sqrt{d}$-far in $L_2$ distance from the uniform distribution.
\end{theorem}

We will also need the following standard moment calculation:

\begin{lemma}[Lemma 6.4 in \cite{chen2021toward}]\label{lem:moment24}
    For Haar-random $\U\in U(2^n)$ and $\rho\in\bH^{2^n\times 2^n}$, let $Z$ denote the random variable $ \sum^{2^n}_{i=1}\left(\bra{i} \U^{\dagger} \M \U \ket{i}\right)^2$. Then \begin{equation}
        \E{Z} = \frac{1}{2^n+1}\left(\Tr(\M)^2 + \norm{\M}^2_{\HS}\right).
    \end{equation} If in addition we have that $\Tr(\M) = 0$, then \begin{equation}
        \E{Z^2} \le \frac{1 + o(1)}{4^n}\norm{\M}^4_{\HS}.
    \end{equation}
\end{lemma}

We are now ready to prove Theorem~\ref{thm:purity_upper}.

\begin{proof}[Proof of Theorem~\ref{thm:purity_upper}]
    Sample a Haar-random basis $\brc{\U\ket{i}}_{i\in[2^n]}$ and measure every copy of $\rho$ in this basis. If $\rho$ is maximally mixed, note that the distribution over outcomes from a single measurement is the uniform distribution $u$ over $[2^n]$. On the other hand, if $\rho$ is a pure state, let $Z$ denote the random variable $\norm{q^{\U} - u}^2_2$, where $q^{\U}$ is the distribution over outcomes from a single measurement. Note that $Z$ is precisely the random variable $Z$ defined in Lemma~\ref{lem:moment24} for $\M = \rho - \rhomm$, so we conclude that $\E{Z} = \frac{1}{2^n + 1}\cdot \norm{\rho - \rhomm}^2_{\HS}$ and $\E{Z^2} \le \frac{1 + o(1)}{16^n}\cdot \norm{\rho - \rhomm}^4_{\HS}$, so by Paley-Zygmund, there is an absolute constant $c > 0$ for which $\Pr{\abs{Z} \ge c\norm{\rho - \rhomm}^2_{\HS}} \ge 9/10$. Note that $\norm{\rho - \rhomm}^2_{\HS} = 1 - 1/2^n$, so with probability at least $9/10$ over the randomness of $\U$, $\norm{q^{\U} - u}_2 \ge \Omega(2^{-n/2})$.
    

    So by Theorem~\ref{thm:ilias}, {\sc TestUniformityL2}($q^{\U},2^n,\Theta(2^{-n/2})$) will take $T = O(2^{n/2})$ samples from $q$ and correctly distinguish whether $q$ is uniform or far from uniform with probability at least $4/5$ over the randomness of the algorithm and of $\U$.
\end{proof}

\section{Exponential separation with bounded quantum memory}
\label{sec:bounded}

In this section, we consider a separation between algorithms with different quantum memory size.
To the end, we will give a shadow tomography task and prove that any learning algorithm with $k$-qubit memory require $\Omega(2^{(n-k)/3})$ copies. In contrast, learning algorithms with $n$-qubit memory can succeed in the task using $\mathcal{O}(n)$ copies.
In particular, the task is to predict the absolute value of the expectation $\Tr(P \rho)$ for all Pauli observables $P$ in an unknown state $\rho$.

\subsection{Lower bound with \texorpdfstring{$k$}{k}-qubit of quantum memory}


We restate the main theorem of this section, originally presented in Section~\ref{sec:intro_bounded}, for convenience:

\paulimemory*

To establish the theorem, we will again utilize the tree structure. The main difference is that each node on the tree is associated with a positive-semidefinite matrix that represents the quantum state of the $k$-qubit memory scaled by the probability of arriving at the node.
In particular, we will work with the following analogue of Definition~\ref{def:treelearnstate} for the bounded quantum memory setting:

\begin{definition}[Tree representation for learning states with bounded quantum memory]\label{def:treebounded}
	Fix an unknown $n$-qubit quantum state $\rho$. A learning algorithm with size-$k$ quantum memory can be expressed as a rooted tree $\calT$ of depth $T$, where each node encodes the current state of the quantum memory in addition to the transcript of measurement outcomes the algorithm has seen so far. Specifically, the tree satisfies the following properties:
	\begin{enumerate}
		\item Each node $u$ is associated with a $k$-qubit unnormalized mixed state $\Sig^{\rho}(u)$ corresponding to the current state of the quantum memory.
		\item For the root $r$ of the tree, $\Sig^{\rho}(r)$ is an initial state denoted by $\Sig_0$.
		\item At each node $u$, we apply a POVM measurement $\brc{F^u_s}_s$ on $\Sig^{\rho}(u)\otimes \rho$ to obtain a classical outcome $s$. Each child node $v$ of $u$ is connected through the edge $e_{u,s}$.
		\item For POVM element $F = M^{\dagger}M$ and $\Sig\in\bH^{2^k\times 2^k}$, define \begin{equation}
			A^{\rho}_{M}(\Sig) \triangleq \Tr_{>k}\left(M(\Sig\otimes \rho)M^{\dagger}\right).
		\end{equation} If $v$ is the child node of $u$ connected through the edge $e_{u,s}$, then
		\begin{equation} \label{eq:recursivesigma}
		\Sig^{\rho}(v) \triangleq A^{\rho}_{M^u_s}(\Sig^{\rho}(u)).    
		\end{equation}
		$A^{\rho}_{M}(\Sig)$ is the first $k$-qubit of the unnormalized post-measurement state from Definition~\ref{eq:POVMpostmeasurement}.
		\item Note that for any node $u$, $p^{\rho}(u)\triangleq \Tr(\Sig^{\rho}(u))$ is the probability that the transcript of measurement outcomes observed by the learning algorithm after $t$ measurements is $u$. And $\Sig^{\rho}(u) / p^{\rho}(u)$ is the state of the $k$-qubit memory at the node $u$.
	\end{enumerate}
\end{definition}

Given $\rho$, we will abuse notation and let $p^{\rho}$ denote the distribution on leaves of $\calT$ given by the probabilities $\brc{p^{\rho}(\ell)}_{\ell}$. Let $P_1,\ldots,P_{4^n - 1}$ denote the collection of non-identity $n$-qubit Pauli observables from Definition~\ref{def:pauli}; in this section, when we refer to a Pauli $P$ we mean one from this collection, and we will use $\E[P]{\cdot}$ to denote expectation with respect to a uniformly random such $P$.

We consider a many-versus-one distinguishing task where we want to distinguish between the completely mixed state $\rhomm$ versus the set of $n$-qubit states $\brc{\rho_P}$ where $P$ ranges over all $n$-qubit Pauli observables not equal to identity, where $\rho_P \triangleq (I + P) / 2^n$.
As in the proof of Theorem~\ref{thm:shadowtomo}, a lower bound for this task immediately translates to one for shadow tomography. For the former, it suffices to show that for $T = o\left(2^{(n - k)/3}\right)$, $\tvd(p^{\rhomm}, \E[P]{p^{\rho_P}}) = o(1)$.

As in the proofs in Section~\ref{sec:learnstate}, the primary technical ingredient for this will be a second moment bound. To formulate this, we will need the following object:
\begin{definition}\label{defn:taup}
	Given a POVM element $F = M^{\dagger}M$ and an unnormalized mixed state $\Sig\in\bH^{2^k\times 2^k}$, $\tau_{M,\Sig}\in\bH^{2^{n+k}\times 2^{n+k}}$ is given by
    \begin{align}
    \label{E:tau}
        \includegraphics[scale=.32, valign = c]{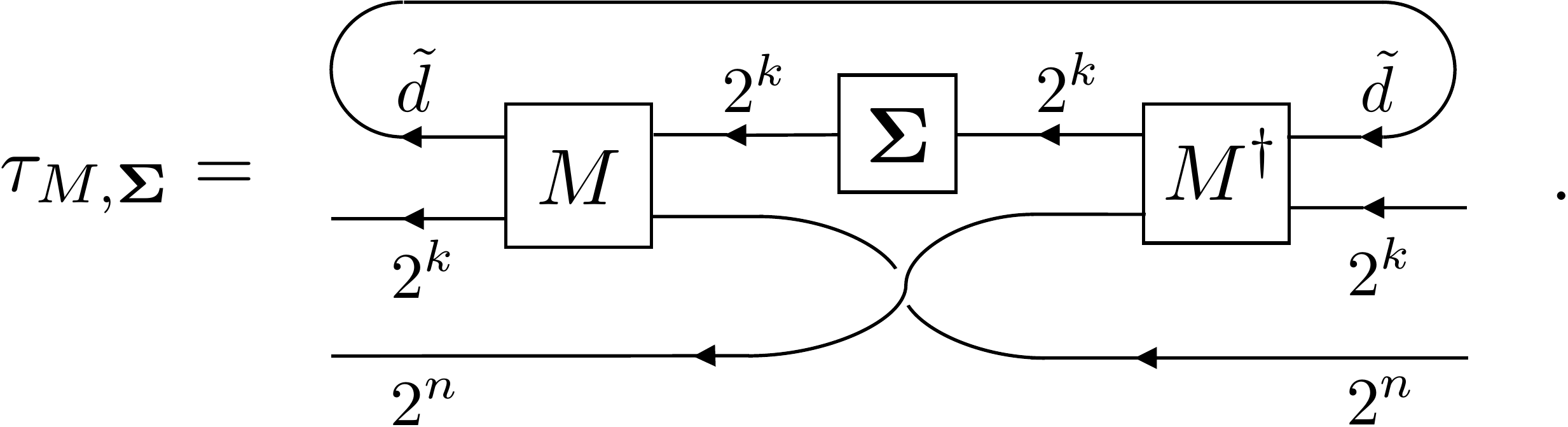}
    \end{align}
    (The dimensions of the Hilbert spaces corresponding to the edges have been labeled.)
\end{definition}
\noindent One can think of the following lemma as bounding a matrix-valued analogue of the quantity defined in \eqref{eq:highincomp-pauli}. Indeed, when $k = 0$, the following specializes up to constant factors to Lemma~\ref{lem:deltapauli}.

\begin{lemma}\label{fact:secondmoment_sizek}
	For any POVM element $M$ and unnormalized mixed state $\Sig\in\bH^{2^k\times 2^k}$,
	\begin{equation}
		\E[P]*{\norm{A^{\rhomm}_{M}(\Sig) - A^{\rho_P}_{M}(\Sig)}^2_{\tr}} \le \frac{1}{2^{n - k}} \cdot \frac{1}{2^{2n} - 1}\cdot \Tr\left(\tau_{M,\Sig}\right)^2. \label{eq:secondmoment_sizek}
	\end{equation}
\end{lemma}

\begin{proof}
	Note that $A^{\rhomm}_{M}(\Sig) - A^{\rho_P}_{M}(\Sig) = -\Tr_{>k}\left(M\left(\frac{P}{2^n} \otimes \Sig\right)M^{\dagger}\right)$, so by the fact that $\norm{X}_{\mathrm{tr}}^2 \leq 2^k \Tr(X^2), \forall X \in \mathbb{H}^{2^k \times 2^k}$,
	the left-hand side of
	\eqref{eq:secondmoment_sizek} is upper bounded by
	\begin{equation}\label{eq:2kfrobeniusnorm}
		2^k \E[P]*{\Tr\left(\left[\Tr_{>k}\left(M\left(\frac{P}{2^n}\otimes \Sig\right)M^{\dagger}\right)\right]^2\right)}.
	\end{equation}
	For fixed $P$, we express the expression inside the expectation diagrammatically as
    \begin{align}
    \label{E:trk}
        \includegraphics[scale=.32, valign = c]{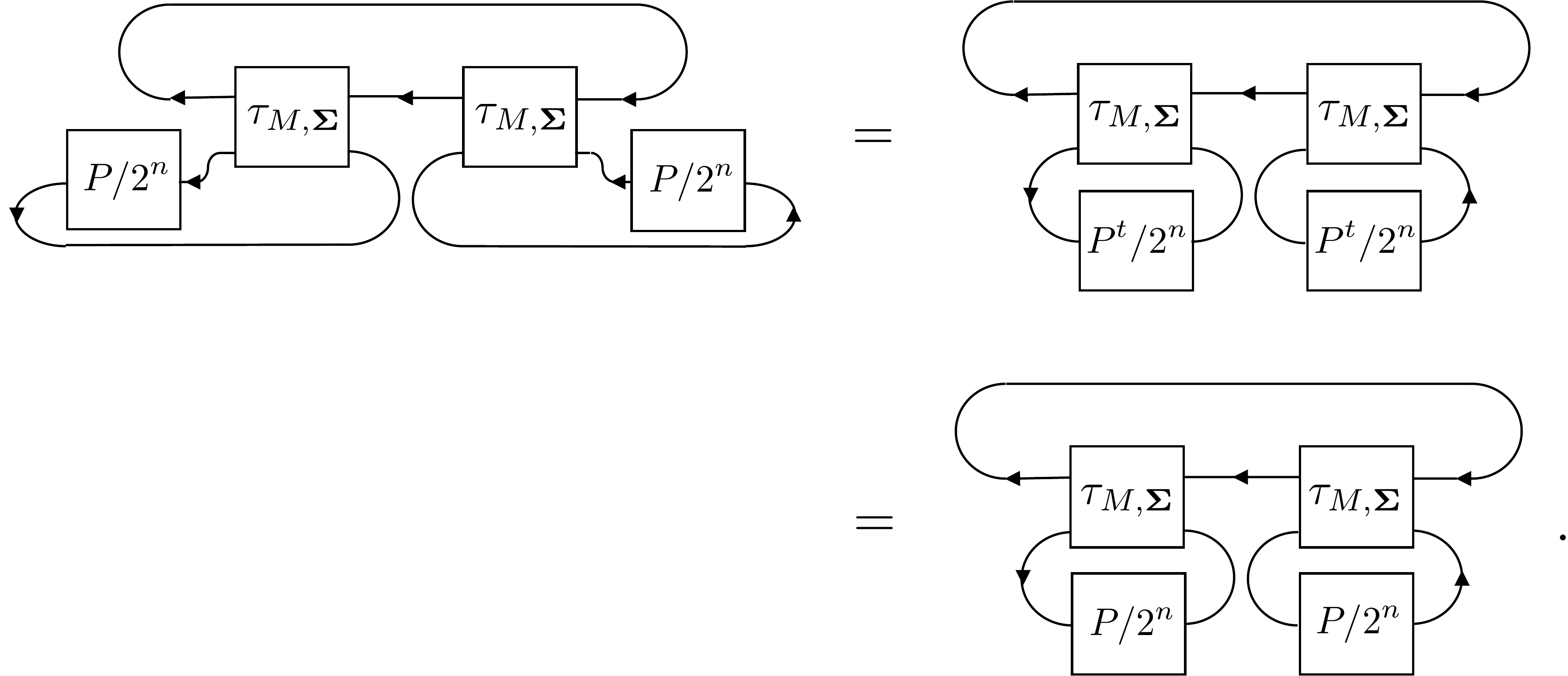}
    \end{align}
	By Lemma~\ref{lem:2design}, averaging~\eqref{E:trk} with respect to $P$ yields 
    \begin{align}
    \label{E:tau2}
        \includegraphics[scale=.32, valign = c]{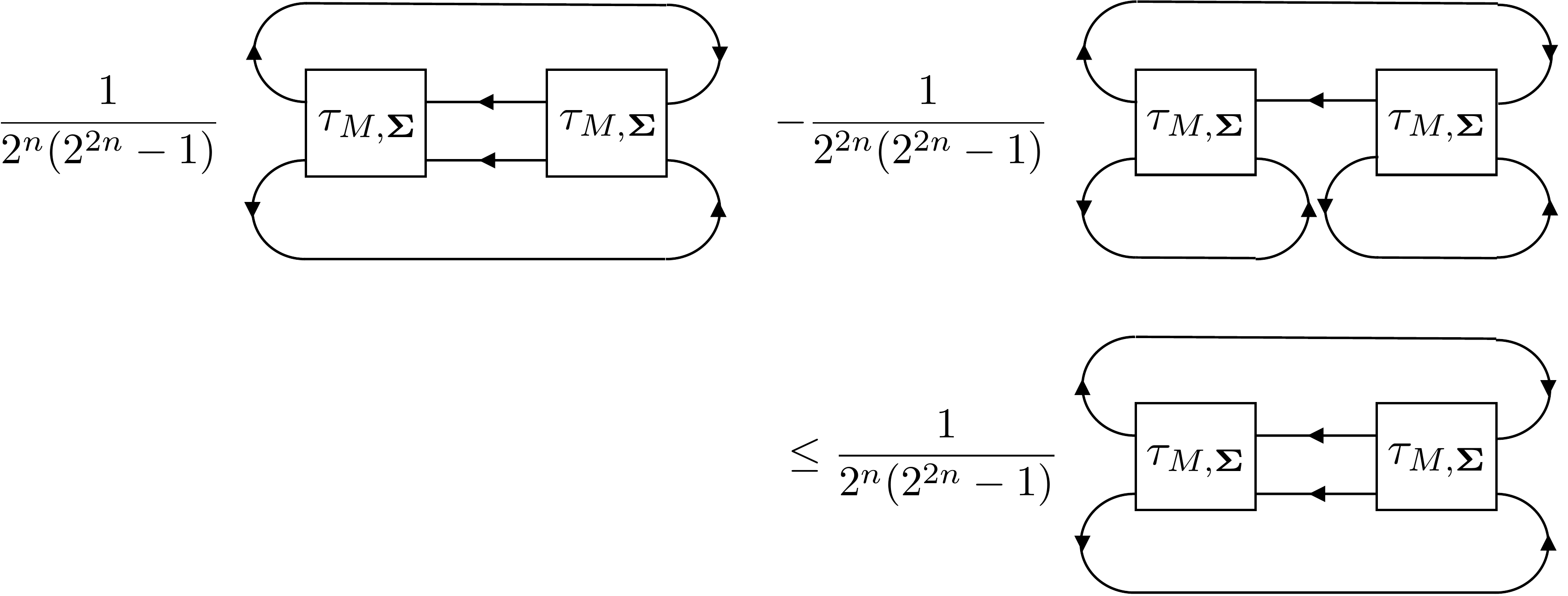}
    \end{align}
 where the inequality follows from the fact that the second term is equal to $\frac{1}{2^{2n} (2^{2n} - 1)} \Tr(\Tr_{> k}(\tau)^2)$ which is non-negative.
	The claim follows from the fact that $\Tr(\tau^2) \le \Tr(\tau)^2$ (as $\tau$ is positive-semidefinite) and utilizing Eqn.~\eqref{eq:2kfrobeniusnorm}.
\end{proof}

We will not be able to make use of the convexity trick present in the proof of Theorem~\ref{thm:shadowtomo} when each node is associated with a probability instead of positive-semidefinite matrix.
Instead, we make use of a careful pruning argument; intuitively, for any leaf $\ell$, we will essentially ignore the contribution to $\tvd(p^{\rhomm}, \E[P]{p^{\rho_P}})$ coming from Paulis $P$ for which $A^{\rho_P}_{M^u_s}$ behaves too differently to $A^{\rhomm}_{M^u_s}$ for some edge $e_{u,s}$ on the path from root to $\ell$.

\begin{definition} \label{def:badgoodPauli}
	A Pauli $P$ is \emph{bad} for an edge $e_{u,s}$ if \begin{equation}
		\norm*{A^{\rhomm}_{M^u_s}(\Sig^{\rhomm}(u)) - A^{\rho_P}_{M^u_s}(\Sig^{\rhomm}(u))}_{\tr} \ge \frac{1}{2^{(n-k)/3}} \cdot \sqrt{\frac{1}{2^{2n} - 1}}\cdot \Tr\left(\tau_{M^u_s,\Sig^{\rhomm}(u)}\right).
	\end{equation} Otherwise we say $P$ is \emph{good} for $e_{u,s}$. Given node $u$, let $P[u]$ denote the set of all Paulis which are good for all edges on the path from root to $u$.
\end{definition}

The following is an immediate consequence of Fact~\ref{fact:secondmoment_sizek} and Markov's:

\begin{fact}\label{fact:fewbad}
	For any edge $e_{u,s}$, there are at most $2^{-(n-k)/3}\cdot (4^n - 1)$ bad Paulis $P\in\bH^{2^n\times 2^n}$. In particular, along any given root-to-leaf path of the learning tree, there are at most $T\cdot 2^{-(n-k)/3}\cdot (4^n - 1)$ Paulis which are bad for some edge along the path.
\end{fact}

Lemma~\ref{fact:fewbad} allows us to bound the $\tvd(p^{\rhomm}, \E[P]{p^{\rho_P}})$ by a small term coming from bad Paulis and a term coming from good ones:

\begin{lemma}\label{lem:goodbad}
	\begin{equation}
		\tvd\left( p^{\rhomm}, \E[P]{p^{\rho_P}} \right) \le T\cdot 2^{-(n-k)/3} + \frac{1}{4^n-1} \sum_{\ell \in\leaf(\calT)} \ \sum_{P \in P[\ell]} \norm*{\Sig^{\rhomm}(\ell) - \Sig^{\rho_P}(\ell)}_{\tr} \label{eq:goodbad}
	\end{equation}
\end{lemma}

\begin{proof}
	Let $\calL$ denote the set of leaves $\ell$ for which $p^{\rhomm}(\ell) \ge \E[P]{p^{\rho_P}(\ell)}$. Then
	\begin{align}
		\tvd\left(p^{\rhomm}, \E[P]{p^{\rho_P}}\right) &= \sum_{\ell\in\calL} p^{\rhomm}(\ell) - \E[P]{p^{\rho_P}(\ell)} \\
		&\le \sum_{\ell\in\calL} \E[P]*{\min(p^{\rhomm}(\ell), {\abs*{p^{\rhomm}(\ell) - p^{\rho_P}(\ell)}})} \\
		&\le \sum_{\ell\in\calL} \E[P]*{\min({p^{\rhomm}(\ell)}, {\norm*{\Sig^{\rhomm}(\ell) - \Sig^{\rho_P}(\ell)}_{\tr}}) } \label{ineq:probtotrace} \\
		&\le \sum_{\ell\in\calL}\brk*{\Pr{P\not\in P[\ell]}\cdot p^{\rhomm}(\ell) + \frac{1}{4^n - 1}\sum_{P\in P[\ell]}\norm*{\Sig^{\rhomm}(\ell) - \Sig^{\rho_P}(\ell)}_{\tr}},
	\end{align}
	The first equality uses the fact that $\tvd(p, q) = \frac{1}{2} \sum_i |p_i - q_i| = \sum_{i: p_i \geq q_i} (p_i - q_i)$.
	Inequality~\eqref{ineq:probtotrace} uses the fact that $\norm*{\Sig^{\rhomm}(\ell) - \Sig^{\rho_P}(\ell)}_{\tr} \geq \Tr(\Sig^{\rhomm}(\ell) - \Sig^{\rho_P}(\ell)) = p^{\rhomm}(\ell) - p^{\rho_P}(\ell)$.
	The lemma follows from Fact~\ref{fact:fewbad} and the fact that $\sum_{\ell} p^{\rhomm}(\ell) \le 1$.
\end{proof}

We are now ready to prove Theorem~\ref{thm:pauli_memory}.

\begin{proof}[Proof of Theorem~\ref{thm:pauli_memory}]
	By Lemma~\ref{lem:goodbad}, it suffices to control the latter term on the right-hand side of Eq.~\eqref{eq:goodbad}. We do so via a hybrid argument. For any leaf $\ell$ with parent $u$ and incoming edge $e_{u,s}$, and any $P\in P[\ell]$, we can upper bound $\norm*{\Sig^{\rhomm}(\ell) - \Sig^{\rho_P}(\ell)}_{\tr}$ by \begin{equation}
		\norm*{A^{\rhomm}_{M^{u}_{s}}(\Sig^{\rhomm}(u)) - A^{\rho_P}_{M^{u}_{s}}(\Sig^{\rhomm}(u))}_{\tr} + \norm*{A^{\rho_P}_{M^{u}_{s}}\left(\Sig^{\rhomm}(u)-\Sig^{\rho_P}(u)\right)}_{\tr}, \label{eq:hybrid}
	\end{equation} where we have used the transition formula in Eq.~\eqref{eq:recursivesigma} and the triangle inequality.
	We can upper bound the former term by \begin{equation}
		\frac{1}{2^{(n-k)/3}}\cdot \sqrt{\frac{2^{2n}}{2^{2n} - 1}}\cdot \Tr\left(\frac{1}{2^n}\tau_{M^{u}_{s},\Sig^{\rhomm}(u)}\right) \label{eq:sumgood1}
	\end{equation} because $P$ is good for edge $e_{u,s}$; see Definition~\ref{def:badgoodPauli} and note that $\sqrt{2^{2n}} \cdot \frac{1}{2^n} = 1$.
	Using Definition~\ref{defn:taup} and the fact that $\{(M_s^u)^{\dagger}M_s^u\}_s$ is a POVM hence $\sum_s (M_s^u)^\dagger M_s^u = \mathds{1}$, we have the following identity.
	\begin{equation}
		\sum_{s} \E[P]*{\Tr\left(\frac{1}{2^n}\tau_{M^{u}_{s},\Sig^{\rhomm}(u)}\right)} =  \E[P]*{\Tr(\Sig^{\rhomm}(u))}, \,\, \forall u: \text{node on} \,\, \mathcal{T}.
	\end{equation}
	Therefore, we have
	\begin{equation}
		\sum_{\substack{u, s: \,\,e_{u,s} \ \text{connected} \\ \text{to a leaf}}} \E[P]*{\Tr\left(\frac{1}{2^n}\tau_{M^{u}_{s},\Sig^{\rhomm}(u)}\right)} =  \E[P]*{\sum_{\substack{u: \ \text{connected} \\ \text{to a leaf}}} \Tr(\Sig^{\rhomm}(u))} = 1.
	\end{equation}
	As for the latter term in Eq.~\eqref{eq:hybrid}, note that for any leaf $\ell$ with parent $u$, $P[u]\subseteq P[\ell]$, so
	\begin{align}
		\sum_{\ell, P \in P[\ell]} \norm*{A^{\rho_P}_{M^{u}_{s}}\left(\Sig^{\rhomm}(u) - \Sig^{\rho_P}(u)\right)}_{\tr} \le \sum_{\substack{u, s: \,\,  e_{u,s} \ \text{connected} \\ \text{to a leaf}, P\in P[u]}}\norm*{A^{\rho_P}_{M^{u}_{s}}\left(\Sig^{\rhomm}(u) - \Sig^{\rho_P}(u)\right)}_{\tr}\\
		= \sum_{\substack{u: \,\,  u \ \text{connected} \\ \text{to a leaf}, P\in P[u]}} \sum_{s} \norm*{A^{\rho_P}_{M^{u}_{s}}\left(\Sig^{\rhomm}(u) - \Sig^{\rho_P}(u)\right)}_{\tr}
		\label{eq:exchange_sum}
	\end{align}
	Note that for any node $u$, the map\begin{equation}
		\calE: X\mapsto \sum_s \ketbra{s}{s}\otimes A^{\rho_P}_{M^{u}_{s}}(X)
	\end{equation} is a quantum channel, so in particular $\norm{\calE(X)}_{\tr} \le \norm{X}_{\tr}$.
	In particular, we can see that
	\begin{equation}
	    \norm*{\calE(\Sig^{\rhomm}(u) - \Sig^{\rho_P}(u))}_{\tr} = \sum_s \norm*{A^{\rho_P}_{M^{u}_{s}}\left(\Sig^{\rhomm}(u) - \Sig^{\rho_P}(u)\right)}_{\tr}.
	\end{equation}
	We may thus upper bound Eq.~\eqref{eq:exchange_sum} by \begin{equation}
		\sum_{\substack{u: \ \text{connected to} \\ \text{a leaf}, P\in P[u]}}\norm*{\Sig^{\rhomm}(u) - \Sig^{\rho_P}(u)}_{\tr} = \sum_{\substack{u: \ \text{at depth} \ T-1 \\ P\in P[u]}}\norm*{\Sig^{\rhomm}(u) - \Sig^{\rho_P}(u)}_{\tr}.\label{eq:sumgood2}
	\end{equation}
	Combining Eq.~\eqref{eq:hybrid}, \eqref{eq:sumgood1}, and \eqref{eq:sumgood2}, we conclude that 
	\begin{equation}
		\sum_{\substack{\ell\in\leaf(\calT) \\ P \in P[\ell]}} \norm*{\Sig^{\rhomm}(\ell) - \Sig^{\rho_P}(\ell)}_{\tr} \le \frac{1}{2^{(n-k)/3}}\cdot \sqrt{\frac{2^{2n}}{2^{2n} - 1}} +\sum_{\substack{u \ \text{at depth} \ T-1 \\ P\in P[u]}}\norm*{\Sig^{\rhomm}(u) - \Sig^{\rho_P}(u)}_{\tr}
	\end{equation}
	We can thus conclude by induction and by Lemma~\ref{lem:goodbad},
	\begin{equation}
		\tvd\left(p^{\rhomm}, \E[P]{p^{\rho_P}}\right) \le T\cdot 2^{-(n-k)/3} + T\cdot 2^{-(n-k)/3}\cdot \sqrt{\frac{2^{2n}}{2^{2n} - 1}}.
	\end{equation}
	In order to achieve the many-versus-one distinguishing task with probability at least $2/3$, we must have $2/3 \leq \tvd\left(p^{\rhomm}, \E[P]{p^{\rho_P}}\right)$ from Le Cam's two point method; see Lemma~\ref{lem:lecam}.
	However for $T = o\left(2^{(n - k)/3}\right)$, $\tvd(p^{\rhomm}, \E[P]{p^{\rho_P}}) = o(1)$.
	This concludes the lower bound that $T \geq \Omega(2^{(n-k) / 3})$.
\end{proof}

\subsection{Upper Bound with \texorpdfstring{$n$}{n}-qubit of quantum memory}

Here we show that with $n$ qubits of quantum memory, we can estimate the absolute values of the expectations in $M$ arbitrary Pauli observables using only $\mathcal{O}(\log M)$ copies of the unknown state.

\begin{theorem}
    Given unknown mixed state $\rho\in\bH^{2^n\times 2^n}$ and any collection of $M$ Pauli observables $P_1,\ldots,P_M$, there is a learning algorithm using $n$ qubits of quantum memory that measures $\mathcal{O}(\log(M)/\epsilon^2)$ copies of $\rho$ and outputs $\epsilon$-accurate estimates for $\abs{\Tr(P_i \rho)}^2$ for all $i\in[M]$ with probability at least $2/3$.
\end{theorem}

\begin{proof}
    We use an $n$-qubit memory to store a single copy of $\rho$ from the previous experiment. In the current experiment, we can combine the quantum memory with the new quantum data $\rho$ to measure the two copies $\rho \otimes \rho$ using an entangled measurement.
    For a pair of copies of $\rho$, if we measured every qubit $j$ of $\rho\otimes \rho$ in the Bell basis to obtain outcome $\ket{\beta_j}\in\brc{\ket{\Psi^+},\ket{\Psi^-},\ket{\Phi^+},\ket{\Phi^-}}$, then note that because $\brc{P_i\otimes P_i}_{i\in[M]}$ is simultaneously diagonalized in the $n$-fold tensor power of the Bell basis, we have \begin{equation}
        \E[\brc{\beta_j}]*{\prod^n_{j=1}\Tr\left(P_{i(j)}\otimes P_{i(j)} \ketbra{\beta_j}{\beta_j}\right)} = \sum_{\brc{\beta_j}}\prod^n_{j=1}\Tr\left(P_{i(j)}\otimes P_{i(j)} \Pi_{\beta_j} (\rho_j \otimes \rho_j) \Pi_{\beta_j}\right) = \Tr(P_i \rho)^2, \label{eq:rhorho}
    \end{equation} where $P_{i(j)}$ denotes the $j$-th qubit of $P_i$, $\rho_j$ denotes the $j$-th qubit of $\rho$, and $\Pi_{\beta}$ denotes the projector to Bell state $\beta$.
    The random variable in the expectation on the left-hand side of \eqref{eq:rhorho} is clearly bounded, so by Chernoff bounds, we can estimate its mean for any fixed $j$ to within error $\epsilon$ with probability at least $1 - 1/3M$ using $\mathcal{O}(\log(M/\delta)/\epsilon^2)$ samples. The claim follows by a union bound.
\end{proof}


\section{Exponential separations in learning quantum channels}

\subsection{Prerequisites}

We begin by generalizing the tree representation for learning quantum states to the setting of learning quantum channels.  First let us state the idea of the definition intuitively before delving into its technical description.  There is some quantum channel $\mathcal{C}$ which we wish to learn about; we have the ability to apply the channel to a state of our choice and then to completely measure the resulting state.  The resulting measurement outcome can be recorded in a classical memory.  The procedure of preparing a state, applying the channel, and then making a measurement is repeated over multiple rounds, wherein the measurement outcomes of previous rounds can inform the states prepared in future rounds, as well as the choice of measurement in future rounds.  That is, the protocol is adaptive.  At the end, we have gained a list of measurement outcomes with which we can judiciously infer properties of the channel $\mathcal{C}$ under investigation.

Now we provide the full technical definition:

\begin{definition}[Tree representation for learning channels]\label{def:treechannel}
Consider a fixed quantum channel $\mathcal{C}$ acting on an $n$-qubit subsystem of a Hilbert space $\mathcal{H} \simeq \mathcal{H}_{\text{\rm main}} \otimes \mathcal{H}_{\text{\rm aux}}$ where $\mathcal{H}_{\text{\rm main}} \simeq (\mathbb{C}^{2})^{\otimes n}$ is the `main system' comprising $n$ qubits and $\mathcal{H}_{\text{\rm aux}} \simeq (\mathbb{C}^2)^{\otimes n'}$ is an `auxiliary system' of $n'$ qubits.  It is convenient to define $d = 2^n$ and $d' = 2^{n'}$.  A learning algorithm without quantum memory can be represented as a rooted tree $\mathcal{T}$ of depth $T$ such that each node encodes all measurement outcomes the algorithm has received thus far.  The tree has the following properties:
\begin{itemize}
    \item Each node $u$ has an associated probability $p^{\mathcal{C}}(u)$.
    \item The root of the tree $r$ has an associated probability $p^{\mathcal{C}}(r) = 1$.
    \item At each non-leaf node $u$, we prepare a state $|\phi_u\rangle$ on $\mathcal{H}$, apply the channel $\mathcal{C}$ onto the $n$-qubit subsystem, and measure a rank-1 POVM $\{\sqrt{w_s^u d d'}\,|\psi_s^u\rangle \langle \psi_s^u|\}_s$ (which can depend on $u$) on the entire system to obtain a classical outcome $s$.  Each child node $v$ of the node $u$ corresponds to a particular POVM outcome $s$ and is connected by the edge $e_{u, s}$. We refer to the set of child node of the node $u$ as $\mathrm{child}(u)$.
    Accordingly, we can relabel the POVM as $\{\sqrt{w_v d d'}\,|\psi_v\rangle \langle \psi_v|\}_{v \in \mathrm{child}(u)}$. 
    \item If $v$ is a child node of $u$, then
    \begin{equation} \label{eq:prob-channel-node}
    p^{\mathcal{C}}(v) = p^{\mathcal{C}}(u) \, w_v d d' \, \langle \psi_v|\,(\mathcal{C} \otimes \mathcal{I}_{\text{\rm aux}})[|\phi_u\rangle \langle \phi_u|] \,|\psi_v\rangle\,.
    \end{equation}
    \item Each root-to-leaf path is of length $T$.  For a leaf of corresponding to node $\ell$, $p^{\mathcal{C}}(\ell)$ is the probability that the classical memory is in state $\ell$ after the learning procedure.
\end{itemize}
\end{definition}

\noindent Each node $u$ in the tree represents the state of the classical memory at one time step of the learning process. The associated probability $p^{\mathcal{C}}(u)$ for a node $u$ is the probability that the classical memory enters the state $u$ during the learning process. Each time we perform one experiment, we transition from a node $u$ to a child node of $u$.

There are several features of the definition which we will remark on.  First and foremost, an important feature of the definition is that we have access to an auxiliary Hilbert space $\mathcal{H}_{\text{aux}}$ for each state preparation and measurement.  In particular, even though the channel $\mathcal{C}$ only acts on $n$ qubits, we can apply it to the first $n$ qubits of a state $|\phi\rangle \in \mathcal{H}_{\text{main}}\otimes \mathcal{H}_{\text{aux}}$ which can be entangled between the $n$ qubits and the auxiliary system.  Moreover, we can measure the resulting state $(\mathcal{C} \otimes \mathcal{I}_{\text{aux}})[|\phi\rangle \langle \phi|]$ using POVM's which are entangled between the $n$ qubits and the auxiliary system.  The presence of the auxiliary system will render our proofs somewhat elaborate; moreover, the presence of the auxiliary system renders our results stronger than previous ones for adaptive incoherent access QUALMs~\cite{aharonov2021quantum}.  In this particular QUALM setting, the notion of learning algorithm is similar, except that there is no auxiliary system.

We consider quantum channel learning tasks which were first studied in the QUALM setting without an auxiliary Hilbert space~\cite{aharonov2021quantum}.  They are:
\begin{definition}[Fixed unitary task]
Suppose that an $n$-qubit quantum channel $\mathcal{C}$ is one of the following with equal probability:
\begin{itemize}
    \item $\mathcal{C}$ is the completely depolarizing channel $\mathcal{D}$.
    \item $\mathcal{C}$ is the unitary channel $\mathcal{C}[\rho] = U \rho U^\dagger$ for $U$ a fixed, Haar-random unitary.
\end{itemize}
The fixed unitary task is to distinguish between the two above possibilities.  We can also consider analogous versions of the problem where $U$ is instead a Haar-random orthogonal matrix, or a Haar-random symplectic matrix.
\end{definition}


\noindent Note that instead of considering the completely depolarizing channel $\mathcal{D}$ can be thought of in a different way which makes the fixed unitary task more illuminating.  Specifically, we can equivalently think of $\mathcal{D}$ as an $n$-qubit unitary channel which applies an i.i.d.~random Haar unitary each time the channel is applied.  From this perspective, $\mathcal{D}$ implements time-dependent random unitary dynamics (i.e.~a new unitary is selected for each application of the channel); the task is then to distinguish this from time-\textit{independent} random unitary dynamics wherein the channel applies a single fixed random unitary.  Said more simply, from this point of view the task is to distinguish a type of time-dependent dynamics from a type of time-independent dynamics.

We also consider another task from~\cite{aharonov2021quantum} with a slightly different flavor:
\begin{definition}[Symmetry distinction task] Suppose that an $n$-qubit quantum channel $\mathcal{C}$ is one of the following with equal probability:
\begin{itemize}
    \item $\mathcal{C}[\rho] = U \rho U^\dagger$ for $U$ a fixed, Haar-random unitary matrix.
    \item $\mathcal{C}[\rho] = O \rho O^\dagger$ for $O$ a fixed, Haar-random orthogonal matrix.
    \item $\mathcal{C}[\rho] = S \rho S^\dagger$ for $S$ a fixed, Haar-random symplectic matrix.
\end{itemize}
The symmetry distinction task is to distinguish between the three above possibilities.
\end{definition}
\noindent Unitary, orthogonal, and symplectic matrices manifest three different forms of what is called time-reversal symmetry~\cite{dyson1962threefold}.  In this terminology, the symmetry distinction task is to determine the time-reversal symmetry class of $\mathcal{C}$.  The task belongs to a class of problems of determining the symmetries of an uncharacterized system, which are important in experimental physics.

In the above distinguishing tasks, we will always reduce them to two-hypothesis distinguishing problem. We define a two-hypothesis distinguishing problem as follows.

\begin{definition}[Two-hypothesis channel distinction task]
The following two events happen with equal probability:
\begin{itemize}
    \item The channel $\mathcal{C}$ is sampled from a probability distribution $D_A$ over channels.
    \item The channel $\mathcal{C}$ is sampled from a probability distribution $D_B$ over channels.
\end{itemize}
The goal is to distinguish whether $\mathcal{C}$ is sampled from $D_A$ or $D_B$.
\end{definition}

\noindent For any two-hypothesis distinguishing problem, we can always apply the two-point method similar to Lemma~\ref{lem:lecam} in learning quantum states.

\begin{lemma}[Le Cam's two-point method, see e.g. Lemma 1 in \cite{yu1997assouad}]\label{lem:lecam-channel}
Consider a learning algorithm without quantum memory that is described by a rooted tree $\mathcal{T}$.
The probability that the learning algorithm solves the two-hypothesis channel distinction task correctly is upper bounded by
\begin{equation}
    \frac{1}{2} \sum_{\ell \in \mathrm{leaf}(\mathcal{T})} \left| \left(\Expect_{\mathcal{C} \sim \mathcal{D}_A} p^{\mathcal{C}}(\ell) \right) - \left( \Expect_{\mathcal{C} \sim \mathcal{D}_B} p^{\mathcal{C}}(\ell) \right) \right|.
\end{equation}
\end{lemma}





\subsection{Review of the Weingarten calculus}


Here we will review Haar measures on unitary, orthogonal, and symplectic matrices, and present several key lemmas that we will leverage in the later proofs.  First we recall the definitions of orthogonal and symplectic matrices.  Denoting the set of unitary matrices on $(\mathbb{C}^2)^{\otimes n}$ by
\begin{equation}
U(d) = \{U \in \text{Mat}_{d \times d}(\mathbb{C}) \, : \, U^\dagger = U^{-1}\}\,,
\end{equation}
the set of orthogonal matrices on is given by
\begin{equation}
O(d) = \{O \in U(d) \, : \, O^t = O^{-1}\}\,,
\end{equation}
and the set of symplectic matrices is given by
\begin{equation}
\text{Sp}(d/2) = \{S \in U(d) \, : \, J S^t J^{-1}  = S^{-1}\}\,.
\end{equation}
Here $J$ is the symplectic form
\begin{equation}
\label{E:Jdef1}
    J = \begin{bmatrix}
    \textbf{0}_{d/2 \times d/2} & \mathds{1}_{d/2 \times d/2} \\
    - \mathds{1}_{d/2 \times d/2} & \textbf{0}_{d/2 \times d/2}
    \end{bmatrix}\,,
\end{equation}
where $\textbf{0}_{d/2 \times d/2}$ is the $d/2 \times d/2$ matrix of all zeroes and $\mathds{1}_{d/2 \times d/2}$ is the $d/2 \times d/2$ identity matrix.  The quantity $J S^t J^{-1}$ is sometimes called the ``symplectic transpose'' and denoted by $S^D$.

Note that $\text{Sp}(d/2)$ is sometimes called the ``symplectic unitary group'' to distinguish it from the group of symplectic matrices which need not be unitary. 
For the orthogonal group, the matrices $O$ will necessarily be real.
For the symplectic unitary group, the matrices $S$ could be complex numbers.
For our purposes, we will adopt standard terminology by dropping the word `unitary' since the context is clear.

Since $U(d), O(d), \text{Sp}(d/2)$ are each compact Lie groups, they admit canonical Haar measures which are right and left-invariant under group multiplication.  For instance, for $U(d)$, the Haar measure satisfies
\begin{equation}
\int_{U(d)} dU \, f(U) = \int_{U(d)} dU \, f(VU) = \int_{U(d)} dU \, f(UV)
\end{equation}
for any $V \in U(d)$ and any $f(U)$.  Analogous expressions hold for the Haar measures corresponding to $O(d)$ and $\text{Sp}(d/2)$.  Such Haar integrals will be essential for our proofs, and so here we catalog important properties.

Now we turn to discussing more detailed properties of the Haar integrals.  Our short overview will be based on~\cite{collins2017weingarten, gu2013moments, matsumoto2013weingarten, aharonov2021quantum}.  Instead of using the integral notation, we will often use expectation values $\mathbb{E}_{U \sim \text{Haar}}[\,\cdot\,]$.

\subsubsection{Haar averaging over \texorpdfstring{$U(d)$}{U(d)}}
\label{sec:haaraverageUd}

For our purposes it will be useful to study moments of the Haar ensemble, in particular
\begin{equation}
\label{E:HaarUint1}
\mathbb{E}_{U \sim \text{Haar}}\!\left[U_{i_1 j_1} U_{i_2 j_2} \cdots U_{i_k j_k} \overline{U_{i_1' j_1'} U_{i_2' j_2'} \cdots U_{i_k' j_k'}}\right] = \sum_{\sigma, \tau \in S_k} \delta_{\sigma(I), I'} \delta_{\tau(J), J'} \text{Wg}^U(\sigma \tau^{-1}, d)\,.
\end{equation}
This equation requires some unpacking.  On the left-hand side, the bar denotes complex conjugation; for instance $\overline{U}_{ij} = U_{ji}^\dagger$.  On the right-hand side, $S_k$ is the symmetric group on $k$ elements, $I$ is a multi-index $I = (i_1,...,i_k)$ and similarly for $I', J, J'$, and
\begin{equation}
\delta_{\sigma(I), I'} := \delta_{i_{\sigma(1)}, i_1'} \delta_{i_{\sigma(2)}, i_2'} \cdots \delta_{i_{\sigma(k)}, i_k'}\,.
\end{equation}
Finally, $\text{Wg}^U( \, \cdot \,,d)$ is a map $S_k \to \mathbb{R}$ called the \textit{unitary Weingarten function} to be specified shortly.  To further compress notation, it will be convenient to fully commit to multi-index notation and write $U_{IJ}^{\otimes k} = U_{i_1 j_1} U_{i_2 j_2} \cdots U_{i_k j_k}$ so that~\eqref{E:HaarUint1} becomes
\begin{equation}
\label{E:HaarUint2}
\mathbb{E}_{U \sim \text{Haar}}\!\left[U_{IJ}^{\otimes k} U_{KL}^{\dagger \, \otimes k}\right] = \sum_{\sigma, \tau \in S_k} \delta_{\sigma(I), I'} \delta_{\tau(J), J'} \text{Wg}^U(\sigma \tau^{-1}, d)\,.
\end{equation}
The utility of~\eqref{E:HaarUint2} is that it allows us to compute $\mathbb{E}_{\text{Haar}}[U^{\otimes k} \otimes U^{\dagger \otimes k}]$, and matrix elements thereof, in terms of data of the symmetric group on $T$ elements.  We remark that $\mathbb{E}_{\text{Haar}}[U^{\otimes k} \otimes U^{\dagger \otimes \ell}]$ vanishes for $k \not = \ell$, so~\eqref{E:HaarUint2} covers all non-trivial cases.

It still remains to specify the unitary Weingarten function $\text{Wg}^U( \, \cdot \,,d)$.  In fact, it can be regarded as the inverse of an easily specified matrix.  To this end, in a slight abuse of notation, we will let permutations $\tau, \sigma$ in $S_k$ also label their representations on $\mathcal{H}^{\otimes k}$.  That is, $\tau$ will denote a unitary $d^k \times d^k$ matrix on $\mathcal{H}^{\otimes k}$ which permutes the $k$ copies of $\mathcal{H}$ according to the permutation specified by the label $\tau$.  Now we can readily define
\begin{equation}
G^U(\sigma \tau^{-1}, d) := \text{tr}(\sigma \tau^{-1}) = d^{\#(\sigma \tau^{-1})}
\end{equation}
such that $\#(\sigma \tau^{-1})$ counts the number of cycles of $\sigma \tau^{-1}$.  Now $\text{Wg}^U(\,\cdot\,,d)$ is defined by the identity
\begin{equation}
\label{E:Haarinverses1}
\sum_{\tau \in S^k} \text{Wg}^U(\sigma^{-1} \tau, d) \,G^U(\tau^{-1} \pi, d) = \delta_{\sigma, \pi}
\end{equation}
for all $\sigma, \pi \in S_k$.  This equation expresses that $\text{Wg}^U$ and $G^U$ are in fact inverses as $k! \times k!$ matrices.  To see this more readily, we can use the notation $\text{Wg}^U(\sigma^{-1} \tau, d) = \text{Wg}^U_{\sigma, \tau}$ and $G^U(\tau^{-1} \pi, d) = G^U_{\tau, \pi}$ so that~\eqref{E:Haarinverses1} is simply $\sum_\tau \text{Wg}^U_{\sigma, \tau}\, G^U_{\tau, \pi} = \delta_{\sigma, \pi}$.

We conclude by presenting a theorem, corollary, and lemma which will be used in our proofs.
\begin{CM3p2}\label{CM3p2}
For any $\sigma \in S_k$ and $d > \sqrt{6} k^{7/4}$,
\begin{align}
\label{eq:bound1}
\frac{1}{1 - \frac{k-1}{d^2}} \leq \frac{(-1)^{k - \#(\sigma)} d^{2k-\#(\sigma)} \text{\rm Wg}^U(\sigma,d)}{\prod_i \frac{(2 \ell_i - 2)!}{(\ell_i-1)! \ell_i!}} \leq \frac{1}{1 - \frac{6 k^{7/2}}{d^2}}
\end{align}
where the left-hand side inequality is valid for any $d \geq k$.  Note that $\sigma \in S_k$ has cycle type $(\ell_1,\ell_2,...)$.
\end{CM3p2}
\noindent An immediate corollary is:
\begin{corollary}
\label{cor:WgU1bd}
$|\text{\rm Wg}^U(\mathds{1}, d) - d^{-k}| \leq \mathcal{O}(k^{7/2} d^{-(k+2)})$\,.
\end{corollary}
\noindent We also recapitulate a useful result from~\cite{aharonov2021quantum}:
\begin{ACQl6}
\label{lem:ACQl6}
$\sum_{\tau \in S_k} |\text{\rm Wg}^U(\tau, d)| = \frac{(d-k)!}{d!}$\,.
\end{ACQl6}

\subsubsection{Haar averaging over \texorpdfstring{$O(d)$}{O(d)}}

Just as Haar averaging over $U(d)$ is intimately related to the permutation group $S_k$, Haar averaging over $O(d)$ (and likewise $\text{Sp}(d/2)$) is intimately related to pair partitions $P_2(2k)$.  Accordingly, we will begin with discussing pair partitions.

Informally, a pair partition on $2k$ is a way of pairing off $2k$ elements (e.g., pairing off people in a ballroom dance class).  There are in fact $\frac{(2k)!}{2^k k!} = (2k-1)!!$ possible pairings of $2k$ elements.  More formally, a pair partition $\mathfrak{m} \in P_2(2k)$ is a function $\mathfrak{m} : [2k] \to [2k]$ satisfying $\mathfrak{m}(2i-1) < \mathfrak{m}(2i)$ for $1 \leq i \leq k$ and $\mathfrak{m}(1) < \mathfrak{m}(3) < \cdots < \mathfrak{m}(2k-1)$. The pair permutation is often notated as
\begin{equation}
    \mathfrak{m} = \{\mathfrak{m}(1), \mathfrak{m}(2)\} \{\mathfrak{m}(3), \mathfrak{m}(4)\} \cdots \{\mathfrak{m}(2k-1), \mathfrak{m}(2k)\}
\end{equation}
where the brackets denote individual pairs, and the constraints on $\mathfrak{m}$ order the pairs in a canonical (and unique) manner.  In words, within each pair the `left' element is always less than the `right' element, and the pairs themselves are ordered according to the `left' element of each pair.

It is natural to endow pair permutations with a group structure so that they form a subgroup $M_{2k}$ of the permutation group $S_{2k}$.  We simply define $M_{2k}$ by
\begin{equation}
M_{2k} := \{\sigma \in S_{2k} \, : \, \sigma(2i-1) < \sigma(2i)\text{ for }1 \leq i \leq k\,,\,\,\sigma(1) < \sigma(3) < \cdots < \sigma(2k-1)\}
\end{equation}
and it is readily checked that $M_{2k}$ forms a group.  We will often leverage the natural bijection between $P_2(2k)$, namely $\mathfrak{m} \mapsto \sigma_{\mathfrak{m}}$ where $\mathfrak{m}(i) = \sigma_{\mathfrak{m}}(i)$ for all $i$.  The identity pairing is denoted by $\mathfrak{e}$, and by the above bijection maps to the identity permutation $\sigma_{\mathfrak{e}} = \mathds{1}$.

Pair permutations also have a notion of cycles, which differs from that of the symmetric group.  That is, the cycle type of $\sigma_{\mathfrak{m}}$ as an element of $M_{2k}$ is different from the cycle type of $\sigma_{\mathfrak{m}}$ thought of as an element of $S_{2k}$\,.  To construct the pair partition cycles $\sigma_{\mathfrak{m}}$ (we will also refer to this as the cycle type of $\mathfrak{m}$), consider the function $f_{\mathfrak{m}} : [2k] \to [2k]$ defined by
\begin{equation}
\label{E:ffrakm1}
f_{\mathfrak{m}}(i) = \begin{cases}
\mathfrak{m}(2j) & \text{if }i = \mathfrak{m}(2j-1) \\
\mathfrak{m}(2j-1) & \text{if }i = \mathfrak{m}(2j)
\end{cases}\,.
\end{equation}
This function maps $i$ to the integer it is paired with under $\mathfrak{m}$.  The function $f_{\mathfrak{e}}$ corresponds to the identity pairing.  We can construct the cycles of $\mathfrak{m}$ as follows.  Consider the sequence
\begin{equation}
(1, f_{\mathfrak{m}}(1), f_{\mathfrak{e}} \circ f_{\mathfrak{m}}(1), f_{\mathfrak{m}} \circ f_{\mathfrak{e}} \circ f_{\mathfrak{m}}(1),...)\,.
\end{equation}
This sequence is periodic, and so we truncate it at its period so that no element is repeated.  We call this truncated list $B_1$, and view it is a cyclically ordered list (i.e., the list is regarded as the same if it is cyclically permuted).  If $B_1$ contains all of $[2k]$, then $\mathfrak{m}$ contains only one cycle, namely $B_1$.  Otherwise, let $j$ be the smallest integer in $[2k]$ with is not in $B_1$, and construct
\begin{equation}
(j, f_{\mathfrak{m}}(j), f_{\mathfrak{e}} \circ f_{\mathfrak{m}}(j), f_{\mathfrak{m}} \circ f_{\mathfrak{e}} \circ f_{\mathfrak{m}}(j),...)\,.
\end{equation}
This is likewise periodic, and we truncate it at its period to get the cyclically ordered list $B_2$.  If $B_1$ and $B_2$ do not contain all of $[2k]$, then we construct a $B_3$, etc.  When the procedure terminates, we have $B_1, B_2, ...$ which contain all of $[2k]$.  The $B_1, B_2, ...$ are the pair partition cycles of $\mathfrak{m}$.  Their corresponding lengths $b_1, b_2,...$ are all even, and the cycle type (also called the coset type) of $\mathfrak{m}$ is given by
\begin{equation}
(\mu_1, \mu_2,...) = (b_1/2, b_2/2,...)
\end{equation}
which is often listed in descending order of cycle size.


Using our notation for pair partitions as well as the multi-index notation we established previously, we have the following formula for a Haar integral over the orthogonal group
\begin{equation}
\mathbb{E}_{O \sim \text{Haar}}\!\left[O_{IJ} O_{J' I'}^t\right] = \sum_{\mathfrak{m},\mathfrak{n} \in P_2(2k)} \Delta_{\mathfrak{m}}(I I') \,\Delta_{\mathfrak{n}}(JJ')\, \text{Wg}^O(\sigma_{\mathfrak{m}}^{-1} \sigma_{\mathfrak{n}})
\end{equation}
where $I I'$ merges the multi-indices $I$ and $I'$ as $I I' = (i_1, i_1',...,i_k, i_k')$ and likewise for $JJ'$.  Letting $I I' = \textbf{I} = (\textbf{i}_1, \textbf{i}_2,...,\textbf{i}_{2k})$ we define
\begin{equation}
\label{E:HaarOint1}
    \Delta_{\mathfrak{m}}(\textbf{I}) := \prod_{s=1}^k \delta_{\textbf{i}_{\mathfrak{m}(2s-1)}, \textbf{i}_{\mathfrak{m}(2s)}}\,.
\end{equation}
Similar to before, $\text{Wg}^O(\,\cdot\,, d)$ is a map $M_{2k} \to \mathbb{R}$ called the orthogonal Weingarten function.  Although we have written~\eqref{E:HaarOint1} in a way that emulates~\eqref{E:HaarUint2}, the formula~\eqref{E:HaarOint1} has a different character to it.  Specifically, examining the left-hand side, we can equivalently write it as $\mathbb{E}_{\text{Haar}}[O_{II', JJ'}]$ where we have simply used the equivalence $O_{J' I'}^t = O_{I' J'}$.  That is,~\eqref{E:HaarOint1} tells us how to compute $\mathbb{E}_{\text{Haar}}[O^{\otimes 2k}]$ and matrix elements thereof for $2k$ even; this integral vanishes if we replace $2k$ by an odd number.  By contrast with the Haar unitary setting where we needed as many $U$'s as $U^\dagger$'s to get a non-trivial integral, here we just need an even number of $O$'s, essentially because $O^\dagger = O^t$.

Analogous to the unitary setting discussed above, we can define the orthogonal Weingarten function $\text{Wg}^O(\,\cdot\,, d)$ in terms of a simpler function
\begin{equation}
\label{E:GOeq1}
G^O(\sigma_{\mathfrak{m}}^{-1} \sigma_{\mathfrak{n}}, d) := d^{\#^O(\sigma_{\mathfrak{m}}^{-1}\sigma_{\mathfrak{n}})}
\end{equation}
where $\#^O(\sigma_{\mathfrak{m}}^{-1}\sigma_{\mathfrak{n}})$ counts the number of $M_{2k}$-cycles of $\sigma_{\mathfrak{m}}^{-1}\sigma_{\mathfrak{n}}$.  
We have the identity
\begin{equation}
\sum_{\mathfrak{n} \in P_2(2k)} \text{Wg}^O(\sigma_{\mathfrak{m}}^{-1} \sigma_{\mathfrak{n}}, d) \, G^O(\sigma_{\mathfrak{n}}^{-1} \sigma_{\mathfrak{p}}, d) = \delta_{\mathfrak{m}, \mathfrak{p}}\,.
\end{equation}
which expresses that $\text{Wg}^O$ and $G^O$ are inverses as $(2k-1)!! \times (2k-1)!!$ matrices.

Finally, we present a theorem, corollary, and lemma which we will leverage in our proofs about the symmetry distinction problem:
\begin{CM4p11}
\label{CM4p11}
For any $\sigma_{\mathfrak{m}} \in M_{2k}$ and $d > 12 k^{7/2}$,
\begin{align}
\label{eq:bound2}
\frac{1 - \frac{24 k^{7/2}}{d}}{1 - \frac{144 k^7}{d^2}} \leq \frac{(-1)^{k - \#^O(\sigma_{\mathfrak{m}})} d^{2k - \#^O(\sigma_{\mathfrak{m}})} \text{\rm Wg}^O(\sigma_{\mathfrak{m}},d)}{\prod_i \frac{(2 \mu_i - 2)!}{(\mu_i-1)! \mu_i!}} \leq \frac{1}{1 - \frac{144 k^{7}}{d^2}}
\end{align}
where $\sigma_{\mathfrak{m}}$ has $M_{2k}$-cycle type $(\mu_1,\mu_2,...)$. 
\end{CM4p11}
\noindent We have the immediate corollary
\begin{corollary}
\label{cor:WgO1bd}
$|\text{\rm Wg}^O(\sigma_{\mathfrak{e}}, d) - d^{-k}| \leq \mathcal{O}(k^{7} d^{-(k+2)})$\,.
\end{corollary}
\noindent Analogous to Lemma 6 of~\cite{aharonov2021quantum} written above, we have~\cite{aharonov2021quantum}:
\begin{ACQl8}
\label{lem:ACQl8}
$\sum_{\mathfrak{m} \in P_2(2k)} |\text{\rm Wg}^O(\sigma_{\mathfrak{m}}, d)| = \frac{(d-2k)!!}{d!!}$\,.
\end{ACQl8}

\subsubsection{Haar averaging over \normalfont \texorpdfstring{$\text{Sp}(d/2)$}{Sp(d/2)}}

Haar averaging over the $\text{Sp}(d/2)$ is very similar to the orthogonal setting.  We have the identity
\begin{equation}
\mathbb{E}_{S \sim \text{Haar}}\!\left[S_{IJ} S_{J' I'}^t\right] = \sum_{\mathfrak{m}, \mathfrak{n} \in P_2(2k)} \Delta_{\mathfrak{m}}'(II') \Delta_{\mathfrak{n}}'(JJ') \, \text{Wg}^{\text{Sp}}(\sigma_{\mathfrak{m}}^{-1} \sigma_{\mathfrak{n}}, d/2)
\end{equation}
where
\begin{equation}
\Delta_{\mathfrak{m}}'(\textbf{I}) := \prod_{s=1}^k J_{\textbf{i}_{\mathfrak{m}(2s-1)}, \textbf{i}_{\mathfrak{m}(2s)}}\,.
\end{equation}
Here $J$ is the canonical symplectic form defined in~\eqref{E:Jdef1}.  The symplectic Weingarten function $\text{Wg}^{\text{Sp}}(\,\cdot\,,d/2)$ taking $M_{2k} \to \mathbb{R}$ is a small modification of the orthogonal Weingarten function, namely
\begin{equation}
\text{Wg}^{\text{Sp}}(\sigma_{\mathfrak{m}}, d/2) = (-1)^k \epsilon(\sigma_{\mathfrak{n}}) \, \text{Wg}^{O}(\sigma_{\mathfrak{m}}, -d)
\end{equation}
where $\epsilon(\sigma_{\mathfrak{m}})$ is the signature of $\sigma_{\mathfrak{m}}$ thought of as an element of $S_{2k}$.

We give a theorem, corollary and lemma analogous to the ones for the orthogonal group above.
\begin{CM4p10}
\label{CM4p10}
For any $\sigma_{\mathfrak{m}} \in M_{2k}$ and $d > 6 k^{7/2}$, we have
\begin{align}
\label{eq:bound3}
\frac{1}{1 - \frac{k-1}{(d/2)^2}} \leq \frac{d^{2k - \#^{\text{\rm Sp}}(\sigma_{\mathfrak{m}})} |\text{\rm Wg}^\text{\rm Sp}(\sigma_{\mathfrak{m}},d/2)|}{\prod_i \frac{(2 \mu_i - 2)!}{(\mu_i-1)! \mu_i!}} \leq \frac{1}{1 - \frac{6 k^{7/2}}{(d/2)^2}}
\end{align}
where $\sigma_{\mathfrak{m}}$ has $M_{2k}$-cycle type $(\mu_1,\mu_2,...)$ and $\#^{\text{\rm Sp}}(\sigma_{\mathfrak{m}}) = \#^O(\sigma_{\mathfrak{m}})$.
\end{CM4p10}
\noindent This has the direct corollary
\begin{corollary}
\label{cor:WgSp1bd}
$|\text{\rm Wg}^\text{\rm Sp}(\sigma_{\mathfrak{e}}, d/2) - d^{-k}| \leq \mathcal{O}(k^{7/2} d^{-(k+2)})$\,.
\end{corollary}
\noindent From~\cite{aharonov2021quantum} we borrow the useful lemma:
\begin{ACQl10}
\label{lem:ACQl10}
$\sum_{\mathfrak{m} \in P_2(2k)} |\text{\rm Wg}^{\text{\rm Sp}}(\sigma_{\mathfrak{m}}, d/2)| = \prod_{j=0}^{k-1} \frac{1}{d + 2j}$\,.
\end{ACQl10}

\subsection{Depolarizing channel versus random unitary}

\subsubsection{Lower bound without quantum memory}

We are now prepared to establish the following results:
\begin{theorem}[Exponential hardness of fixed unitary task without quantum memory]\label{thm:channelhard1}
Any learning algorithm without quantum memory requires
\begin{equation}
\label{E:TOmega1}
    T \geq \Omega\left( d^{1/3} \right),
\end{equation}
to correctly distinguish between the completely depolarizing channel $\mathcal{D}$ on $n$ qubits from a fixed, Haar-random unitary channel $\mathcal{U}[\rho] = U \rho U^\dagger$ on $n$ qubits with probability at least $2/3$.
\end{theorem}

\begin{theorem}[Exponential hardness of fixed orthogonal matrix task without quantum memory]\label{thm:Ochannelhard1}
Any learning algorithm without quantum memory requires
\begin{equation}
\label{E:TOmega2}
    T \geq \Omega\left( d^{2/7} \right),
\end{equation}
to correctly distinguish between the completely depolarizing channel $\mathcal{D}$ on $n$ qubits from a fixed, Haar-random orthogonal matrix channel $\mathcal{U}[\rho] = O \rho O^t$ on $n$ qubits with probability at least $2/3$.
\end{theorem}

\begin{theorem}[Exponential hardness of fixed symplectic matrix task without quantum memory]\label{thm:Spchannelhard1}
Any learning algorithm without quantum memory requires
\begin{equation}
\label{E:TOmega3}
    T \geq \Omega\left( d^{1/3} \right),
\end{equation}
to correctly distinguish between the completely depolarizing channel $\mathcal{D}$ on $n$ qubits from a fixed, Haar-random symplectic matrix channel $\mathcal{U}[\rho] = S \rho S^D$ on $n$ qubits with probability at least $2/3$.
\end{theorem}

\noindent Hence, we established an exponential lower bound when the algorithms do not have a quantum memory.
The proofs of Theorems~\ref{thm:channelhard1},~\ref{thm:Ochannelhard1}, and~\ref{thm:Spchannelhard1} have many similarities; however, the first involves heavy use of the combinatorics of permutations, whereas the latter two involve heavy use of the combinatorics of pair permutations.  As such, we will prove Theorems~\ref{thm:channelhard1} first, followed by a simultaneous proof of Theorems~\ref{thm:Ochannelhard1} and~\ref{thm:Spchannelhard1}.

We now turn to a proof of Theorem's~\ref{thm:channelhard1},~\ref{thm:Ochannelhard1}, and~\ref{thm:Spchannelhard1} which are our main result about the fixed unitary problem.  We note that a special case of these theorems were proved in~\cite{aharonov2021quantum}, namely where the learning protocol (see Definition~\ref{def:treechannel}) does not have an auxiliary system, i.e.~$\mathcal{H}_{\text{aux}}$ is a trivial Hilbert space.  The inclusion of an $\mathcal{H}_{\text{aux}}$ of arbitrary size $d' = 2^{n'}$ is our main technical contribution here; our proof will follow the same contours as that of~\cite{aharonov2021quantum}, but with substantive modifications and generalizations.  Indeed, the original proof strategy leads to $T$ bounds like~\eqref{E:TOmega1},~\eqref{E:TOmega2}, and~\eqref{E:TOmega3}, but reduced by factors of $d'^T$, rendering the bounds useless even when $d' \sim \mathcal{O}(1)$ (but non-zero).

\subsubsection{Proof of Theorem~\ref{thm:channelhard1}}

Let us begin with a proof of Theorem~\ref{thm:channelhard1}:

\begin{proof}
The proof begins by utilizing Lemma~\ref{lem:lecam-channel}, which gives
\begin{equation}
    \frac{2}{3} \leq \frac{1}{2} \sum_{\ell \in \mathrm{leaf}(\mathcal{T})} \left| p^{\mathcal{D}}(\ell) - \left( \Expect_{\mathcal{U}} p^{\mathcal{U}}(\ell) \right) \right|.
\end{equation}
The goal now would be to obtain an upper bound right hand side.
It is convenient to establish some notation.  Each root-to-leaf path in $\mathcal{T}$ is specified by a sequence of vertices $v_0, v_1,..., v_T$ where $v_0 = r$ is the root and $v_T = \ell$ is a leaf.  Moreover, the leaf $\ell$ determines the entire root-to-leaf path: it is the shortest path from that leaf to the root.  So knowing $\ell$ is the same as knowing the entire path $v_0 = r, v_1,...,v_{T-1}, v_T = \ell$. 
Recall from Eq.~\eqref{eq:prob-channel-node} and following the root-to-leaf path $v_0 = r, v_1,...,v_{T-1}, v_T = \ell$, the probability of the leaf $\ell$ under the channel $\mathcal{C}$ is
\begin{equation}
    p^{\mathcal{C}}(\ell) = \prod_{t=1}^T \left(w_{v_t} d d' \, \langle \psi_{v_t}|\,(\mathcal{C} \otimes \mathcal{I}_{\text{\rm aux}})[|\phi_{v_{t-1}}\rangle \langle \phi_{v_{t-1}}|] \,|\psi_{v_t}\rangle \right),
\end{equation}
where each $\ket{\phi_{v_{t-1}}}, \ket{\psi_{v_{t-1}}}$ lives in $(d d')$-dimensional Hilbert space $\mathcal{H} \simeq \mathcal{H}_{\text{\rm main}} \otimes \mathcal{H}_{\text{\rm aux}}$ from Definition~\ref{def:treechannel}.
To analyze the probability, we let
\begin{align}
\label{E:Phibig1}
|\Phi_\ell\rangle &= |\phi_{v_0}\rangle  \otimes |\phi_{v_1}\rangle \otimes \cdots \otimes |\phi_{v_{T-1}}\rangle \\
\label{E:Psibig1}
|\Psi_\ell\rangle &= |\psi_{v_1}\rangle  \otimes |\psi_{v_2}\rangle \otimes \cdots \otimes |\psi_{v_T}\rangle  \\
\label{E:Wbig1}
W_\ell &= w_{v_1} w_{v_2} \cdots w_{v_T}
\end{align}
where $v_0 = r$ and $v_T = \ell$.  Notice that $|\Phi_\ell\rangle$ and $|\Psi_\ell\rangle$ each live in $\mathcal{H}^{\otimes T}$.
Recall from Definition~\ref{def:treechannel}, for any node $u$, the set $\{\sqrt{w_v d d'}\,|\psi_v\rangle \langle \psi_v|\}_{v \in \mathrm{child}(u)}$ is a POVM.
Hence, we have $\sum_{v \in \mathrm{child}(u)} w_v d d' \,|\psi_v\rangle \langle \psi_v| = \mathds{1}_{\text{main}} \otimes \mathds{1}_{\text{aux}}$.
It is not hard to use this fact to derive the following identity
\begin{align}
\label{E:Widentity1}
\sum_{\ell \,\in\, \text{leaf}(\mathcal{T})} (d d')^T W_\ell \, |\Psi_\ell\rangle \langle \Psi_\ell| &= (\mathds{1}_{\text{main}} \otimes \mathds{1}_{\text{aux}})^{\otimes T}
\end{align}
and so accordingly $\sum_{\ell \,\in\, \text{leaf}(\mathcal{T})} W_\ell = 1$.

With these notations at hand, we can write $p^{\mathcal{D}}(\ell)$ diagrammatically as
\begin{align}
\label{E:PS1}
    \includegraphics[scale=.32, valign = c]{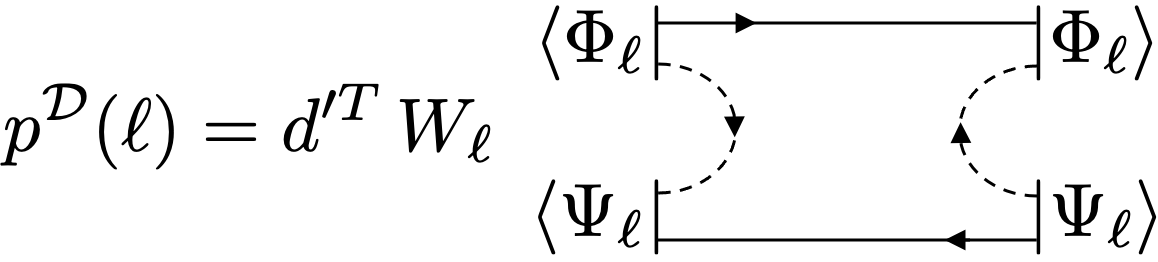}.
\end{align}
For $\mathbb{E}_{\text{\rm Haar}}[p^{\mathcal{U}}(\ell)]$, we utilize the Haar averaging of unitary group discussed in Section~\ref{sec:haaraverageUd} to obtain
\begin{align}
\label{E:QS1}
    \includegraphics[scale=.32, valign = c]{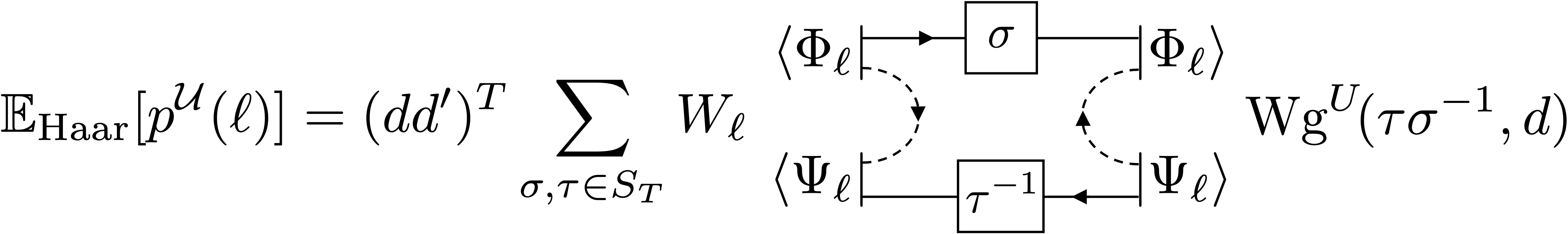}
\end{align}
where the solid lines correspond to $\mathcal{H}_{\text{main}}^{\otimes T}$ and the dotted lines correspond to $\mathcal{H}_{\text{aux}}^{\otimes T}$\,.  It is convenient to let $p_{\sigma, \tau}(\ell)$ denote the summand of~\eqref{E:QS1}.
We now use the triangle inequality in combination with the Cauchy-Schwarz inequality to write
\begin{align}
\label{E:firstineq1}
&\frac{1}{2}\sum_{\ell \,\in\,\text{leaf}(\mathcal{T})} |p^{\mathcal{D}}(\ell) - \mathbb{E}_{\text{Haar}}[p^{\mathcal{U}}(\ell)]| \nonumber \\
&\qquad \leq \frac{1}{2}\sum_{\ell \,\in\,\text{leaf}(\mathcal{T})} |p^{\mathcal{D}}(\ell) - p_{\mathds{1},\mathds{1}}(\ell)| + \frac{1}{2}\sum_{\ell \,\in\,\text{leaf}(\mathcal{T})} \sum_{\sigma \not = \mathds{1}}|p_{\sigma, \mathds{1}}(\ell)| +  \frac{1}{2}\sum_{\ell \,\in\,\text{leaf}(\mathcal{T})} \sum_{\tau \not = \mathds{1}, \,\sigma}|p_{\sigma, \tau}(\ell)| \,.
\end{align}
We will bound each term in turn.

\subsubsection*{First term}
Applying Cauchy-Schwarz to the first term in~\eqref{E:firstineq1} we find that it is less than or equal to
\begin{align}
\label{E:11bound}
    \includegraphics[scale=.32, valign = c]{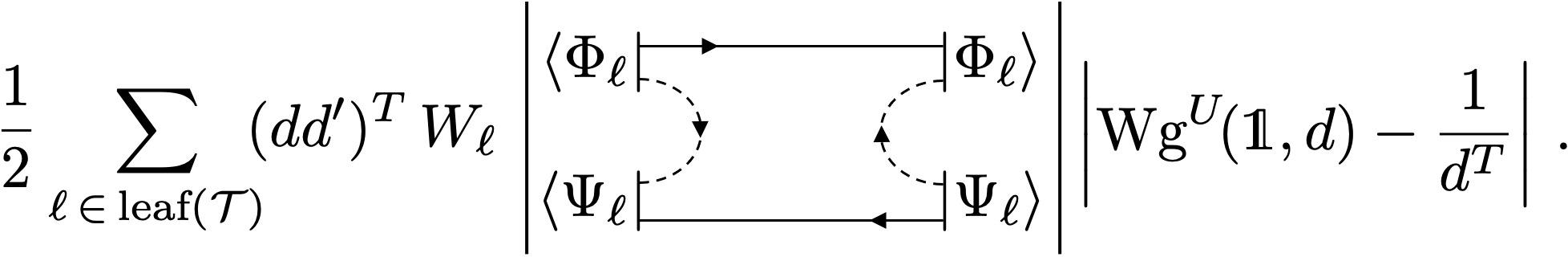}
\end{align}
We can remove the absolute values on the diagrammatic term since it is strictly positive; this enables us to perform the sum over leaves $\sum_{\ell \in \mathrm{leaf}(\mathcal{T})}$ and via the identity given in~\eqref{E:Widentity1}. The first term in~\eqref{E:firstineq1} can now be written as
\begin{equation}
\label{E:Wg1remain1}
\frac{d^T}{2}\,\left|\text{Wg}^U(\mathds{1},d) - \frac{1}{d^T}\right|\,.
\end{equation}
Since by Corollary~\ref{cor:WgU1bd} we have $|\text{Wg}^U(\mathds{1},d) - \frac{1}{d^T}| \leq \mathcal{O}(T^{7/2}/d^{T+2})$ for $T < \left(\frac{d}{\sqrt{6}}\right)^{4/7}$, Eqn.~\eqref{E:Wg1remain1} is $\mathcal{O}(T^{7/2}/d^{2})$ and so
\begin{equation}
\frac{1}{2}\sum_{\ell \,\in\,\text{leaf}(\mathcal{T})} |p^{\mathcal{D}}(\ell) - p_{\mathds{1},\mathds{1}}(\ell)| \leq \mathcal{O}\!\left(\frac{T^{7/2}}{d^2}\right)\,.
\end{equation}

\subsubsection*{Second term}
Next we treat the second term in~\eqref{E:firstineq1}, namely $\frac{1}{2}\sum_{\ell \,\in\,\text{leaf}(\mathcal{T})} \sum_{\sigma \not = \mathds{1}}|p_{\sigma, \mathds{1}}(\ell)|$.  Utilizing the Cauchy-Schwarz inequality, this term is less than or equal to
\begin{align}
\label{E:cs2ndterm}
    \includegraphics[scale=.32, valign = c]{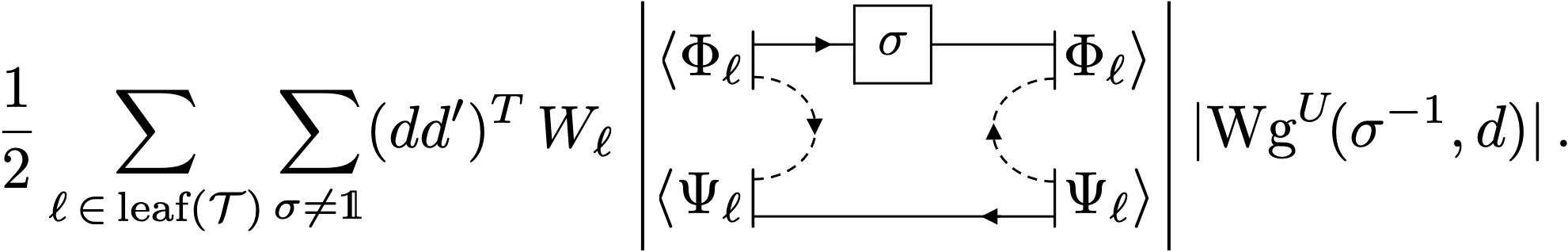}
\end{align}
We can dissect the middle term in the summand using the H\"{o}lder inequality
\begin{align}
    \includegraphics[scale=.32, valign = c]{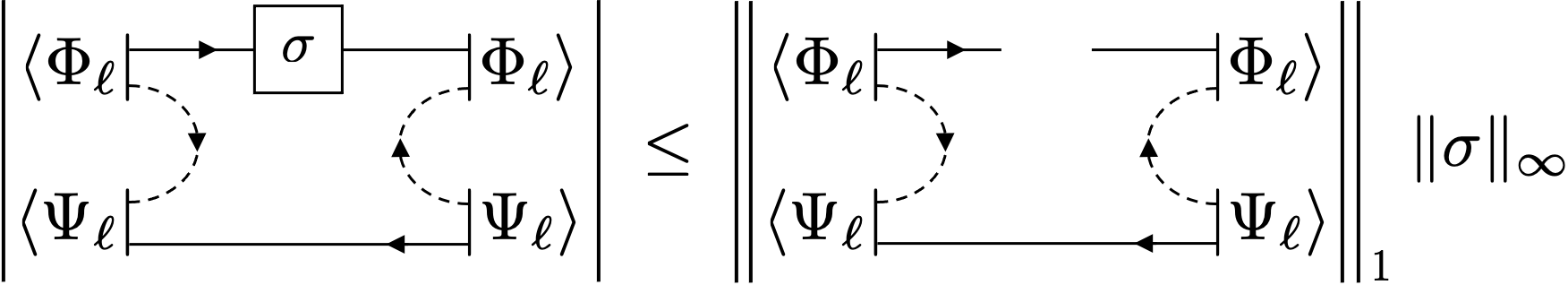}
\end{align}
and notice that $\|\sigma\|_\infty = 1$.  Furthermore, using the fact that the matrix inside the 1-norm on the right-hand side is positive semi-definite, we can replace the 1-norm with the trace and find the bound
\begin{align}
    \includegraphics[scale=.32, valign = c]{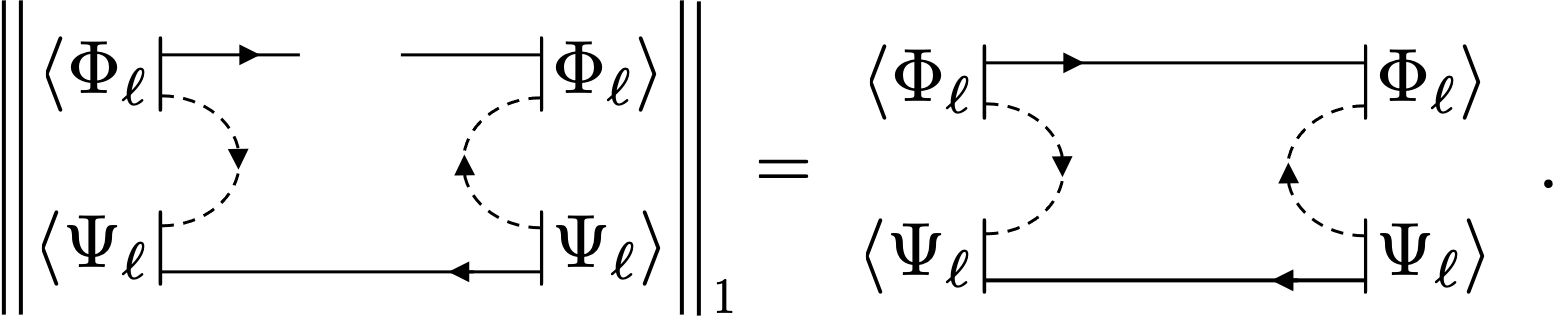}
\end{align}
Thus~\eqref{E:cs2ndterm} is upper bounded by
\begin{align}
    \includegraphics[scale=.32, valign = c]{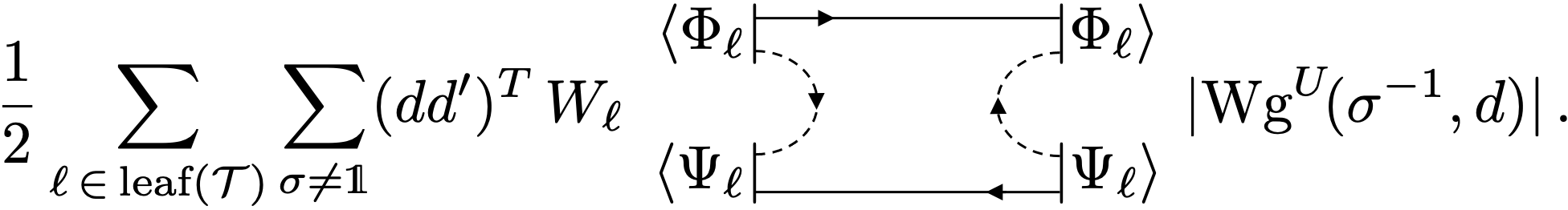}
\end{align}
Then applying the identity~\eqref{E:Widentity1} we are left with
\begin{equation}
\frac{d^T}{2} \sum_{\sigma \not = \mathds{1}} |\text{Wg}^U(\sigma^{-1},d)|
\end{equation}
which is less than or equal to $\mathcal{O}(T^2/d)$ by Lemma 6 of~\cite{aharonov2021quantum}.  In summary, we have
\begin{equation}
\frac{1}{2}\sum_{\ell \,\in\,\text{leaf}(\mathcal{T})} \sum_{\sigma \not = \mathds{1}}|p_{\sigma, \mathds{1}}(\ell)| \leq \mathcal{O}\!\left(\frac{T^2}{d}\right)\,.
\end{equation}

\subsubsection*{Third term}
Finally we treat the third term in~\eqref{E:firstineq1}, namely $\frac{1}{2}\sum_{\ell \,\in\,\text{leaf}(\mathcal{T})} \sum_{\tau \not = \mathds{1}, \,\sigma}|p_{\sigma, \tau}(\ell)|$, which is the most difficult case.  Leveraging the Cauchy-Schwarz inequality, this term is upper bounded by
\begin{align}
\label{E:thirdterm1_pre}
    \includegraphics[scale=.32, valign = c]{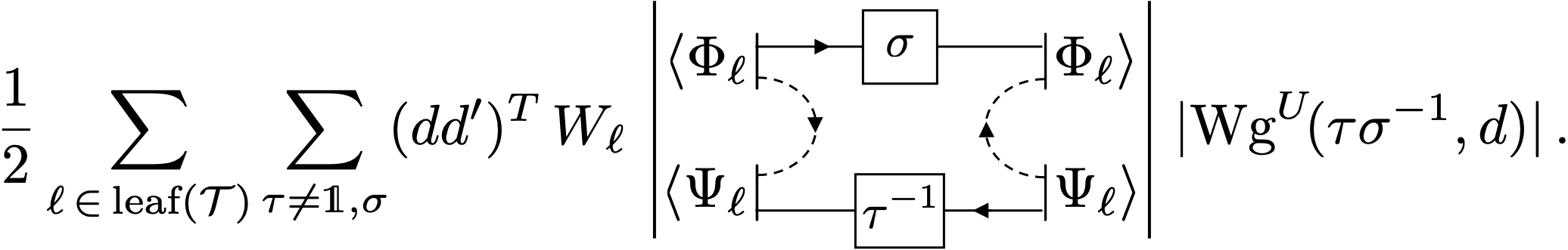}
\end{align}
Applying the H\"{o}lder inequality to the second term in the summand as before, we find
\begin{align}
\label{E:Holder2}
    \includegraphics[scale=.32, valign = c]{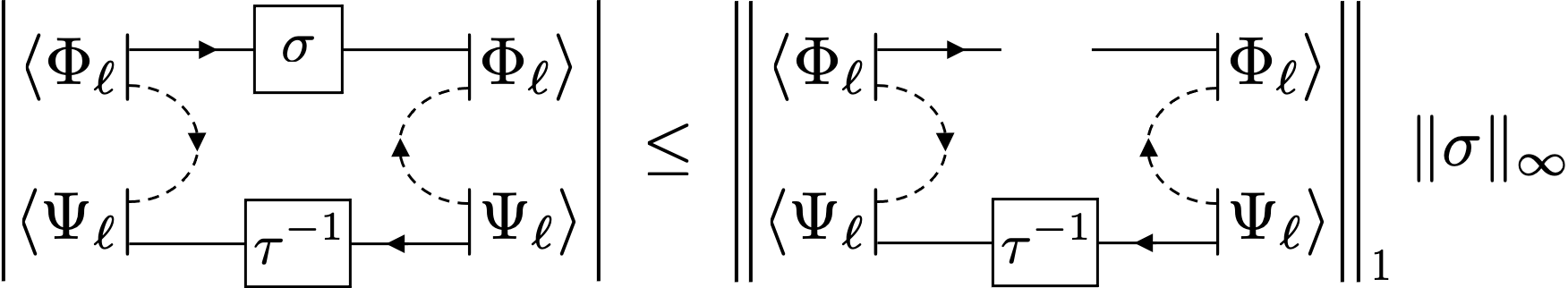}
\end{align}
where $\|\sigma\|_\infty = 1$.  We also use the convenient inequality
\begin{align}
\label{E:thirdterm1}
\includegraphics[scale=.32, valign = c]{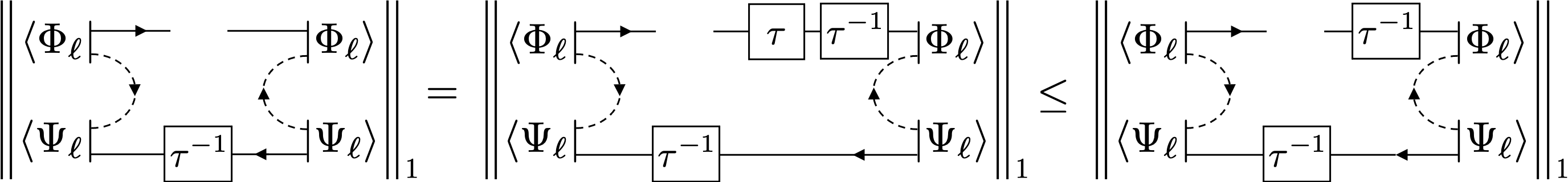}
\end{align}
to reorganize the order of the tensor legs; we have used the H\"{o}lder inequality to go from the middle term to the last term, and the fact that $\|\tau\|_\infty = 1$.

For a fixed permutation $\tau^{-1}$, we can decompose it into cycles $C_1 C_2 \cdots C_{\#(\tau^{-1})}$.  We say that $i \to j$ is in the $m$th cycle $C_m$ if $C_m = (\cdots ij \cdots)$.  Using this notation, for fixed $\tau^{-1} = C_1 C_2 \cdots C_{\#(\tau^{-1})}$ and letting $v_0 = r, v_1,...,v_{T-1}, v_T = \ell$ be the root-to-leaf path terminating in $\ell$, we have

\begin{lemma}
    We have the following identity represented using tensor network diagrams.
    \begin{align}
    \label{E:cycledecomp1}
        \includegraphics[scale=.32, valign = c]{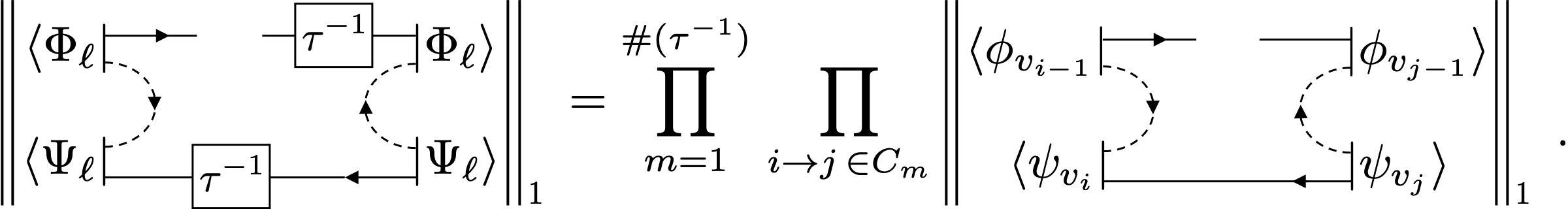}
    \end{align}
\end{lemma}

\begin{proof}
    Take any cycle $C_m$ and any $i\to j\in C_m$. The solid leg of $\bra{\phi_{v_{i - 1}}}$ is dangling and the dotted leg connects to $\bra{\psi_{v_i}}$. Similarly, the solid leg of $\ket{\phi_{v_{j-1}}}$ is dangling and the dotted leg connects to $\ket{\psi_{v_j}}$. Lastly, the solid leg of $\bra{\psi_{v_i}}$ connects to $\ket{\psi_{v_j}}$. This accounts for all legs among $\phi_{v_{i-1}}$, $\psi_{v_i}$, $\phi_{v_{j - 1}}$, and $\psi_{v_j}$, so the part of the diagram on the left of \eqref{E:cycledecomp1} that corresponds to these four states is not connected to the rest of the diagram. In this fashion, we conclude that the diagram on the left is a tensor product of the diagrams on the left for all $m$ and $i\to j\in C_m$, from which the lemma follows.
\end{proof}

We would like to convert the product of trace norms in \eqref{E:cycledecomp1} into a product of traces. We do so via the following basic estimate. First, for any $i\in[T]$, define the unnormalized density operator $\widetilde{\rho}_{v_i} \in \text{Mat}_{d \times d}(\mathbb{C})$ by
\begin{align}
\label{E:rhotildedef}
    \includegraphics[scale=.32, valign = c]{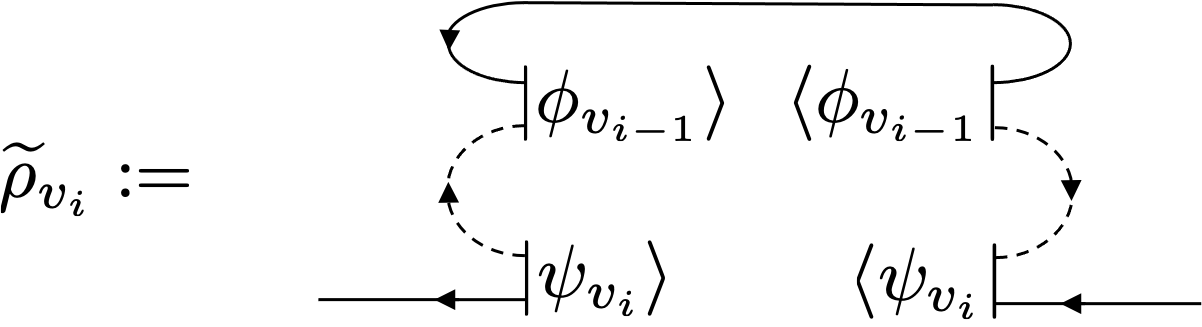}
\end{align}

\begin{lemma}
    For any $i\to j\in C_m$, the corresponding term on the right-hand side of \eqref{E:cycledecomp1} is upper bounded by $\Tr(\wt{\rho}_{v_i}\wt{\rho}_{v_j})$. In particular, \eqref{E:cycledecomp1} is at most \begin{equation}\label{E:productwithsqrt1}
        \prod^{\#(\tau^{-1})}_{m = 1}\prod_{i\to j\in C_m} \sqrt{\Tr(\wt{\rho}_{v_i}\wt{\rho}_{v_j})}.
    \end{equation}
\end{lemma}

\begin{proof}
Using the relations $\|A\|_1 = \|A\otimes A^\dagger\|_1^{1/2} \leq \|\text{SWAP}  \cdot (A\otimes A^\dagger)\|_1^{1/2}$, we find
\begin{align}
    \includegraphics[scale=.32, valign = c]{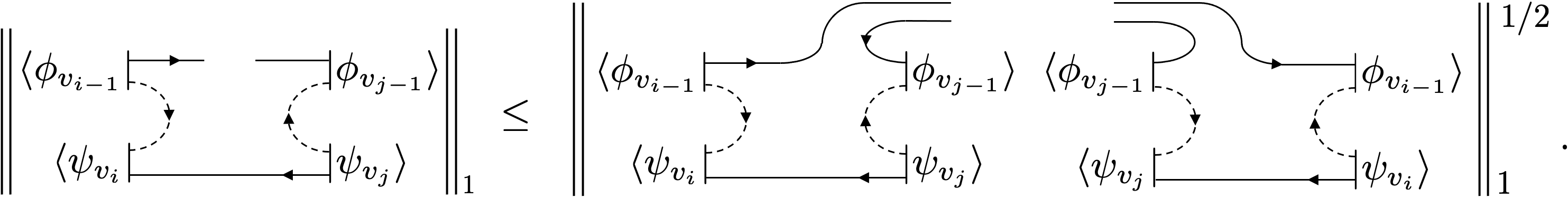} \nonumber
\end{align}
Since the operator inside the 1-norm on the right-hand side is clearly positive semi-definite, we can replace the 1-norm with a trace, namely
\begin{align}
\label{E:1normproperty2}
    \includegraphics[scale=.32, valign = c]{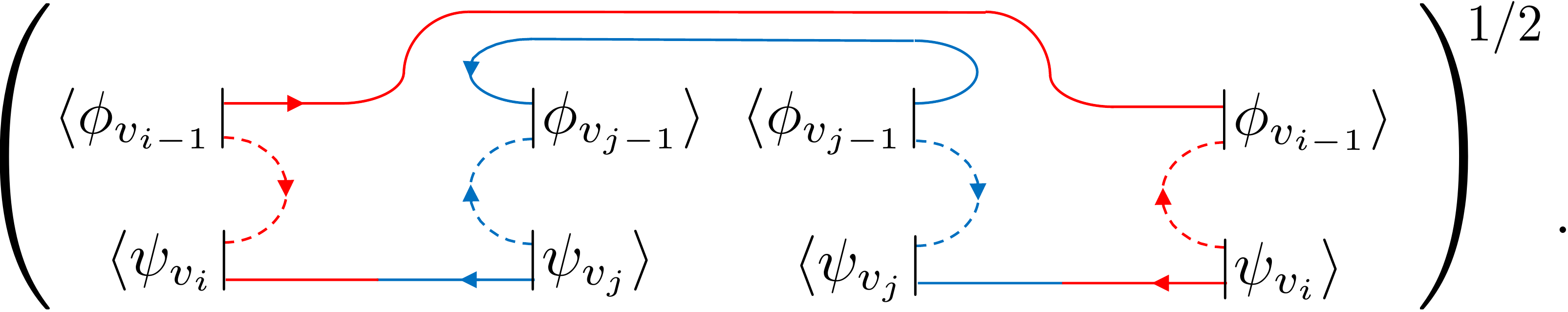}
\end{align}
Observe that we can equivalently rewrite~\eqref{E:1normproperty2} as $\sqrt{\text{tr}(\widetilde{\rho}_{v_i} \widetilde{\rho}_{v_j})}$\,, where $\wt{\rho}_{v_i}$ is the component of the diagram denoted in red and $\wt{\rho}_{v_j}$ is the one in blue, yielding the first part of the lemma. The second follows from plugging this into the right-hand side of Eqn.~\eqref{E:cycledecomp1}.
\end{proof} 

To bound \eqref{E:productwithsqrt1}, it is convenient to develop some notation for cycles $C_m$\,.  The usual notation for a cycle of length $p$ (which for us is less than or equal to $T$) is $C_m = (a_1 a_2 \cdots a_p)$ where in our setting $\{a_1, a_2,...,a_p\} \subseteq \{1,2,...,T\}$.  We will decorate each $a_i$ by an additional subscript as $a_{m,i}$ to remember that it that belongs to the $m$th cycle $C_m$. Similarly, we will sometimes write $p = |C_m|$ to remind ourselves that it depends on $m$.  In this notation, we can write \eqref{E:productwithsqrt1} as a product of
\begin{equation}
\label{E:cycletobound1}
\sqrt{\text{tr}(\widetilde{\rho}_{v_{a_{m,1}}} \widetilde{\rho}_{v_{a_{m,2}}})}  \,  \sqrt{\text{tr}(\widetilde{\rho}_{v_{a_{m,2}}} \widetilde{\rho}_{v_{a_{m,3}}})} \,\cdots \, \sqrt{\text{tr}(\widetilde{\rho}_{v_{a_{m,p-1}}} \widetilde{\rho}_{v_{a_{m,p}}})} \,  \sqrt{\text{tr}(\widetilde{\rho}_{v_{a_{m,p}}} \widetilde{\rho}_{v_{a_{m,1}}})}
\end{equation}
over the different $m = 1, \ldots, \#(\tau^{-1})$.
We can further process the above expression for a given $m$.
We proceed by analyzing two cases: (i) $p = |C_m|$ is even, and (ii) $p = |C_m|$ is odd. \\ \\
\textbf{Case 1:} \textit{$p$ is even.} Each term in Eqn.~\eqref{E:cycletobound1} has the form $\sqrt{\text{tr}(\widetilde{\rho}_{v_{a_{m,i}}} \widetilde{\rho}_{v_{a_{m,i+1}}})}$ except for the last term; however, if we treat the $i$ subscripts of $a_{m,i}$ modulo $p$, then we can write the last term as $\sqrt{\text{tr}(\widetilde{\rho}_{v_{a_{m,p}}} \widetilde{\rho}_{v_{a_{m,p+1}}})}$\,.  We elect to use this notation.  Then we can rearrange and group the terms in Eqn.~\eqref{E:cycletobound1} as follows
\begin{equation}
\left(\prod_{i\text{ odd}} \sqrt{\text{tr}(\widetilde{\rho}_{v_{a_{m,i}}} \widetilde{\rho}_{v_{a_{m,i+1}}})}\right) \left(\,\prod_{j\text{ even}} \sqrt{\text{tr}(\widetilde{\rho}_{v_{a_{m,j}}} \widetilde{\rho}_{v_{a_{m,j+1}}})}\right)\,.
\end{equation}
Using the inequality $ab \leq \frac{1}{2}(a^2 + b^2)$, the above is upper bounded by
\begin{equation}
\frac{1}{2} \prod_{i\text{ odd}} \text{tr}(\widetilde{\rho}_{v_{a_{m,i}}} \widetilde{\rho}_{v_{a_{m,i+1}}}) + \frac{1}{2}\prod_{j\text{ even}} \text{tr}(\widetilde{\rho}_{v_{a_{m,j}}} \widetilde{\rho}_{v_{a_{m,j+1}}})\,.
\end{equation}
We will call the first term $\frac{1}{2}\,R_{m,\,-}$ and the second term $\frac{1}{2}\,R_{m,\,+}$\,. \\ \\
\textbf{Case 2:} \textit{$p$ is odd.} Again consider the product in Eqn.~\eqref{E:cycletobound1}.  We can rearrange and group terms as
\begin{equation}
\label{E:oddsplit1}
\sqrt{\text{tr}(\widetilde{\rho}_{v_{a_{m,p}}} \widetilde{\rho}_{v_{a_{m,1}}})}\left(\prod_{\substack{i\text{ odd} \\ 1 \leq i \leq p - 2}} \sqrt{\text{tr}(\widetilde{\rho}_{v_{a_{m,i}}} \widetilde{\rho}_{v_{a_{m,i+1}}})}\right) \left(\,\prod_{j\text{ even}} \sqrt{\text{tr}(\widetilde{\rho}_{v_{a_{m,j}}} \widetilde{\rho}_{v_{a_{m,j+1}}})}\right)\,.
\end{equation}
If two matrices $A,B$ are Hermitian and positive semi-definite, then using Cauchy-Schwarz combined with operator norm inequalities we have $\text{tr}(AB) \leq \|A\|_2 \, \|B\|_2 \leq \|A\|_1 \, \|B\|_1 \leq \text{tr}(A) \, \text{tr}(B)$.  Accordingly, we have $\sqrt{\text{tr}(\widetilde{\rho}_{v_{a_{m,p}}} \widetilde{\rho}_{v_{a_{m,1}}})} \leq \sqrt{\text{tr}(\widetilde{\rho}_{v_{a_{m,p}}}) \, \text{tr}(\widetilde{\rho}_{v_{a_{m,1}}})}$ and so~\eqref{E:oddsplit1} is upper bounded by
\begin{equation}
\left(\sqrt{\text{tr}(\widetilde{\rho}_{v_{a_{m,p}}})} \prod_{\substack{i\text{ odd} \\ 1 \leq i \leq p - 2}} \sqrt{\text{tr}(\widetilde{\rho}_{v_{a_{m,i}}} \widetilde{\rho}_{v_{a_{m,i+1}}})}\right) \left(\sqrt{\text{tr}(\widetilde{\rho}_{v_{a_{m,1}}})}\,\prod_{j\text{ even}} \sqrt{\text{tr}(\widetilde{\rho}_{v_{a_{m,j}}} \widetilde{\rho}_{v_{a_{m,j+1}}})}\right)\,.
\end{equation}
Again using the inequality $ab \leq \frac{1}{2}(a^2 + b^2)$, we have the upper bound
\begin{equation}
\frac{1}{2} \,\text{tr}(\widetilde{\rho}_{v_{a_{m,p}}})\prod_{\substack{i\text{ odd} \\ 1 \leq i \leq p - 2}} \text{tr}(\widetilde{\rho}_{v_{a_{m,i}}} \widetilde{\rho}_{v_{a_{m,i+1}}}) + \frac{1}{2}\, \text{tr}(\widetilde{\rho}_{v_{a_{m,1}}})\prod_{j\text{ even}} \text{tr}(\widetilde{\rho}_{v_{a_{m,j}}} \widetilde{\rho}_{v_{a_{m,j+1}}})
\end{equation}
where we analogously call the first term $\frac{1}{2}\,R_{m,\,-}$ and the second term $\frac{1}{2}\,R_{m,\,+}$\,. 

In both cases, note that $R_{m,\,-}$ and $R_{m,\,+}$ implicitly depend on the leaf $\ell$, so we will denote them by $R^{\ell}_{m,\,-}$ and $R^{\ell}_{m,\,+}$ when we wish to make this dependence explicit.\\ \\
\indent Putting together our notation and bounds from Case 1 and Case 2, we find that Eqn.~\eqref{E:productwithsqrt1} is upper bounded by
\begin{align}
\label{E:multipleTs}
\prod_{m=1}^{\#(\tau^{-1})} \,\prod_{i \to j \, \in C_m} \sqrt{\text{tr}(\widetilde{\rho}_{v_i} \widetilde{\rho}_{v_j})} &\leq \frac{1}{2^{\#(\tau^{-1})}} \prod_{m=1}^{\#(\tau^{-1})} \left(R_{m,-} + R_{m,+} \right) \nonumber \\
&= \frac{1}{2^{\#(\tau^{-1})}} \sum_{i_1,...,i_{\#(\tau^{-1})} = \pm} R_{1,i_1} R_{2,i_2} \cdots R_{\#(\tau^{-1}), i_{\#(\tau^{-1})}}\,.
\end{align}
Each term in the sum in the last line of~\eqref{E:multipleTs} is a product of terms like $\text{tr}(\widetilde{\rho}_{v_i})$ and $\text{tr}(\widetilde{\rho}_{v_j} \widetilde{\rho}_{v_k})$.  But the key point is that we have arranged the equations so that each term in the sum has $\widetilde{\rho}_{v_i}$ for each $i = 1,...,T$ appear \textit{exactly once}.  This has the following highly useful consequence.

\begin{lemma}\label{lem:collapse}
    Fix any $i_1,...,i_{\#(\tau^{-1})} \in \{+,-\}$. Then
    \begin{equation}
        \label{E:sboundhard1}
        \sum_{\ell \,\in\,\text{\rm leaf}(\mathcal{T})} (d d')^T \,W_\ell \, R^{\ell}_{1,i_1} R^{\ell}_{2,i_2} \cdots R^{\ell}_{\#(\tau^{-1}), i_{\#(\tau^{-1})}} \le d^{\frac{T}{2} - \left\lfloor \frac{L(\tau^{-1})}{2}\right\rfloor}
    \end{equation}
where $L(\tau^{-1})$ is the length of the longest cycle in $\tau^{-1}$.
\end{lemma}

\begin{proof}
    For ease of notation, we will let $R^{\ell}_j\triangleq R^{\ell}_{j,i_j}$.
    Recall from~\eqref{E:Wbig1} that $W_\ell = w_{v_1} w_{v_2} \cdots w_{v_T}$ and note that
    \begin{equation}
        \sum_{v: \, \text{depth}(v) = i} dd'\,w_{v} \, \widetilde{\rho}_{v} = \langle \phi_{\text{parent}(v)} | \phi_{\text{parent}(v)}\rangle\, \mathds{1}_{d \times d} = \mathds{1}_{d \times d}\,.
    \end{equation}
    Accordingly we have that for any $\rho\in\text{Mat}_{d\times d}(\mathbb{C})$,
    \begin{align}
        \label{E:2tracebound1}
        \sum_{v: \, \text{depth}(v) = i} dd'\, w_{v} \, \text{tr}(\widetilde{\rho}_{v} \, \rho ) = \text{tr}(\rho)\,,
    \end{align}
    and in particular, for $\rho = \mathds{1}$,
    \begin{align}
        \label{E:1tracebound1}
        \sum_{v: \, \text{depth}(v) = i} dd'\,w_{v} \, \text{tr}(\widetilde{\rho}_{v}) &= d.
    \end{align}
    We now turn to bounding the left-hand side of \eqref{E:sboundhard1}. Recalling that $\prod_j R^{\ell}_j$ as a product of terms like $\Tr(\wt{\rho}_{v_i})$ and $\Tr(\wt{\rho}_{v_i}\wt{\rho}_{v_{i'}})$, we will define some sets of indices encoding this data. Let $S^{(T)}_1\subseteq[T]$ denote the indices $i$ for which $\Tr(\wt{\rho}_{v_i})$ appears in $\prod_j R^{\ell}_j$, and let $S^{(T)}_2\subseteq[T]\times[T]$ denote the set of (unordered) pairs $(i,i')$ for which $\Tr(\wt{\rho}_{v_i}\wt{\rho}_{v_{i'}})$ appears, so that for any root-to-leaf path in $\calT$ consisting of nodes $v_1,\ldots,v_T = \ell$, we have
    \begin{equation}
        \label{eq:prodRs}
        \prod_j R^{\ell}_j = \prod_{i\in S^{(T)}_1} \Tr(\wt{\rho}_{v_i}) \cdot \prod_{(i,i')\in S^{(T)}_2} \Tr(\wt{\rho}_{v_i}\wt{\rho}_{v_{i'}})
    \end{equation} by definition. Now construct $S^{(t)}_1\subseteq[t]$ and $S^{(t)}_2\subseteq[t]\times[t]$ for $1 \le t < T$ inductively as follows. If $t\in S^{(t)}_1$ then define $S^{(t-1)}_1 \triangleq S^{(t)}_1 \backslash \brc{t}$ and $S^{(t-1)}_2\triangleq S^{(t)}_2$. Otherwise if $(t,t')\in S^{(t)}_2$ for some $t' < t$, then define $S^{(t-1)}_1 \triangleq S^{(t-1)}_1 \cup \brc{t'}$ and $S^{(t-1)}_2\triangleq S^{(t)}_2 \backslash \brc{(t,t')}$. We collect some basic observations about these two set sequences:
    
    \begin{observation}\label{obs:singleton}
        For every $i\in S^{(T)}_1$, we have that $i\in S^{(i)}_1$.
    \end{observation}
    
    \begin{observation}\label{obs:pair}
        For every $(i,i')\in S^{(T)}_2$, if $i\le i'$ then $i\in S^{(i)}_1$ while $i'\not\in S^{(i')}_1$.
    \end{observation}
    
    The reason for defining these set sequences is that we can extract $\wt{\rho}_{v_T}$ from the product on the right-hand side of \eqref{eq:prodRs} and apply \eqref{E:1tracebound1} (resp. \eqref{E:2tracebound1}) if $T\in S^{(T)}_1$ (resp. $(T,t')\in S^{(T)}_2$ for some $t' < T$) to obtain \begin{equation}
        \label{eq:induct_prodRs}
        \sum_{\ell\in\leaf(\calT)} (dd')^T W_{\ell} \prod_j R^{\ell}_j = d^{\bone{T\in S^{(T)}_1}}\sum_{u: \,\text{depth}(u) = T - 1} (dd')^{T-1} W_u \prod_{i\in S^{(T-1)}_1} \Tr(\wt{\rho}_{v_i}) \cdot \prod_{(i,i') \in S^{(T-1)}_2} \Tr(\wt{\rho}_{v_i}\wt{\rho}_{v_{i'}}),
    \end{equation} where $W_u = w_{v_1}\cdots w_{v_{T-1}}$ if the path from root to $u$ in $\calT$ consists of $v_1,\ldots,v_{T-1} = u$.
    Proceeding inductively, we can express the right-hand side of \eqref{eq:induct_prodRs} as \begin{equation}
        d^{\sum^T_{t=1} \bone{t\in S^{(t)}_1}}.
    \end{equation} By Observations~\ref{obs:singleton} and \ref{obs:pair}, $\sum^T_{t=1}\bone{t\in S^{(t)}_1} = |S^{(T)}_1| + |S^{(T)}_2|$. Because every even cycle $C_m$ of $\tau^{-1}$ contributes $|C_m|/2$ pairs to $S^{(T)}_2$, and every odd cycle $C_m$ contributes $\floor{|C_m|/2}$ pairs to $S^{(T)}_2$ and one element to $S^{(T)}_1$, we conclude that
    \begin{align}
        \label{E:sboundhard2}
        \sum_{\ell\in\leaf(\calT)} (dd')^T W_{\ell} \prod_j R^{\ell}_j &= d^{\sum_{m=1}^{\#(\tau^{-1})} \left\lceil \frac{|C_m|}{2} \right\rceil} \nonumber \\
        &= d^{\frac{T}{2} - \sum_{m=1}^{\#(\tau^{-1})} \left\lfloor \frac{|C_m|}{2} \right\rfloor} \nonumber \\
        &\leq d^{\frac{T}{2} - \left\lfloor \frac{L(\tau^{-1})}{2} \right\rfloor}
    \end{align}
    as claimed.
\end{proof}

Putting all of our previous analysis together, in particular by combining Eqn.'s~\eqref{E:Holder2}, \eqref{E:thirdterm1}, \eqref{E:cycledecomp1}, \eqref{E:productwithsqrt1}, \eqref{E:multipleTs}, and \eqref{E:sboundhard1}, we arrive at
\begin{equation}
\sum_{\ell \, \in \, \text{leaf}(\mathcal{T})} \sum_{\tau \not = \mathds{1}, \,\sigma}|p_{\sigma, \tau}(s)| \leq d^{T} \sum_{\sigma} |\text{Wg}^U(\sigma^{-1}, d)| \sum_{\tau \not = \mathds{1}} d^{- \left\lfloor \frac{L(\tau^{-1})}{2} \right\rfloor} \,.
\end{equation}
The first sum on the right-hand side is bounded by $d^T \sum_{\sigma} |\text{Wg}^U(\sigma^{-1}, d)| \leq 1 + \mathcal{O}\!\left( \frac{T^2}{d}\right)$.  Considering the second sum on the right-hand side, let $N(T,\ell)$ be the number of permutations in $S_T$ where the length of the longest cycle is $\ell$.  Then the second sum can be written as
\begin{equation}
\label{E:sumNTell}
\sum_{\ell = 2}^T N(T,\ell) \, d^{- \left\lfloor \frac{\ell}{2} \right\rfloor}
\end{equation}
where we omit $k = 1$ from the sum since it corresponds to the identity permutation.  Since $N(T,\ell) \leq \binom{T}{\ell} \, \ell! = \frac{T!}{(T - \ell)!} < T^\ell$,~\eqref{E:sumNTell} is upper bounded by
\begin{equation}
\sum_{\ell = 2}^\infty T^\ell  \, d^{- \left\lfloor \frac{\ell}{2} \right\rfloor} = \frac{(1 + T) \frac{T^2}{d}}{1 - \frac{T^2}{d}} = \frac{T^3}{d} + \frac{T^2}{d} + \mathcal{O}\!\left(\frac{T^5}{d^2}\right)\,.
\end{equation}
Now if $T \le o(d^{1/3})$, then this quantity is $o(1)$ for some absolute constant $c > 0$. Altogether, we find
\begin{equation}
\frac{1}{2}\sum_{\ell\,\in\,\text{leaf}(\mathcal{T})} \sum_{\tau \not = \mathds{1}, \,\sigma}|p_{\sigma, \tau}(\ell)| \leq o\!\left(1\right)\,.
\end{equation}

\end{proof}

\subsubsection{Proof of Theorems~\ref{thm:Ochannelhard1} and~\ref{thm:Spchannelhard1}}

As discussed above, we will present proof of Theorems~\ref{thm:Ochannelhard1} and~\ref{thm:Spchannelhard1}, making heavy use of pair partitions.

\begin{proof}

The probability distribution $p^{\mathcal{D}}(\ell)$ is notated the same way as before.
We can depict $\mathbb{E}_{\text{\rm Haar}}[p^{\mathcal{O}}(\ell)]$ diagrammatically by
\begin{align}
\label{E:OQS1}
    \includegraphics[scale=.32, valign = c]{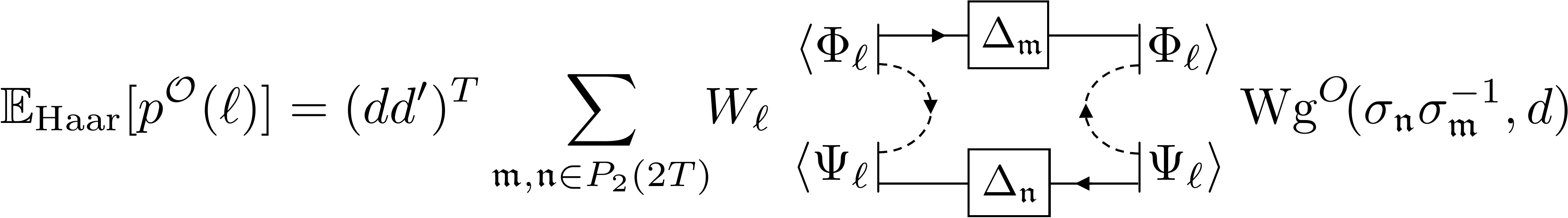}
\end{align}
where $p_{\mathfrak{m}, \mathfrak{n}}^{\mathcal{O}}(\ell)$ denotes the summand.  Similarly $\mathbb{E}_{\text{\rm Haar}}[p^{\mathcal{S}}(\ell)]$ is given by
\begin{align}
\label{E:SpQS1}
    \includegraphics[scale=.32, valign = c]{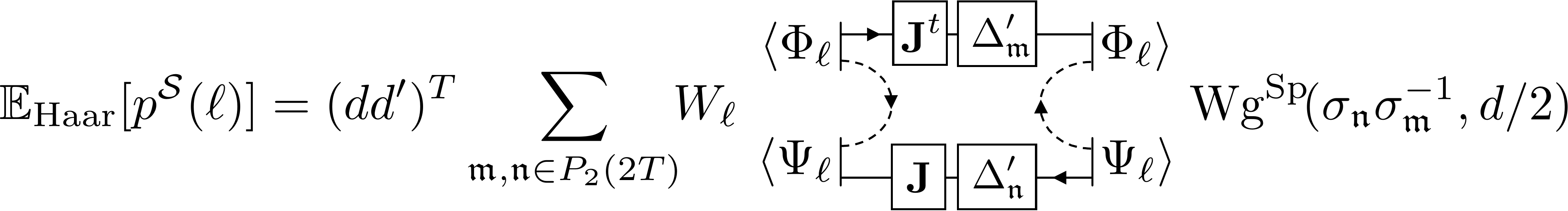}
\end{align}
where here $p_{\mathfrak{m}, \mathfrak{n}}^{\mathcal{S}}(\ell)$ denotes the summand.  Moreover, $\textbf{J} := J^{\otimes t}$.  As before, we use the triangle inequality in combination with the Cauchy-Schwarz inequality to write the two inequalities
\begin{align}
\label{E:Ofirstineq1}
&\frac{1}{2}\sum_{\ell \,\in\,\text{leaf}(\mathcal{T})} |p^{\mathcal{D}}(\ell) - \mathbb{E}_{\text{Haar}}[p^{\mathcal{O}}(\ell)]| \nonumber \\
&\qquad \leq \frac{1}{2}\sum_{\ell \,\in\,\text{leaf}(\mathcal{T})} |p^{\mathcal{D}}(\ell) - p_{\mathfrak{e}, \mathfrak{e}}^{\mathcal{O}}(\ell)| + \frac{1}{2}\sum_{\ell \,\in\,\text{leaf}(\mathcal{T})} \sum_{\mathfrak{m} \not = \mathfrak{e}}|p_{\mathfrak{m}, \mathfrak{e}}^{\mathcal{O}}(\ell)| +  \frac{1}{2}\sum_{\ell \,\in\,\text{leaf}(\mathcal{T})} \sum_{\mathfrak{m} \not = \mathfrak{e}, \,\mathfrak{n}}|p_{\mathfrak{m}, \mathfrak{n}}^{\mathcal{O}}(\ell)| \,.
\end{align}
and
\begin{align}
\label{E:Spfirstineq1}
&\frac{1}{2}\sum_{\ell \,\in\,\text{leaf}(\mathcal{T})} |p^{\mathcal{D}}(\ell) - \mathbb{E}_{\text{Haar}}[p^{\mathcal{S}}(\ell)]| \nonumber \\
&\qquad \leq \frac{1}{2}\sum_{\ell \,\in\,\text{leaf}(\mathcal{T})} |p^{\mathcal{D}}(\ell) - p_{\mathfrak{e}, \mathfrak{e}}^{\mathcal{S}}(\ell)| + \frac{1}{2}\sum_{\ell \,\in\,\text{leaf}(\mathcal{T})} \sum_{\mathfrak{m} \not = \mathfrak{e}}|p_{\mathfrak{m}, \mathfrak{e}}^{\mathcal{S}}(\ell)| +  \frac{1}{2}\sum_{\ell \,\in\,\text{leaf}(\mathcal{T})} \sum_{\mathfrak{m} \not = \mathfrak{e}, \,\mathfrak{n}}|p_{\mathfrak{m}, \mathfrak{n}}^{\mathcal{S}}(\ell)| \,.
\end{align}
We will bound the right-hand sides of~\eqref{E:Ofirstineq1} and~\eqref{E:Spfirstineq1} term by term.

\subsubsection*{First term for $O(d)$ case}
We apply the Cauchy-Schwarz inequality to the first term in~\eqref{E:Ofirstineq1} to find the upper bound
\begin{align}
\label{E:O11bound}
    \includegraphics[scale=.32, valign = c]{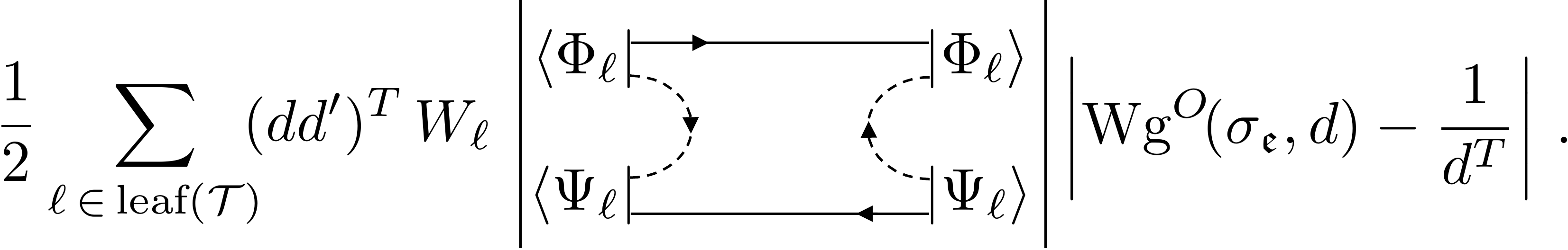}
\end{align}
As in the unitary setting, we remove the absolute values on the diagrammatic term by virtue of its positivity, and sum over leaves to get
\begin{equation}
\label{E:OWg1remain1}
\frac{d^T}{2}\,\left|\text{Wg}^O(\sigma_{\mathfrak{e}},d) - \frac{1}{d^T}\right|\,.
\end{equation}
Using Corollary~\ref{cor:WgO1bd} which gives us $|\text{Wg}^O(\sigma_{\mathfrak{e}},d) - \frac{1}{d^T}| \leq \mathcal{O}(T^{7}/d^{T+2})$ for $T < \left(\frac{d}{12}\right)^{2/7}$, we immediately find
\begin{equation}
\frac{1}{2}\sum_{\ell \,\in\,\text{leaf}(\mathcal{T})} |p^{\mathcal{D}}(\ell) - p_{\mathfrak{e}, \mathfrak{e}}^{\mathcal{O}}(\ell)| \leq \mathcal{O}\!\left(\frac{T^{7}}{d^2}\right)\,.
\end{equation}

\subsubsection*{First term for \normalfont $\text{Sp}(d/2)$ \normalfont \textbf{case}}

We recapitulate the same manipulations in the symplectic case.  Applying the Cauchy-Schwarz inequality to the first term in~\eqref{E:Spfirstineq1} gives us the upper bound
\begin{align}
\label{E:Sp11bound}
    \includegraphics[scale=.32, valign = c]{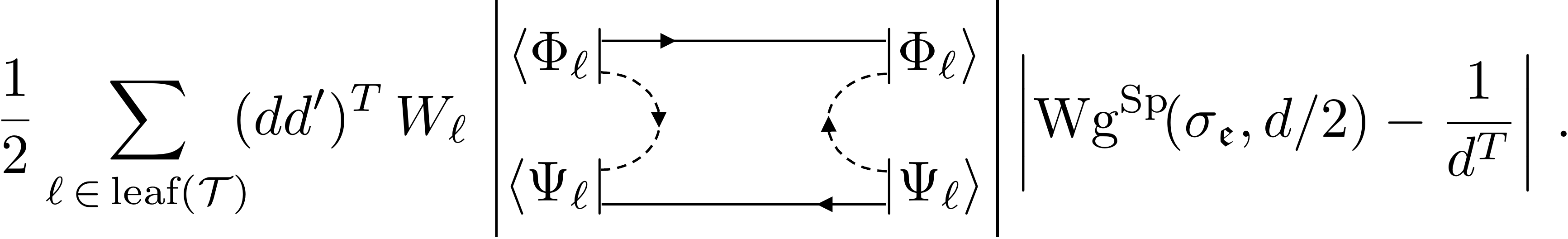}
\end{align}
We again remove the absolute values around the diagrammatic term, and sum over leaves to find
\begin{equation}
\label{E:SpWg1remain1}
\frac{d^T}{2}\,\left|\text{Wg}^{\text{Sp}}(\sigma_{\mathfrak{e}},d/2) - \frac{1}{d^T}\right|\,.
\end{equation}
Using the analogous Corollary~\ref{cor:WgSp1bd} which provides the bound $|\text{Wg}^O(\sigma_{\mathfrak{e}},d) - \frac{1}{d^T}| \leq \mathcal{O}(T^{7/2}/d^{T+2})$ for $T < \left(\frac{d}{6}\right)^{2/7}$, we have
\begin{equation}
\frac{1}{2}\sum_{\ell \,\in\,\text{leaf}(\mathcal{T})} |p^{\mathcal{D}}(\ell) - p_{\mathfrak{e}, \mathfrak{e}}^{\mathcal{S}}(\ell)| \leq \mathcal{O}\!\left(\frac{T^{7/2}}{d^2}\right)\,.
\end{equation}

\subsubsection*{Second term for $O(d)$ case}
Applying Cauchy-Schwarz to the second term in~\eqref{E:Ofirstineq1}, we find the upper bound
\begin{align}
\label{E:Ocs2ndterm}
    \includegraphics[scale=.32, valign = c]{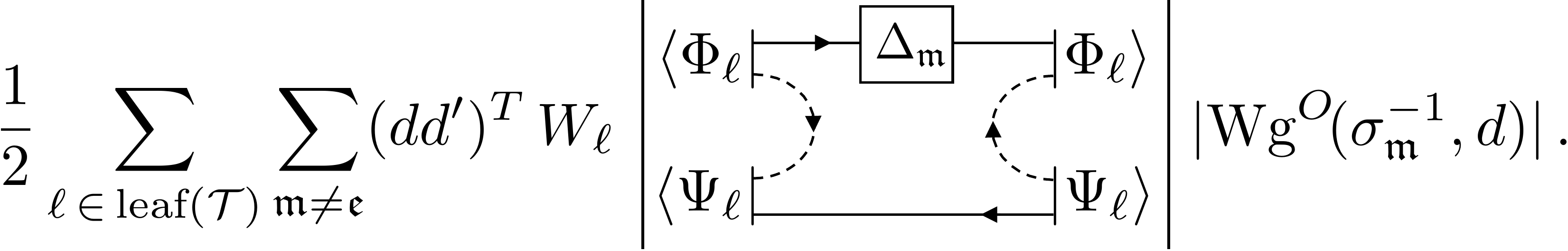}
\end{align}
Let us define the matrix
\begin{align}
\label{E:Cdef1}
    \includegraphics[scale=.32, valign = c]{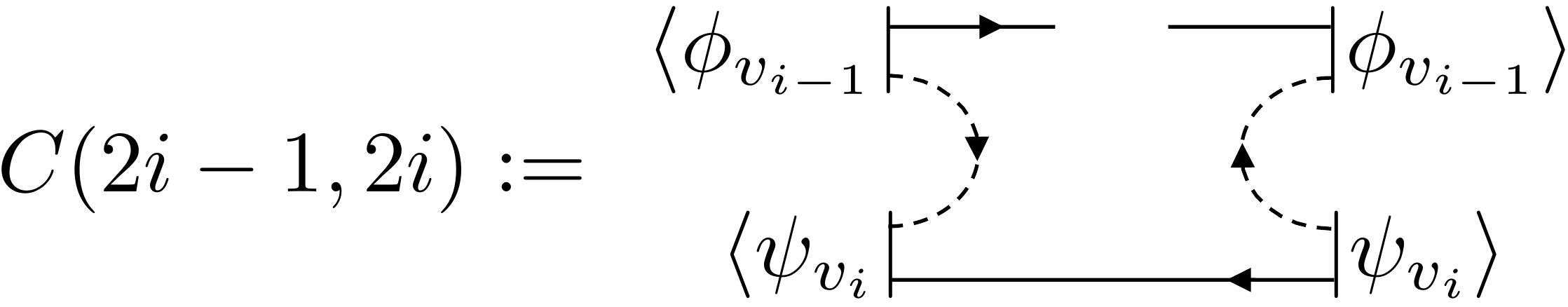}
\end{align}
where the $2i-1$ and $2i$ are just labels (i.e.~they are not matrix indices).  We will further define $C(2i,2i-1) := C(2i-1,2i)^t$.  Then the diagrammatic term in~\eqref{E:Ocs2ndterm} is equivalent to
\begin{align}
\left|\text{tr}\!\left(\Delta_{\mathfrak{m}}\, \bigotimes_{i=1}^T C(2i-1,2i) \right)\right|\,.
\end{align}
This can be expressed more explicitly as the absolute value of a product of traces of the $C$'s, namely
\begin{align}
\label{E:Ctraces1}
&\bigg| \text{tr}\big(C(2,1) C(f_\mathfrak{m}(1), f_\mathfrak{e}\circ f_\mathfrak{m}(1)) C(f_\mathfrak{m} \circ f_\mathfrak{e} \circ f_\mathfrak{m}(1), f_\mathfrak{e}\circ f_\mathfrak{m} \circ f_\mathfrak{e} \circ f_\mathfrak{m}(1))\cdots C(f_{\mathfrak{e}} \circ f_\mathfrak{m^{-1}}(2), f_\mathfrak{m^{-1}}(2))\big)  \nonumber \\
&\qquad \qquad \qquad \qquad \qquad \qquad \qquad \qquad \qquad \qquad \qquad \qquad \qquad  \qquad \qquad \qquad \qquad \cdot \text{tr}\big(\cdots\big) \cdots \text{tr}\big(\cdots\big) \bigg|
\end{align}
where we have used the definition of $f_{\mathfrak{m}}$ and $f_{\mathfrak{e}}$ as per~\eqref{E:ffrakm1}.  Each trace corresponds to a particular $M_{2T}$-cycle of $\mathfrak{m}$.  Using the 1-norm inequality $\|A_1 A_2 \cdots A_k\|_1 \leq \prod_{i=1}^k \|A_i\|_1$\,, Eqn.~\eqref{E:Ctraces1} is upper bounded by
\begin{equation}
\prod_{i=1}^T \|C(2i-1,2i)\|_1
\end{equation}
where we have used $\|C(2i-1,2i)\|_1 = \|C(2i,2i-1)\|_1$.  Since each $C(2i-1,2i)$ is positive semi-definite, $\|C(2i-1,2i)\|_1 = \text{tr}(C(2i-1,2i))$
and so~\eqref{E:Ocs2ndterm} has the upper bound
\begin{align}
    \includegraphics[scale=.32, valign = c]{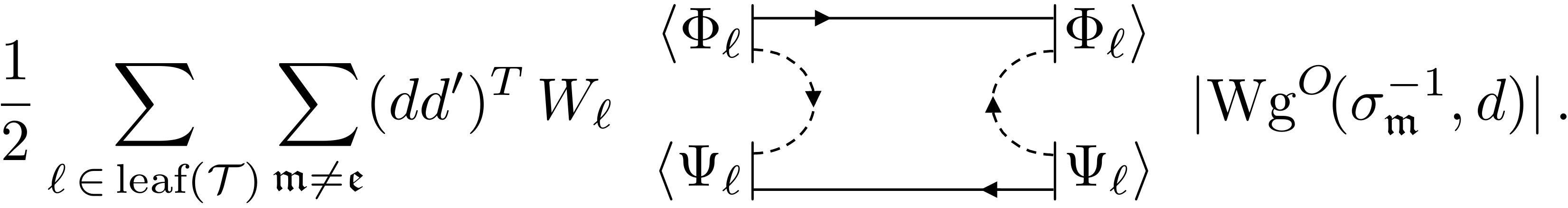}
\end{align}
Summing over leaves we arrive at
\begin{equation}
\frac{d^T}{2} \sum_{\mathfrak{m} \not = \mathfrak{e}} |\text{Wg}^{O}(\sigma_{\mathfrak{m}}^{-1},d)|
\end{equation}
which is upper bounded by $\mathcal{O}(T^7/d^2) + \mathcal{O}(T^2/d)$ using Corollary~\ref{cor:WgO1bd} in combination with Lemma 8 of~\cite{aharonov2021quantum}.  The ultimate result is
\begin{equation}
\frac{1}{2}\sum_{\ell \,\in\,\text{leaf}(\mathcal{T})} \sum_{\mathfrak{m} \not = \mathds{1}}|p_{\mathfrak{m}, \mathfrak{e}}^{\mathcal{O}}(\ell)| \leq \mathcal{O}\!\left(\frac{T^7}{d^2}\right) + \mathcal{O}\!\left(\frac{T^2}{d}\right)\,.
\end{equation}

\subsubsection*{Second term for \normalfont $\text{Sp}(d/2)$ \normalfont \textbf{case}}

A similar proof holds in the symplectic setting. We likewise apply Cauchy-Schwarz to the second term in~\eqref{E:Spfirstineq1} to obtain the upper bound
\begin{align}
\label{E:Spcs2ndterm}
    \includegraphics[scale=.32, valign = c]{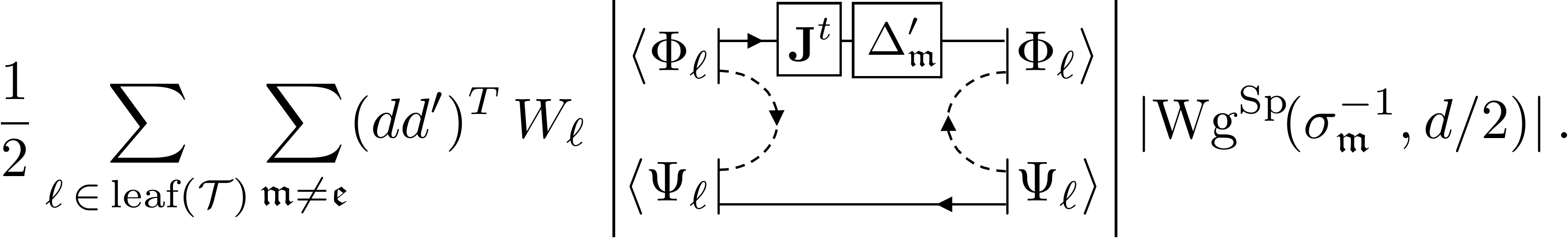}
\end{align}
Using the same notation for $C(2i-1,2i)$ as in Eqn.~\eqref{E:Ctraces1} above, we further define
\begin{equation}
\widetilde{C}(2i-1,2i) := C(2i-1,2i)\cdot J^t
\end{equation}
and similarly $\widetilde{C}(2i,2i-1) := \widetilde{C}(2i-1,2i)^t$.  Then the diagrammatic term in Eqn.~\eqref{E:Spcs2ndterm} can be written as
\begin{align}
\left|\text{tr}\!\left(\Delta_{\mathfrak{m}}'\, \bigotimes_{i=1}^T \widetilde{C}(2i-1,2i) \right)\right|\,.
\end{align}
This can be expanded analogously to~\eqref{E:Ctraces1}, namely as
\begin{equation}
\label{E:Ctraces2}
\bigg| \text{tr}\big(\widetilde{C}(2,1) \,J\, C(f_\mathfrak{m}(1), f_\mathfrak{e}\circ f_\mathfrak{m}(1)) \,J \cdots J\,\widetilde{C}(f_{\mathfrak{e}} \circ f_\mathfrak{m^{-1}}(2), f_\mathfrak{m^{-1}}(2))\,J\big) \,\text{tr}\big(\cdots\big) \cdots \text{tr}\big(\cdots\big) \bigg|\,.
\end{equation}
Using the same $1$-norm inequality as the orthogonal case, Eqn.~\eqref{E:Ctraces2} is upper bounded by
\begin{align}
\prod_{i=1}^T \|\widetilde{C}(2i-1, 2i) \,J\|_1 &\leq \prod_{i=1}^T \| C(2i-1,2i)\|_1 \, \|J\|_\infty \,\|J^t\|_\infty \nonumber \\
&= \prod_{i=1}^T \text{tr}(C(2i-1,2i))\,.
\end{align}
In the first line we have used the H\"{o}lder inequality, and in the second line we used $\|J\|_\infty = \|J^t\|_\infty = 1$ as well as $\|C(2i-1,2i)\|_1 = \text{tr}(C(2i-1,2i))$.   Thus we arrive at an upper bound for~\eqref{E:Spcs2ndterm}:
\begin{align}
    \includegraphics[scale=.32, valign = c]{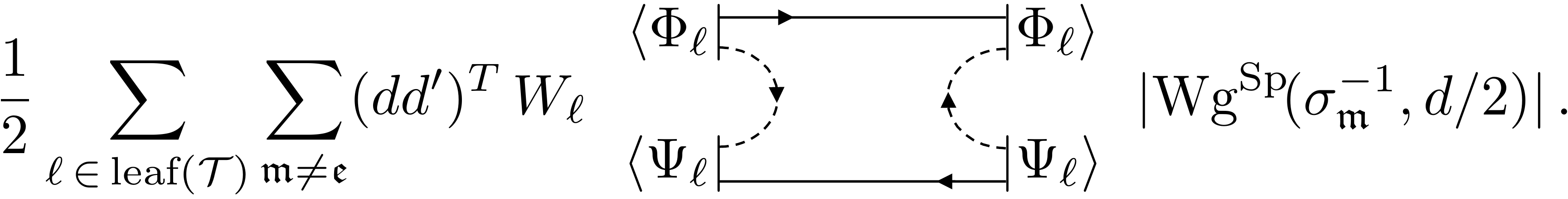}
\end{align}
As before we sum over leaves, giving us
\begin{equation}
\frac{d^T}{2} \sum_{\mathfrak{m} \not = \mathfrak{e}} |\text{Wg}^{\text{Sp}}(\sigma_{\mathfrak{m}}^{-1},d/2)|\,.
\end{equation}
This is is upper bounded by $\mathcal{O}(T^{7/2}/d^2) + \mathcal{O}(T^2/d)$ using Corollary~\ref{cor:WgSp1bd} in combination with Lemma 10 of~\cite{aharonov2021quantum}, and so in the end we obtain
\begin{equation}
\frac{1}{2}\sum_{\ell \,\in\,\text{leaf}(\mathcal{T})} \sum_{\mathfrak{m} \not = \mathds{1}}|p_{\mathfrak{m}, \mathfrak{e}}^{\mathcal{S}}(\ell)| \leq \mathcal{O}\!\left(\frac{T^{7/2}}{d^2}\right) + \mathcal{O}\!\left(\frac{T^2}{d}\right)\,.
\end{equation}

\subsubsection*{Third term for $O(d)$ case}
The third term in~\eqref{E:Ofirstineq1} is $\frac{1}{2}\sum_{\ell \,\in\,\text{leaf}(\mathcal{T})} \sum_{\mathfrak{n} \not = \mathfrak{e}, \,\mathfrak{m}}|p_{\mathfrak{m},\mathfrak{n}}^{\mathcal{O}}(\ell)|$.  Applying the Cauchy-Schwarz inequality we obtain the upper bound
\begin{align}
\label{E:Othirdterm1}
    \includegraphics[scale=.32, valign = c]{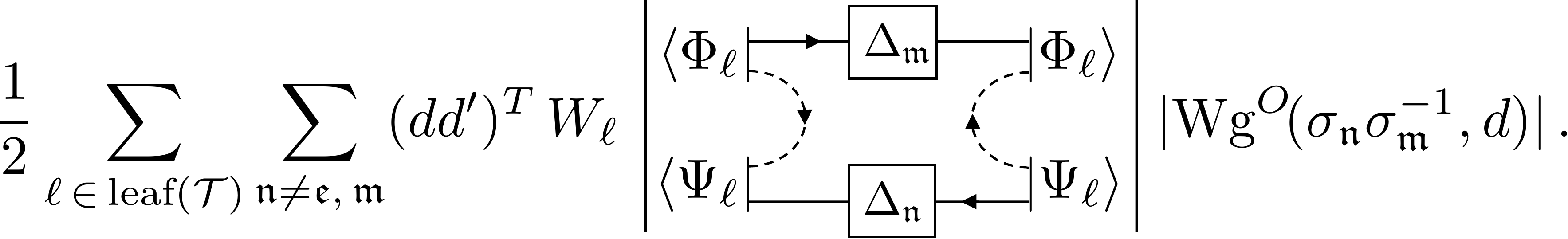}
\end{align}
Here we generalize our notation for the $C$ matrices by writing
\begin{align}
\label{E:Cdef3}
    \includegraphics[scale=.32, valign = c]{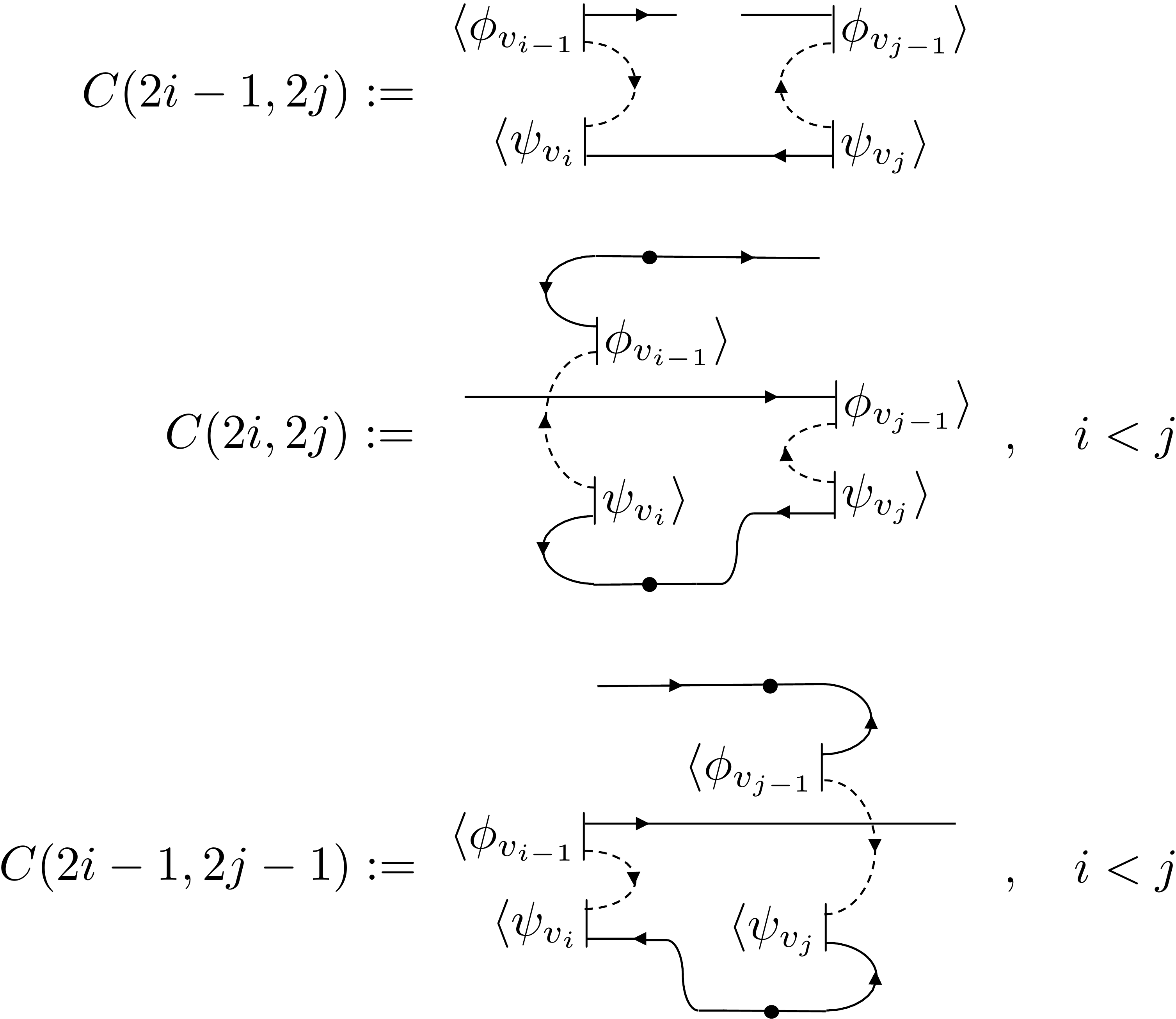}
\end{align}
where as before $C(j,i) := C(i,j)^t$.  Then the diagrammatic term in~\eqref{E:Othirdterm1} can be written as
\begin{align}
\label{E:Ctraces1long}
&\bigg| \text{tr}\big(C(f_\mathfrak{n} \circ f_{\mathfrak{e}}(1),1) C(f_\mathfrak{m}(1), f_{\mathfrak{n}}\circ f_\mathfrak{e}\circ f_\mathfrak{m}(1)) C(f_\mathfrak{m}\circ f_{\mathfrak{n}}\circ f_\mathfrak{e}\circ f_\mathfrak{m}(1), f_{\mathfrak{n}} \circ f_{\mathfrak{e}}\circ f_\mathfrak{m}\circ f_{\mathfrak{n}}\circ f_\mathfrak{e}\circ f_\mathfrak{m}(1)) \cdots \big)  \nonumber \\
&\qquad \qquad \qquad \qquad \qquad \qquad \qquad \qquad \qquad \qquad \qquad \qquad \qquad  \qquad \qquad \qquad \qquad \cdot \text{tr}\big(\cdots\big) \cdots \text{tr}\big(\cdots\big) \bigg|
\end{align}
and so using the same 1-norm bound as before we obtain the upper bound
\begin{align}
\label{E:OCproduct1}
\prod_{i=1}^T \|C(\mathfrak{n}(2i), \mathfrak{n}(2i-1))\|_1\,.
\end{align}

To simplify~\eqref{E:OCproduct1}, we define the unnormalized density operator $\widetilde{\rho}_{v_i} \in \text{Mat}_{d \times d}(\mathbb{C})$:
\begin{align}
    \includegraphics[scale=.32, valign = c]{rhotildedef.png}
\end{align}
for any $i \in [T]$.  Then we have the following Lemma:
\begin{lemma}\label{lemm:Cbound1} For any $i,j \in [T]$,
\begin{align}
\|C(2i-1, 2j)\|_1 &\leq \sqrt{\text{\rm tr}(\widetilde{\rho}_{v_i} \widetilde{\rho}_{v_j})} \\
\|C(2i, 2j)\|_1 &\leq \sqrt{\text{\rm tr}(\widetilde{\rho}_{v_i} \widetilde{\rho}_{v_j}^{\,t})} \\
\|C(2i-1, 2j-1)\|_1 &\leq \sqrt{\text{\rm tr}(\widetilde{\rho}_{v_i} \widetilde{\rho}_{v_j}^{\,t})}\,.
\end{align}
\end{lemma}
\begin{proof}
First consider $\|C(2i-1, 2j)\|_1$.  Since $\|A\|_1 = \|A \otimes A^\dagger \|_1^{1/2} \leq \|\text{SWAP} \cdot (A \otimes A^\dagger)\|_1^{1/2}$, we have
\begin{align}
    \includegraphics[scale=.32, valign = c]{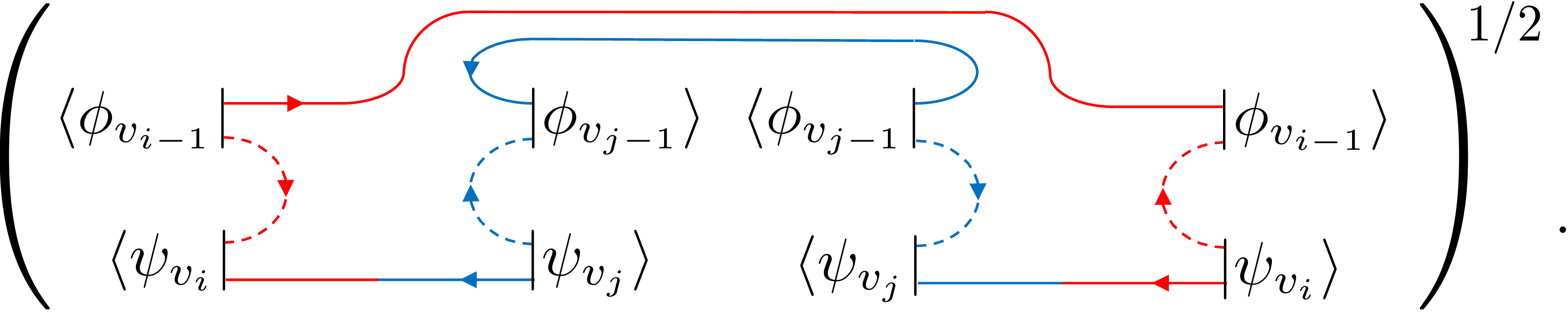} \nonumber
\end{align}
Since the tensor network in the trace is positive semi-definite, we can compute the 1-norm by taking to trace and find
\begin{align}
\label{E:O1normproperty2}
\includegraphics[scale=.32, valign = c]{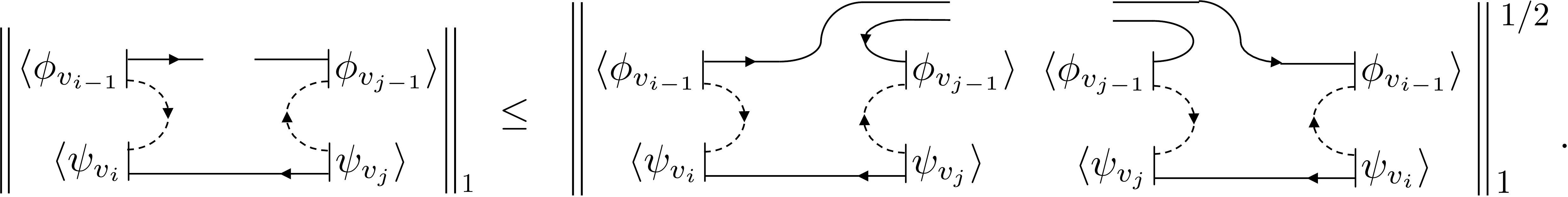}
\end{align}
We have colored the tensor diagram suggestively so that it is transparent how to rewrite it as $\sqrt{\text{tr}(\widetilde{\rho}_{v_{i}} \widetilde{\rho}_{v_{j}})}$\,.

Now consider $\|C(2i,2j)\|_1$.  Using the same inequality $\|A\|_1 = \|A \otimes A^\dagger \|_1^{1/2} \leq \|\text{SWAP} \cdot (A \otimes A^\dagger)\|_1^{1/2}$, we find the upper bound
\begin{align}
\label{E:O1normproperty3}
\includegraphics[scale=.32, valign = c]{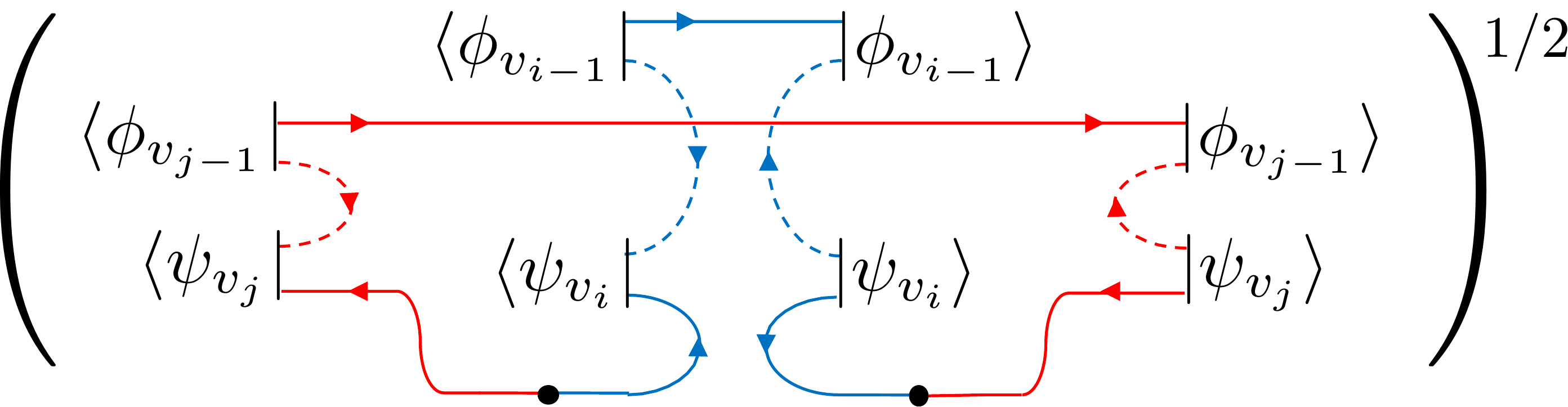}
\end{align}
which is clearly equal to $\sqrt{\text{\rm tr}(\widetilde{\rho}_{v_i} \widetilde{\rho}_{v_j}^{\,t})}$.  The upper bound on $\|C(2i-1,2j-1)\|_1$ is given by an identical argument.  
\end{proof}
It will be convenient to define the underline notation
\begin{equation}
\underline{i} := \begin{cases} \frac{i+1}{2}&\text{for }i\text{ odd} \\
\frac{i}{2}&\text{for }i\text{ even}
\end{cases}
\end{equation}
and analogously for $\underline{j}$\,.  We will say that $i,j$ have same parity if they are equal modulo $2$, and have different parity otherwise.  Then using the above Lemma, we can upper bound~\eqref{E:OCproduct1} by
\begin{equation}\label{E:Oproductwithsqrt1}
        \prod^{\#^O(\mathfrak{n})}_{m = 1}\left(\prod_{\substack{i\leftrightarrow j\in B_m \\ i,j\,\,\text{\rm opposite parity}}} \sqrt{\Tr(\wt{\rho}_{v_{\underline{i}}}\wt{\rho}_{v_{\underline{j}}})} \prod_{\substack{k\leftrightarrow \ell\in B_m \\ k,\ell\,\,\text{\rm same parity}}} \sqrt{\Tr(\wt{\rho}_{v_{\underline{k}}}\wt{\rho}_{v_{\underline{\ell}}}^{\,t})}\right)\,.
\end{equation}

Now we turn to bounding~\eqref{E:Oproductwithsqrt1}.  We begin by developing some notation for cycles $B_m$\,.  Standard notation for an $M_{2T}$ cycle of ``length $p$'' is $B_m = (a_1 a_2 \cdots a_{2p})$ where for us $\{a_1, a_2,...,a_{2p}\} \subseteq  [2T]$.  Since the number of elements in an $M_{2T}$-cycle is always even, by convention we define the length as half the number of elements (i.e.~$p$ instead of $2p$).  We decorate each $a_i$ by an additional subscript as $a_{m,i}$ to remember that it that belongs to the $m$th cycle $B_m$. We sometimes write $p = |B_m|$ to remind ourselves that it depends on $m$.  Let us also define
\begin{align}
\text{tr}(\widetilde{\rho}_{v_{\underline{i}}} \widetilde{\rho}_{v_{\underline{j}}}^{\,t(i,j)}) := \begin{cases}
\text{tr}(\widetilde{\rho}_{v_{\underline{i}}} \widetilde{\rho}_{v_{\underline{j}}}) & \text{if }i,j\text{ have different parities}\\
\text{tr}(\widetilde{\rho}_{v_{\underline{i}}} \widetilde{\rho}_{v_{\underline{j}}}^{\,t}) & \text{if }i,j\text{ have the same parity}\\
\end{cases}\,.
\end{align}
With these notations in mind, we can write~\eqref{E:Oproductwithsqrt1} as a product over
\begin{equation}
\label{E:Ocycletobound1}
\sqrt{\text{tr}(\widetilde{\rho}_{v_{\underline{a_{m,1}}}} \widetilde{\rho}_{v_{\underline{a_{m,2}}}}^{\,t(a_{m,1}, a_{m,2})})} \,\cdots \, \sqrt{\text{tr}(\widetilde{\rho}_{v_{\underline{a_{m,2p-1}}}} \widetilde{\rho}_{v_{\underline{a_{m,2p}}}}^{\,t(a_{m,2p-1}, a_{m,2p})})} \,  \sqrt{\text{tr}(\widetilde{\rho}_{v_{\underline{a_{m,2p}}}} \widetilde{\rho}_{v_{\underline{a_{m,1}}}}^{\,t(a_{m,2p}, a_{m,1})})}
\end{equation}
over $m = 1, \ldots, \#(\mathfrak{n})$; here we drop the $O$ superscript on $\#^O$ since we are only discussing pair permutation here.  We will further analyze the above for fixed $m$ in two cases: (i) $p = |B_m|$ is even, and (ii) $p = |B_m|$ is odd.

It will be convenient to prove a Lemma which is slightly more general than what we need for the orthogonal case; the advantage of this generality is that it will immediately apply to the symplectic case.  The Lemma is as follows:

\begin{lemma}
\label{lemm:Rdecomp}
Let $\text{\rm tr}(\widetilde{\rho}_{v_{\underline{i}}} \widetilde{\rho}_{v_{\underline{j}}}^{\,f(i,j)})$ equal either $\text{\rm tr}(\widetilde{\rho}_{v_{\underline{i}}} \widetilde{\rho}_{v_{\underline{j}}})$, $\text{\rm tr}(\widetilde{\rho}_{v_{\underline{i}}} \widetilde{\rho}_{v_{\underline{j}}}^{\,t})$, or $\text{\rm tr}(\widetilde{\rho}_{v_{\underline{i}}} \textbf{J}\,\widetilde{\rho}_{v_{\underline{j}}}^{\,t} \textbf{J}^{-1})$, depending on the value of $i,j$.  Defining
\begin{align}
R_{m,-} &:= \text{\rm tr}(\widetilde{\rho}_{v_{\underline{a_{m,p}}}})\prod_{\substack{i\,\,\text{\rm odd} \\ 1 \leq i \leq 2p - 2}} \text{\rm tr}(\widetilde{\rho}_{v_{\widetilde{a}_{m,i}}} \widetilde{\rho}_{v_{\underline{a_{m,i+1}}}}^{\,f(a_{m,i},a_{m,i+1})}) \\
R_{m,+} &:= \text{\rm tr}(\widetilde{\rho}_{v_{\underline{a_{m,1}}}})\prod_{j\,\,\text{\rm even}} \text{\rm tr}(\widetilde{\rho}_{v_{\underline{a_{m,j}}}} \widetilde{\rho}_{v_{\underline{a_{m,j+1}}}}^{\,f(a_{m,j},a_{m,j+1})})\,,
\end{align}
we have the inequality
\begin{align}
\label{E:lemmamultipleTs}
\prod_{m=1}^{\#(\mathfrak{n})} \,\prod_{i \leftrightarrow j \, \in B_m} \sqrt{\text{\rm tr}(\widetilde{\rho}_{v_{\underline{i}}} \widetilde{\rho}_{v_{\underline{j}}}^{\,f(i,j)})} &\leq \frac{1}{2^{\#(\mathfrak{n})}} \sum_{i_1,...,i_{\#(\mathfrak{n})} = \pm} R_{1,i_1} R_{2,i_2} \cdots R_{\#(\mathfrak{n}), i_{\#(\mathfrak{n})}}\,.
\end{align}
\end{lemma}

\begin{proof}
Similar to the unitary setting, the argument proceeds in two cases. \\ \\
\textbf{Case 1:} \textit{$p$ is even.} Every term in Eqn.~\eqref{E:Ocycletobound1} has the form $\sqrt{\text{tr}(\widetilde{\rho}_{v_{\underline{a_{m,i}}}} \widetilde{\rho}_{v_{\underline{a_{m,i+1}}}}^{\,f(\underline{a_{m,i}},\underline{a_{m,i+1}})})}$ except the last term.  The $i$ subscripts of $\widetilde{a}_{m,i}$ will be treated modulo $2p$, and  $\sqrt{\text{tr}(\widetilde{\rho}_{v_{\underline{a_{m,2p}}}} \widetilde{\rho}_{v_{\underline{a_{m,2p+1}}}}^{\,f(\underline{a_{m,2p}},\underline{a_{m,2p+1}})})}$\,.  With this notation at hand, we organize Eqn.~\eqref{E:Ocycletobound1} as
\begin{equation}
\left(\prod_{i\text{ odd}} \sqrt{\text{tr}(\widetilde{\rho}_{v_{\underline{a_{m,i}}}} \widetilde{\rho}_{v_{\underline{a_{m,i+1}}}}^{\,f(\underline{a_{m,i}},\underline{a_{m,i+1}})})}\right) \left(\,\prod_{j\text{ even}} \sqrt{\text{tr}(\widetilde{\rho}_{v_{\underline{a_{m,j}}}} \widetilde{\rho}_{v_{\underline{a_{m,j+1}}}}^{\,f(\underline{a_{m,j}},\underline{a_{m,j+1}})})}\right)\,.
\end{equation}
Since $ab \leq \frac{1}{2}(a^2 + b^2)$, the expression above is upper bounded by
\begin{equation}
\frac{1}{2} \prod_{i\text{ odd}} \text{tr}(\widetilde{\rho}_{v_{\underline{a_{m,i}}}} \widetilde{\rho}_{v_{\underline{a_{m,i+1}}}}^{\,f(\underline{a_{m,i}},\underline{a_{m,i+1}})}) + \frac{1}{2}\prod_{j\text{ even}} \text{tr}(\widetilde{\rho}_{v_{\underline{a_{m,j}}}} \widetilde{\rho}_{v_{\underline{a_{m,j+1}}}}^{\,f(\underline{a_{m,j}},\underline{a_{m,j+1}})})\,.
\end{equation}
We will call the first term $\frac{1}{2}\,R_{m,\,-}$ and the second term $\frac{1}{2}\,R_{m,\,+}$\,. \\ \\
\textbf{Case 2:} \textit{$p$ is odd.} In this setting we can arrange the terms in Eqn.~\eqref{E:cycletobound1} as
\begin{equation}
\label{E:Ooddsplit1}
\sqrt{\text{tr}(\widetilde{\rho}_{v_{\underline{a_{m,p}}}} \widetilde{\rho}_{\underline{v_{a_{m,1}}}}^{f(a_{m,p},a_{m,1})})}\left(\prod_{\substack{i\text{ odd} \\ 1 \leq i \leq 2p - 2}} \sqrt{\text{tr}(\widetilde{\rho}_{v_{\underline{a_{m,i}}}} \widetilde{\rho}_{v_{\underline{a_{m,i+1}}}}^{f(a_{m,i}, a_{m,i+1})})}\right) \left(\,\prod_{j\text{ even}} \sqrt{\text{tr}(\widetilde{\rho}_{\underline{v_{a_{m,j}}}} \widetilde{\rho}_{v_{\underline{a_{m,j+1}}}}^{f(a_{m,j},a_{m,j+1})})}\right)\,.
\end{equation}
Since for $A,B$ Hermitian and positive semi-definite we have $\text{tr}(AB) \leq \|A\|_2 \, \|B\|_2 \leq \|A\|_1 \, \|B\|_1 \leq \text{tr}(A) \, \text{tr}(B)$ it follows that
\begin{equation}
\sqrt{\text{tr}(\widetilde{\rho}_{v_{\underline{a_{m,p}}}} \widetilde{\rho}_{v_{\underline{a_{m,1}}}}^{f(a_{m,p},a_{m,1})})} \leq \sqrt{\text{tr}(\widetilde{\rho}_{v_{\underline{a_{m,p}}}})\,\text{tr}(\widetilde{\rho}_{v_{\underline{a_{m,1}}}}^{f(a_{m,p},a_{m,1})})} = \sqrt{\text{tr}(\widetilde{\rho}_{v_{\underline{a_{m,p}}}})\,\text{tr}(\widetilde{\rho}_{v_{\underline{a_{m,1}}}})}\,.    
\end{equation}
Using the above inequality,~\eqref{E:Ooddsplit1} has the upper bound
\begin{equation}
\left(\sqrt{\text{tr}(\widetilde{\rho}_{v_{\underline{a_{m,p}}}})} \prod_{\substack{i\text{ odd} \\ 1 \leq i \leq 2p - 2}} \sqrt{\text{tr}(\widetilde{\rho}_{v_{\underline{a_{m,i}}}} \widetilde{\rho}_{v_{\underline{a_{m,i+1}}}}^{f(a_{m,i},a_{m,i+1})})}\right) \left(\sqrt{\text{tr}(\widetilde{\rho}_{v_{\underline{a_{m,1}}}})}\,\prod_{j\text{ even}} \sqrt{\text{tr}(\widetilde{\rho}_{v_{\underline{a_{m,j}}}} \widetilde{\rho}_{v_{\underline{a_{m,j+1}}}}^{f(a_{m,j}, a_{m,j+1})})}\right)\,.
\end{equation}
Further using $ab \leq \frac{1}{2}(a^2 + b^2)$, we find the upper bound
\begin{equation}
\frac{1}{2} \,\text{tr}(\widetilde{\rho}_{v_{\underline{a_{m,p}}}})\prod_{\substack{i\text{ odd} \\ 1 \leq i \leq 2p - 2}} \text{tr}(\widetilde{\rho}_{v_{\underline{a_{m,i}}}} \widetilde{\rho}_{v_{\underline{a_{m,i+1}}}}^{f(a_{m,i}, a_{m,i+1})}) + \frac{1}{2}\, \text{tr}(\widetilde{\rho}_{v_{a_{m,1}}})\prod_{j\text{ even}} \text{tr}(\widetilde{\rho}_{\underline{v_{a_{m,j}}}} \widetilde{\rho}_{v_{\underline{a_{m,j+1}}}}^{f(a_{m,j}, a_{m,j+1}})
\end{equation}
where similarly the first term is called $\frac{1}{2}\,R_{m,\,-}$ and the second term is called $\frac{1}{2}\,R_{m,\,+}$\,. 

Since $R_{m,\,-}$ and $R_{m,\,+}$ implicitly depend on the leaf $\ell$, sometimes we will denote them by $R^{\ell}_{m,\,-}$ and $R^{\ell}_{m,\,+}$ to be explicit.\\ \\
\indent Taking Case 1 and Case 2 together, we find that Eqn.~\eqref{E:Oproductwithsqrt1} has the upper bound
\begin{align}
\label{E:OmultipleTs}
\prod_{m=1}^{\#(\mathfrak{n})} \,\prod_{i \leftrightarrow j \, \in B_m} \sqrt{\text{tr}(\widetilde{\rho}_{v_{\underline{i}}} \widetilde{\rho}_{v_{\underline{j}}}^{f(i,j)})} &\leq \frac{1}{2^{\#(\mathfrak{n})}} \prod_{m=1}^{\#(\mathfrak{n})} \left(R_{m,-} + R_{m,+} \right) \nonumber \\
&= \frac{1}{2^{\#(\mathfrak{n})}} \sum_{i_1,...,i_{\#(\mathfrak{n})} = \pm} R_{1,i_1} R_{2,i_2} \cdots R_{\#(\mathfrak{n}), i_{\#(\mathfrak{n})}}\,.
\end{align}
This is the desired bound.
\end{proof}
Observe that each term in the sum in the last line of~\eqref{E:OmultipleTs} is a product of terms like $\text{tr}(\widetilde{\rho}_{v_{\underline{i}}})$ and $\text{tr}(\widetilde{\rho}_{v_{\underline{j}}} \widetilde{\rho}_{v_{\underline{k}}}^{f(j,k)})$.  As before, the key point is that we arranged the equations so that each term in the sum has $\widetilde{\rho}_{v_i}$ for each $i = 1,...,T$ appear \textit{exactly once}.  This allows us to prove the following lemma, which is akin to Lemma~\ref{lem:collapse} above:
\begin{lemma}\label{lem:fcollapse}
    Fix any $i_1,...,i_{\#(\mathfrak{n})} \in \{+,-\}$. Then
    \begin{equation}
        \label{E:Osboundhard1}
        \sum_{\ell \,\in\,\text{\rm leaf}(\mathcal{T})} (d d')^T \,W_\ell \, R^{\ell}_{1,i_1} R^{\ell}_{2,i_2} \cdots R^{\ell}_{\#(\tau^{-1}), i_{\#(\mathfrak{n})}} \le d^{\frac{T}{2} - \left\lfloor \frac{L(\mathfrak{n})}{2}\right\rfloor}
    \end{equation}
where $L(\mathfrak{n})$ is the length of the longest cycle in $\mathfrak{n}$ (where we recall that the length of an $M_{2T}$ cycle is defined as half the number of integers in that cycle).
\end{lemma}

\begin{proof}
    To ease notation, let $R^{\ell}_j\triangleq R^{\ell}_{j,i_j}$.
    Recall that $W_\ell = w_{v_1} w_{v_2} \cdots w_{v_T}$ and note that
    \begin{align}
        \sum_{v: \, \text{depth}(v) = i} dd'\,w_{v} \, \widetilde{\rho}_{v} &= \langle \phi_{\text{parent}(v)} | \phi_{\text{parent}(v)}\rangle\, \mathds{1}_{d \times d} = \mathds{1}_{d \times d} \\ \nonumber \\
         \sum_{v: \, \text{depth}(v) = i} dd'\,w_{v} \, \widetilde{\rho}_{v}^{\,t} &= \langle \phi_{\text{parent}(v)} | \phi_{\text{parent}(v)}\rangle\, \mathds{1}_{d \times d} = \mathds{1}_{d \times d} \\ \nonumber \\
          \sum_{v: \, \text{depth}(v) = i} dd'\,w_{v} \, J \widetilde{\rho}_{v}^{\,t} J^{-1} &= \langle \phi_{\text{parent}(v)} | \phi_{\text{parent}(v)}\rangle\, \mathds{1}_{d \times d} = \mathds{1}_{d \times d}\,.
    \end{align}
Taking the traces of the above equations against any $\rho\in\text{Mat}_{d\times d}(\mathbb{C})$, we find
    \begin{align}
        \label{E:O2tracebound1}
 \sum_{v: \, \text{depth}(v) = i} dd'\,w_{v} \, \text{tr}(\rho \,\widetilde{\rho}_{v}) &= \text{tr}(\rho) \\ \nonumber \\
         \sum_{v: \, \text{depth}(v) = i} dd'\,w_{v} \, \text{tr}(\rho\,\widetilde{\rho}_{v}^{\,t}) &= \text{tr}(\rho) \\ \nonumber \\
          \sum_{v: \, \text{depth}(v) = i} dd'\,w_{v} \, \text{tr}(\rho\,J\widetilde{\rho}_{v}^{\,t} J^{-1}) &= \text{tr}(\rho)\,.
    \end{align}
In particular, for $\rho = \mathds{1}$ we have
    \begin{align}
        \label{E:O1tracebound1}
        \sum_{v: \, \text{depth}(v) = i} dd'\,w_{v} \, \text{tr}(\widetilde{\rho}_{v}) &= d.
    \end{align}
    With these various identities in mind, we can now turn to bounding~\eqref{E:Osboundhard1}. As we discussed, above $\prod_j R^{\ell}_j$ is a product of terms like $\Tr(\wt{\rho}_{v_{\underline{i}}})$ and $\Tr(\wt{\rho}_{v_{\underline{i}}}\wt{\rho}_{v_{\underline{i'}}}^{f(i,i')})$.  It is convenient to define some sets of indices to encode this data. Let $S^{(T)}_1\subseteq[T]$ denote the indices $\underline{i}$ for which $\Tr(\wt{\rho}_{v_{\underline{i}}})$ appears in $\prod_j R^{\ell}_j$, and also let $S^{(T)}_2\subseteq[T]\times[T]$ denote the set of (unordered) pairs $(\underline{i},\underline{i'})$ for which $\Tr(\wt{\rho}_{v_{\underline{i}}}\wt{\rho}_{v_{\underline{i'}}}^{f(i,i')})$ appears, so that for any root-to-leaf path in $\calT$ consisting of nodes $v_1,\ldots,v_T = \ell$, we have
    \begin{equation}
        \label{eq:OprodRs}
        \prod_j R^{\ell}_j = \prod_{\underline{i}\in S^{(T)}_1} \Tr(\wt{\rho}_{v_{\underline{i}}}) \cdot \prod_{(\underline{i},\underline{i'})\in S^{(T)}_2} \Tr(\wt{\rho}_{v_{\underline{i}}}\wt{\rho}_{v_{\underline{i'}}}^{f(i,i')})\,.
    \end{equation}
    Next we construct $S^{(t)}_1\subseteq[t]$ and $S^{(t)}_2\subseteq[t]\times[t]$ for $1 \le t < T$ by the following inductive procedure. If $t\in S^{(t)}_1$ then we define $S^{(t-1)}_1 \triangleq S^{(t)}_1 \backslash \brc{t}$ and $S^{(t-1)}_2\triangleq S^{(t)}_2$. Otherwise if $(t,t')\in S^{(t)}_2$ for some $t' < t$, then we define $S^{(t-1)}_1 \triangleq S^{(t-1)}_1 \cup \brc{t'}$ and $S^{(t-1)}_2\triangleq S^{(t)}_2 \backslash \brc{(t,t')}$. We recall two key observations about such sequences, which we also leveraged in Lemma~\ref{lem:collapse}:  
    \begin{observation}\label{Oobs:singleton}
        For every $i\in S^{(T)}_1$, we have that $i\in S^{(i)}_1$.
    \end{observation}
    
    \begin{observation}\label{Oobs:pair}
        For every $(i,i')\in S^{(T)}_2$, if $i\le i'$ then $i\in S^{(i)}_1$ while $i'\not\in S^{(i')}_1$.
    \end{observation}
    
    We have defined this set of sequences in order to extract $\wt{\rho}_{v_T}$ from the product on the right-hand side of \eqref{eq:OprodRs} and apply \eqref{E:O1tracebound1} (respectively \eqref{E:O2tracebound1}) if $T\in S^{(T)}_1$ (respectively $(T,t')\in S^{(T)}_2$ for some $t' < T$) to obtain \begin{equation}
        \label{eq:Oinduct_prodRs}
        \sum_{\ell\in\leaf(\calT)} (dd')^T W_{\ell} \prod_j R^{\ell}_j = d^{\bone{T\in S^{(T)}_1}}\sum_{u: \,\text{depth}(u) = T - 1} (dd')^{T-1} W_u \prod_{\underline{i}\in S^{(T-1)}_1} \Tr(\wt{\rho}_{v_{\underline{i}}}) \cdot \prod_{(i,i') \in S^{(T-1)}_2} \Tr(\wt{\rho}_{v_{\underline{i}}}\wt{\rho}_{v_{\underline{i'}}}^{f(i,i')})\,,
    \end{equation}
    with $W_u = w_{v_1}\cdots w_{v_{T-1}}$ if the path from root to $u$ in $\calT$ is given by $v_1,\ldots,v_{T-1} = u$.
    We can proceed inductively by expressing the right-hand side of \eqref{eq:Oinduct_prodRs} as \begin{equation}
        d^{\sum^T_{t=1} \bone{t\in S^{(t)}_1}}.
    \end{equation}
    By virture of Observations~\ref{Oobs:singleton} and \ref{Oobs:pair}, $\sum^T_{t=1}\bone{t\in S^{(t)}_1} = |S^{(T)}_1| + |S^{(T)}_2|$. Since every $M_{2T}$-cycle $B_m$ of $\mathfrak{n}$ contributes with $|B_m|$ off contributes $|B_m|/2$ pairs to $S^{(T)}_2$, and every $M_{2T}$-cycle $B_m$ with $|B_m|$ even contributes $\floor{|B_m|/2}$ pairs to $S^{(T)}_2$ and one element to $S^{(T)}_1$, we conclude the formula
    \begin{align}
        \label{E:sboundhard2_v2}
        \sum_{\ell\in\leaf(\calT)} (dd')^T W_{\ell} \prod_j R^{\ell}_j &= d^{\sum_{m=1}^{\#(\mathfrak{n})} \left\lceil \frac{|B_m|}{2} \right\rceil} \nonumber \\
        &= d^{\frac{T}{2} - \sum_{m=1}^{\#(\mathfrak{n})} \left\lfloor \frac{|B_m|}{2} \right\rfloor} \nonumber \\
        &\leq d^{\frac{T}{2} - \left\lfloor \frac{L(\mathfrak{n})}{2} \right\rfloor}
    \end{align}
as we desired.
\end{proof}
    
Putting all our previous bounds together, in particular Eqn.'s~\eqref{E:Othirdterm1},~\eqref{E:Oproductwithsqrt1},~\eqref{E:OmultipleTs}, and~\eqref{E:Osboundhard1}, we arrive at
\begin{equation}
\sum_{\ell \, \in \, \text{leaf}(\mathcal{T})} \sum_{\mathfrak{n} \not = \mathfrak{e}, \,\mathfrak{m}}|p_{\sigma, \tau}^{\mathcal{O}}(\ell)| \leq d^{T} \sum_{\mathfrak{m}} |\text{Wg}^O(\sigma_{\mathfrak{m}}^{-1}, d)| \sum_{\mathfrak{n} \not = \mathfrak{e}} d^{- \left\lfloor \frac{L^O(\mathfrak{n})}{2} \right\rfloor} \,.
\end{equation}
The first sum on the right-hand side is bounded by $d^T \sum_{\mathfrak{m}} |\text{Wg}^O(\sigma_{\mathfrak{m}}^{-1}, d)| \leq 1 + \mathcal{O}\!\left( \frac{T^2}{d}\right)$.  Considering the second sum on the right-hand side, let $N^O(T,\ell)$ be the number of permutations in $S_T$ where the length of the longest $M_{2T}$-cycle is $2\ell$.  Then the second sum can be written as
\begin{equation}
\label{E:sumNTell_v2}
\sum_{\ell = 2}^T N^O(T,\ell) \, d^{- \left\lfloor \frac{\ell}{2} \right\rfloor}
\end{equation}
where we omit $\ell = 1$ from the sum since it corresponds to the identity pair permutation.  Since $N^O(T,\ell) \leq \binom{2T}{2\ell}(2\ell - 1)!! = \frac{(2T)!}{(2T - 2\ell)! \,2^\ell \ell!} < T^\ell$,~\eqref{E:sumNTell_v2} is upper bounded by
\begin{equation}
\sum_{\ell = 2}^\infty T^\ell  \, d^{- \left\lfloor \frac{\ell}{2} \right\rfloor} = \frac{(1 + T) \frac{T^2}{d}}{1 - \frac{T^2}{d}} = \frac{T^3}{d} + \frac{T^2}{d} + \mathcal{O}\!\left(\frac{T^5}{d^2}\right)\,.
\end{equation}
Now if $T \le o(d^{1/3})$, then this quantity is $o(1)$ for some absolute constant $c > 0$. Altogether, we find
\begin{equation}
\frac{1}{2}\sum_{\ell\,\in\,\text{leaf}(\mathcal{T})} \sum_{\mathfrak{n}\not = \mathfrak{e}, \,\mathfrak{m}}|p_{\mathfrak{m}, \mathfrak{n}}^{\mathcal{O}}(\ell)| \leq o\!\left(1\right)\,.
\end{equation}

\subsubsection*{Third term for \normalfont $\text{Sp}(d/2)$ \normalfont \textbf{case}}
The third term in~\eqref{E:Spfirstineq1} is $\frac{1}{2}\sum_{\ell \,\in\,\text{leaf}(\mathcal{T})} \sum_{\mathfrak{n} \not = \mathfrak{e}, \,\mathfrak{m}}|p_{\mathfrak{m},\mathfrak{n}}^{\mathcal{S}}(\ell)|$.  Applying the Cauchy-Schwarz inequality we obtain the upper bound
\begin{align}
\label{E:Spthirdterm1}
    \includegraphics[scale=.32, valign = c]{Othirdterm1.pdf}
\end{align}
To bound the diagrammatic term, we introduce the tilde notation $\widetilde{C}(2i-1,2j) := C(2i-1,2j)$, $\widetilde{C}(2i-1,2j-1) := C(2i-1,2j-1)$ as well as
\begin{align}
\label{E:CtildeJ}
    \includegraphics[scale=.32, valign = c]{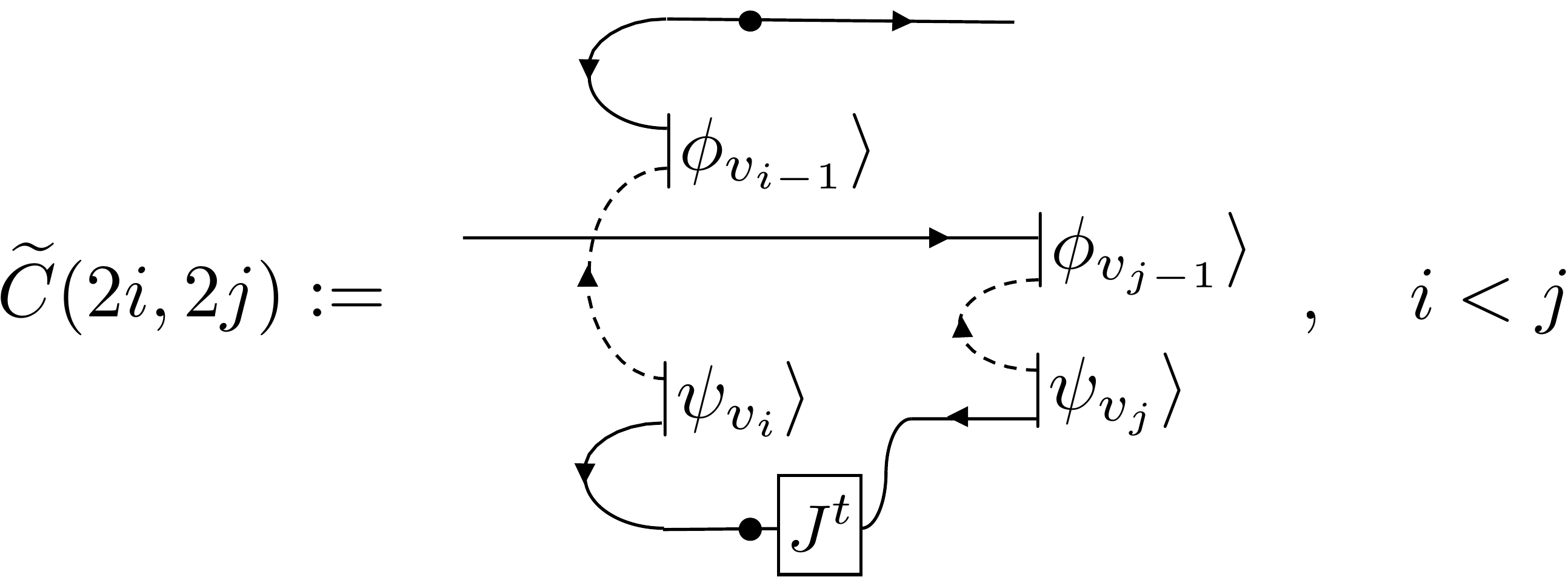}
\end{align}
where as usual $\widetilde{C}(j,i) = \widetilde{C}(i,j)^t$.  Then the diagrammatic term in~\eqref{E:Spthirdterm1} can be written as
\begin{align}
\label{E:Ctraces1longJ}
&\bigg| \text{tr}\big(C(f_\mathfrak{n} \circ f_{\mathfrak{e}}(1),1) \,J\,C(f_\mathfrak{m}(1), f_{\mathfrak{n}}\circ f_\mathfrak{e}\circ f_\mathfrak{m}(1)) \,J\,C(f_\mathfrak{m}\circ f_{\mathfrak{n}}\circ f_\mathfrak{e}\circ f_\mathfrak{m}(1), f_{\mathfrak{n}} \circ f_{\mathfrak{e}}\circ f_\mathfrak{m}\circ f_{\mathfrak{n}}\circ f_\mathfrak{e}\circ f_\mathfrak{m}(1)) \,J \cdots \big)  \nonumber \\
&\qquad \qquad \qquad \qquad \qquad \qquad \qquad \qquad \qquad \qquad \qquad \qquad \qquad  \qquad \qquad \qquad \qquad \cdot \text{tr}\big(\cdots\big) \cdots \text{tr}\big(\cdots\big) \bigg|
\end{align}
which is upper bounded in the 1-norm by
\begin{align}
\label{E:SCproduct1}
\prod_{i=1}^T \|\widetilde{C}(\mathfrak{n}(2i), \mathfrak{n}(2i-1)\,J)\|_1 &\leq \prod_{i=1}^T \|\widetilde{C}(\mathfrak{n}(2i), \mathfrak{n}(2i-1))\|_1\,\|J\|_\infty \nonumber \\
&\leq \prod_{i=1}^T \|\widetilde{C}(\mathfrak{n}(2i), \mathfrak{n}(2i-1))\|_1\,.
\end{align}
We now have the following Lemma, analogous to Lemma~\ref{lemm:Cbound1}:
\begin{lemma}\label{lemm:Cbound2} For any $i,j \in [T]$,
\begin{align}
\|\widetilde{C}(2i-1, 2j)\|_1 &\leq \sqrt{\text{\rm tr}(\widetilde{\rho}_{v_i} \widetilde{\rho}_{v_j})} \\
\|\widetilde{C}(2i, 2j)\|_1 &\leq \sqrt{\text{\rm tr}(\widetilde{\rho}_{v_i} \widetilde{\rho}_{v_j}^{\,D})} \\
\|\widetilde{C}(2i-1, 2j-1)\|_1 &\leq \sqrt{\text{\rm tr}(\widetilde{\rho}_{v_i} \widetilde{\rho}_{v_j}^{\,t})}
\end{align}
where we recall that $A^D := J A^t J^{-1}$ is the symplectic transpose.
\end{lemma}
\begin{proof}
The first and third inequalities follow from Lemma~\ref{lemm:Cbound1} since $\widetilde{C}(2i-1, 2j) = C(2i-1,2j)$ and $\widetilde{C}(2i-1,2j-1) = C(2i-1,2j-1)$.  For the second inequality, we again use $\|A\|_1 = \|A \otimes A^\dagger \|_1^{1/2} \leq \|\text{SWAP}\cdot(A \otimes A^\dagger)\|_1^{1/2}$ to find the upper bound
\begin{align}
    \includegraphics[scale=.32, valign = c]{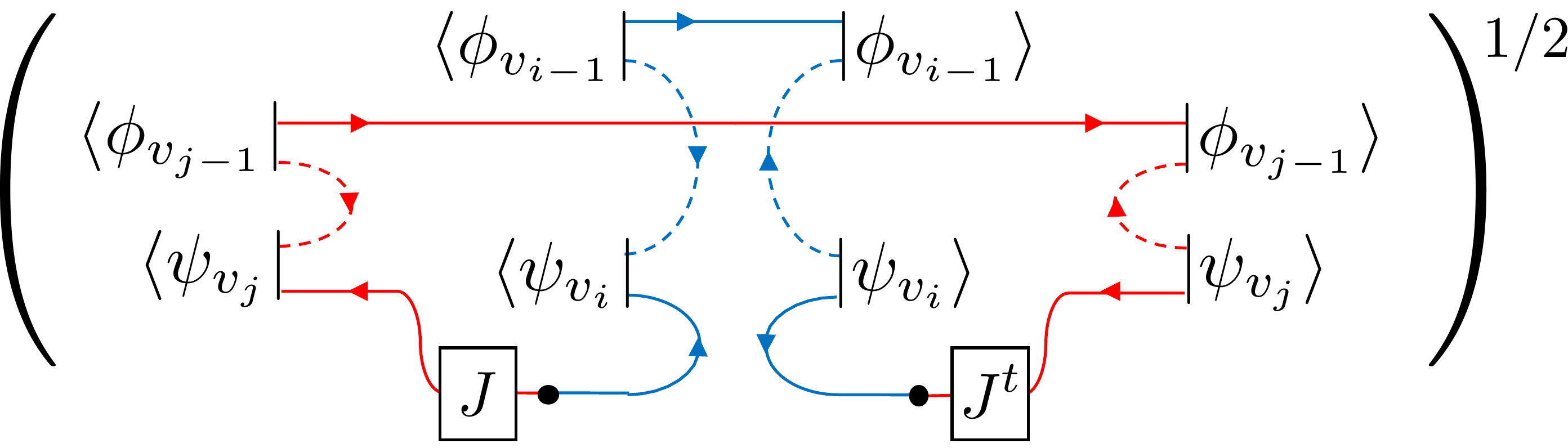} \nonumber
\end{align}
which evidently equals $\sqrt{\text{\rm tr}(\widetilde{\rho}_{v_i} \widetilde{\rho}_{v_j}^{\,D})}$.
\end{proof}
Leveraging the above Lemma, we can upper bound~\eqref{E:SCproduct1} by
\begin{equation}\label{E:Spproductwithsqrt1}
        \prod^{\#^O(\mathfrak{n})}_{m = 1}\left(\prod_{\substack{i\leftrightarrow j\in B_m \\ i,j\,\,\text{\rm opposite parity}}} \sqrt{\Tr(\wt{\rho}_{v_i}\wt{\rho}_{v_j})} \prod_{\substack{k\leftrightarrow \ell\in B_m \\ k,\ell\,\,\text{\rm both odd}}} \sqrt{\Tr(\wt{\rho}_{v_k}\wt{\rho}_{v_\ell}^{\,t})} \prod_{\substack{p\leftrightarrow q\in B_m \\ p,q\,\,\text{\rm both even}}} \sqrt{\Tr(\wt{\rho}_{v_p}\wt{\rho}_{v_q}^{\,D})}\right)\,.
    \end{equation}
Defining
\begin{align}
\text{tr}(\widetilde{\rho}_{v_{\underline{i}}} \widetilde{\rho}_{v_{\underline{j}}}^{\,D(i,j)}) := \begin{cases}
\text{tr}(\widetilde{\rho}_{v_{\underline{i}}} \widetilde{\rho}_{v_{\underline{j}}}) & \text{if }i,j\text{ have different parities}\\
\text{tr}(\widetilde{\rho}_{v_{\underline{i}}} \widetilde{\rho}_{v_{\underline{j}}}^{\,t}) & \text{if }i,j\text{ are both odd}\\
\text{tr}(\widetilde{\rho}_{v_{\underline{i}}} \widetilde{\rho}_{v_{\underline{j}}}^{\,D}) & \text{if }i,j\text{ are both even}\\
\end{cases}\,,
\end{align}
we can write~\eqref{E:Spproductwithsqrt1} as a product over
\begin{equation}
\label{E:Spcycletobound1}
\sqrt{\text{tr}(\widetilde{\rho}_{v_{\underline{a_{m,1}}}} \widetilde{\rho}_{v_{\underline{a_{m,2}}}}^{\,D(a_{m,1}, a_{m,2})})}  \,\cdots \, \sqrt{\text{tr}(\widetilde{\rho}_{v_{\underline{a_{m,p-1}}}} \widetilde{\rho}_{v_{\underline{a_{m,p}}}}^{\,D(a_{m,p-1}, a_{m,p})})} \,  \sqrt{\text{tr}(\widetilde{\rho}_{v_{\underline{a_{m,p}}}} \widetilde{\rho}_{v_{\underline{a_{m,1}}}}^{\,D(a_{m,p}, a_{m,p-1})})}
\end{equation}
over $m = 1, \ldots, \#^{\text{Sp}}(\mathfrak{n})$.  Leveraging Lemmas~\ref{lemm:Rdecomp} and~\ref{lem:fcollapse} in the same exact was as in the orthogonal case, we obtain
\begin{equation}
\sum_{\ell \, \in \, \text{leaf}(\mathcal{T})} \sum_{\mathfrak{n} \not = \mathfrak{e}, \,\mathfrak{m}}|p_{\sigma, \tau}^{\mathcal{S}}(\ell)| \leq d^{T} \sum_{\mathfrak{m}} |\text{Wg}^{\text{Sp}}(\sigma_{\mathfrak{m}}^{-1}, d/2)| \sum_{\mathfrak{n} \not = \mathfrak{e}} d^{- \left\lfloor \frac{L(\mathfrak{n})}{2} \right\rfloor} \,.
\end{equation}
The first sum on the right-hand side is bounded by $d^T \sum_{\mathfrak{m}} |\text{Wg}^{\text{Sp}}(\sigma_{\mathfrak{m}}^{-1}, d/2)| \leq 1 + \mathcal{O}\!\left( \frac{T^2}{d}\right)$.  Considering the second sum on the right-hand side, let $N^{\text{Sp}}(T,\ell)$ be the number of permutations in $S_T$ where the length of the longest $M_{2T}$-cycle is $2\ell$.  Since $N^{\text{Sp}}(T,\ell) = N^O(T,\ell)$,  the second sum can be written as
\begin{equation}
\label{E:sumNTell_v3}
\sum_{\ell = 2}^T N^{\text{Sp}}(T,\ell) \, d^{- \left\lfloor \frac{\ell}{2} \right\rfloor} \leq \frac{T^3}{d} + \frac{T^2}{d} + \mathcal{O}\!\left(\frac{T^5}{d^2}\right)\,.
\end{equation}
Now if $T \le o(d^{1/3})$, then this quantity is $o(1)$ for some absolute constant $c > 0$. Altogether, we find
\begin{equation}
\frac{1}{2}\sum_{\ell\,\in\,\text{leaf}(\mathcal{T})} \sum_{\mathfrak{n}\not = \mathfrak{e}, \,\mathfrak{m}}|p_{\mathfrak{m}, \mathfrak{n}}^{\mathcal{S}}(\ell)| \leq o\!\left(1\right)\,.
\end{equation}

\end{proof}

\subsubsection{Corollaries involving state distinction}

Here we remark on some immediate corollaries of Theorems~\ref{thm:channelhard1},~\ref{thm:Ochannelhard1}, and~\ref{thm:Spchannelhard1}.  We begin with a corollary essentially identical to one from~\cite{aharonov2021quantum}:
\begin{corollary}
\label{cor:state1}
Any learning algorithm without quantum memory requires
\begin{equation}
    T \geq \Omega\left( 2^{n/3} \right),
\end{equation}
to correctly distinguish between the maximally mixed state $\mathds{1}/2^n$ on $n$ qubits and a fixed, Haar-random state $|\Psi\rangle\langle\Psi|$ on $n$ qubits with probability at least $2/3$.
\end{corollary}
\begin{proof}
Suppose by contradiction that a learning algorithm could distinguish between $\mathds{1}/d$ and a fixed, Haar-random $|\Psi\rangle\langle\Psi|$ with probability at least $2/3$ using $T < \mathcal{O}(d^{1/3})$.  Then we could use this learning algorithm to solve the unitary distinction problem by taking the channel $\mathcal{C}$ and applying it to $|0\rangle^{\otimes n}$; if $\mathcal{C} = \mathcal{D}$ then we would get the maximally mixed state, and if $\mathcal{C} = \mathcal{U}$ we would get a fixed Haar random state.  Moreover we could distinguish the two cases in using $T < \mathcal{O}(d^{1/3})$ by running the alleged state distinction algorithm.  But this is impossible since it contradicts Theorem~\ref{thm:channelhard1}, so such a state distinction algorithm can not exist.
\end{proof}

While the above corollary is weaker than Theorem~\ref{thm:purity_lower} for which $T \geq \Omega(d^{1/2})$, it is emblematic of a general strategy for using learning bounds on channel distinction problems to derive corresponding learning bounds on state distinction problems.  Further leveraging this strategy, we can prove the following two additional corollaries:
\begin{corollary}
\label{cor:realstate}
Any learning algorithm without quantum memory requires
\begin{equation}
    T \geq \Omega\left( 2^{n/3} \right),
\end{equation}
to correctly distinguish between the maximally mixed state $\mathds{1}/2^n$ on $n$ qubits and a fixed, real Haar random state $|\Psi\rangle\langle\Psi|$ on $n$ qubits with probability at least $2/3$.
\end{corollary}
\noindent To prove this corollary, we observe that $|\Psi\rangle \langle \Psi| = O \left(|0\rangle \langle 0|\right)^{\otimes n} O^t$ for a fixed, Haar-random orthogonal matrix; alternatively, it is also the case that $|\Psi\rangle \langle \Psi| = S \left(|0\rangle \langle 0|\right)^{\otimes n} S^D$ for a fixed, Haar-random symplectic matrix.  Leveraging the symplectic version (in particular since Theorem~\ref{thm:Ochannelhard1} has a stronger bound than Theorem~\ref{thm:Spchannelhard1}), the corollary follows by the same arguments as the proof of Corollary~\ref{cor:state1} in combination with Theorem~\ref{thm:Spchannelhard1}.  Corollary~\ref{cor:realstate} also follow from the results in~\cite{aharonov2021quantum}, although the corollary was not stated there.

\subsubsection{Upper bound without quantum memory}

The upper bound without quantum memory can be obtained by reducing the problem to a purity testing and utilize Theorem~\ref{thm:purity_upper}. This results in the following corollary.

\begin{corollary}\label{thm:depo_vs_scram_upper}
There is a learning algorithm without quantum memory which takes $T = O(2^{n/2})$ accesses to the unknown quantum channel $\mathcal{C}$ to distinguish between whether $\mathcal{C}$ is a fixed Haar-random unitary channel or a completely depolarizing channel $\mathcal{D}$.
\end{corollary}
\begin{proof}
We perform $T$ repeated experiments given by the following.
Input the all-zero state $\ket{0^n}$ to the unknown quantum channel $\mathcal{C}$ to obtain the output state $\rho_{\mathrm{out}}$.
When $\mathcal{C}$ is a scrambling unitary channel, $\rho_{\mathrm{out}}$ is a fixed pure state.
When $\mathcal{C}$ is a completely depolarizing channel, $\rho_{\mathrm{out}}$ is the completely mixed state.
We then measure the output state $\rho_{\mathrm{out}}$ in the computational basis to obtain the classical data.
The collection of classical data given by the computational basis measurements can be used to classify if $\rho_{\mathrm{out}}$ is a fixed pure state or the completely mixed state.
Theorem~\ref{thm:purity_upper} tells us that $T = O(2^{n/2})$ is sufficient to distinguish between the two cases. Hence, we can distinguish between whether $\mathcal{C}$ is a scrambling unitary channel or a completely depolarizing channel using $T = O(2^{n/2})$.
\end{proof}

\subsubsection{Upper bound with quantum memory}

The exponential lower bound for algorithms without quantum memory is in stark contrast to those with quantum memory.
The following result from~\cite{aharonov2021quantum} states that a linear number of channel applications $T$ and quantum gates suffices if we use a learning algorithm with quantum memory.

\begin{theorem}[Fixed unitary task is easy with an $n$ qubit quantum memory~\cite{aharonov2021quantum}]
There exists a learning algorithm with $n$ qubits of quantum memory which, with constant probability, can distinguish a completely depolarizing channel $\mathcal{D}$ from a fixed, Haar-random $\mathcal{U}$ (either unitary, orthogonal, or symplectic) using only $T = \mathcal{O}(1)$ applications of the channel.  Moreover, the algorithm is gate efficient and has $\mathcal{O}(n)$ gate complexity.
\end{theorem}
\noindent The protocol is simply a swap test; the basic idea is that for a pure state $|\phi\rangle$ on $n$ qubits we have $\text{tr}(\mathcal{D}[|\phi\rangle\!\langle \phi|]^2) = \frac{1}{d}$ whereas $\text{tr}(\mathcal{U}[|\phi\rangle\!\langle \phi|]^2) = 1$.  The ability to obtain quantum interference between $\mathcal{D}[|\phi\rangle\!\langle \phi|]$ and a copy of itself $\mathcal{D}[|\phi\rangle\!\langle \phi|]$ (or $\mathcal{U}[|\phi\rangle\!\langle \phi|]$ and a copy of itself $\mathcal{U}[|\phi\rangle\!\langle \phi|]$) is enabled by the $n$ qubit quantum memory which can store a single copy of the state.

\subsection{Symmetry distinction problem}

\subsubsection{Lower bound}

Using Theorem~\ref{thm:channelhard1},~\ref{thm:Ochannelhard1},~and~\ref{thm:Spchannelhard1}, we can show the hardness of distinguishing between unitary channel, orthogonal matrix channel, and symplectic matrix channel for learning algorithms without quantum memory.
This multiple-hypothesis distinguishing task is equivalent to uncovering what symmetry is encoded in a quantum evolution.
Orthogonal matrix channel and symplectic matrix channel are quantum evolutions with different type of time-reversal symmetry, while unitary channel is a general evolution without additional symmetry.

\begin{theorem}\label{thm:channelhard2}
Any learning algorithm without quantum memory requires
\begin{equation}
\label{E:TOmega3.5}
    T \geq \Omega\left( 2^{2n/7} \right),
\end{equation}
to correctly distinguish between a fixed, Haar-random unitary channel $\mathcal{C}^U$, orthogonal matrix channel $\mathcal{C}^O$, or symplectic matrix channel $\mathcal{C}^S$ on $n$ qubits with probability at least $2/3$.
\end{theorem}
\begin{proof}
Given a tree representation $\mathcal{T}$ of the learning algorithm without quantum memory.
The probability that the algorithm correctly identifies the class of channels is equal to
\begin{equation}
    \frac{1}{3} \sum_{\mathcal{C} \in \{\mathcal{C}^U, \mathcal{C}^O, \mathcal{C}^S\}} \sum_{\ell \in \mathrm{leaf}(\mathcal{T})} p^{\mathcal{C}}(\ell) \,\, \mathrm{I}[\Upsilon(\ell) = \mathcal{C}],
\end{equation}
where $\Upsilon(\ell)$ is an element in the set $\{\mathcal{C}^U, \mathcal{C}^O, \mathcal{C}^S\}$ equal to the output when the algorithm arrives at the leaf $\ell$. It is not hard to see that the success probability is upper bounded by
\begin{align}
    &\frac{1}{3} \sum_{\ell \in \mathrm{leaf}(\mathcal{T})} \max\left(p^{\mathcal{C}^U}(\ell), p^{\mathcal{C}^O}(\ell), p^{\mathcal{C}^S}(\ell)\right) \\
    &\leq \frac{1}{3} \sum_{\ell \in \mathrm{leaf}(\mathcal{T})} p^{\mathcal{D}}(\ell) +  \max_{\mathcal{C} \in \{\mathcal{C}^U, \mathcal{C}^O, \mathcal{C}^S\}}\left( \left|p^{\mathcal{C}}(\ell) - p^{\mathcal{D}}(\ell)\right| \right)\\
    &\leq \frac{1}{3} + \frac{1}{3} \sum_{\mathcal{C} \in \{\mathcal{C}^U, \mathcal{C}^O, \mathcal{C}^S\}}\left( \sum_{\ell \in \mathrm{leaf}(\mathcal{T})} \left|p^{\mathcal{C}}(\ell) - p^{\mathcal{D}}(\ell)\right| \right).
\end{align}
From the proof of Theorem~\ref{thm:channelhard1},~\ref{thm:Ochannelhard1},~and~\ref{thm:Spchannelhard1}, we have when $T = o(2^{2n/7})$, we have
\begin{equation}
    \sum_{\ell \in \mathrm{leaf}(\mathcal{T})} \left|p^{\mathcal{C}}(\ell) - p^{\mathcal{D}}(\ell)\right| = o(1), \quad \forall \mathcal{C} \in \{\mathcal{C}^U, \mathcal{C}^O, \mathcal{C}^S\}.
\end{equation}
Therefore, when $T = o(2^{2n/7})$, the success probability will be upper bounded by $1/3 + o(1)$.
Hence to achieve a success probability of at least $2/3$, we must have $T = \Omega(2^{2n/7})$.
\end{proof}

\subsubsection{Upper bound}

We present an upper bound for algorithms without quantum memory. There is still a gap between the lower and upper bounds, which we leave as an open question.

\begin{theorem}\label{thm:symmetry_tomo}
There is an algorithm without quantum memory that uses
\begin{equation}
\label{E:TOmega4}
    T = \mathcal{O}\left( 2^{n} \right),
\end{equation}
to correctly distinguish between a fixed, Haar-random unitary channel $\mathcal{C}^U$, orthogonal matrix channel $\mathcal{C}^O$, or symplectic matrix channel $\mathcal{C}^S$ on $n$ qubits with probability at least $2/3$.
\end{theorem}

We will perform three sets of experiments.
Each set of experiments is a quantum state tomography based on random Clifford measurements \cite{kueng2017low, guctua2020fast} on the output state $\mathcal{C}(\ketbra{\psi_i}{\psi_i})$ for an input pure state $\ket{\psi_i}$, for $i = 1,2,3$. 
We will take $\ket{\psi_1} = \ket{0^n}$, $\ket{\psi_2} = \frac{1}{\sqrt{2}}(\ket{0} + \ket{1}) \otimes \ket{0^{n-1}}$, and $\ket{\psi_3} = \frac{1}{\sqrt{2}}(\ket{0} - \ket{1}) \otimes \ket{0^{n-1}}$.
Note that for each $i$, $\mathcal{C}(\ketbra{\psi_i}{\psi_i})$ is a pure state, which we will denote by $\phi_i \in \mathbb{C}^{2^n}$.

We will use the following guarantee:

\begin{lemma}[State tomography, see \cite{kueng2017low,guctua2020fast}]\label{lem:tomo}
    There is an algorithm which for any $\epsilon > 0$, given copies of an unknown pure state $\ket{\phi}$ with density matrix $\rho\in\mathbb{H}^{2^n\times 2^n}$, makes $\mathcal{O}(2^n/\epsilon^2)$ random Clifford measurements and outputs the density matrix $\wh{\rho}$ of some pure state for which $\norm{\rho - \wh{\rho}}_{\tr} \le \epsilon$ with probability at least $14/15$.
\end{lemma}

\begin{corollary}\label{cor:triangle}
    Suppose $\wh{\rho_i} = \ketbra{\wh{\phi_i}}{\wh{\phi_i}}$ is the output of applying the algorithm in Lemma~\ref{lem:tomo} to $\rho = \ketbra{\phi_i}{\phi_i}$. Then provided the algorithm succeeds, we have that for any matrix $M\in\mathbb{C}^{2^n\times 2^n}$ and any $i,j\in\brc{1,2,3}$ \begin{equation}
        \abs*{\abs*{\wh{\phi}^t_i M \wh{\phi}_j} - \abs*{\phi_i^t M \phi}} \le 2\epsilon\norm{M}_{\infty}.
    \end{equation}
\end{corollary}

\begin{proof}
    Because $\norm{\rho_i - \wh{\rho_i}}_F\le \norm{\rho_i - \wh{\rho_i}}_{\tr} \le \epsilon$, we have that $\norm{\phi_i - \zeta\cdot\wh{\phi}_i} \le \epsilon$ for some choice of phase $\zeta_i\in\mathbb{C}$. As $\abs{\zeta_i\zeta_j\cdot \wh{\phi}^t_i M \wt{\phi}_j} = \abs{\wh{\phi}^t_i M \wt{\phi}_j}$, we may assume without loss of generality that $\zeta_i,\zeta_j = 1$. Now note that \begin{equation}
        \abs{\wh{\phi}^t_i M \wh{\phi}_j} \le \abs{\phi^t_i M\wh{\phi}_j} + \epsilon\norm{M}_{\infty} \le \abs{\phi^t_i M \phi_j} + 2\epsilon\norm{M}_{\infty}
    \end{equation}  by triangle inequality, from which the claim follows.
\end{proof}

Henceforth, let $\wh{\phi}_i$ denote the pure state obtained by applying state tomography to copies of $\phi_i$ with $\epsilon = 1/5$. We collect the following basic facts about $\wh{\phi}_i$ which will allow us to distinguish among the three types of channels.

\begin{lemma}\label{lem:z2Jz3}
    Condition on the outcome of Lemma~\ref{lem:tomo}. If $\mathcal{C}$ is a Haar-random symplectic matrix channel, then $\abs{\wh{\phi}^t_2 J \wh{\phi}_3} > 1/2$. If $\mathcal{C}$ is a Haar-random unitary or orthogonal channel, then $\abs{\wh{\phi}^t_2 J \wh{\phi}_3} < 1/2$ with probability at least $14/15$ over the randomness of the channel.
\end{lemma}

\begin{proof}
    By definition of the symplectic matrix channel, $\phi^t_2 J \phi_3 = \psi^t_2 S^t J S \psi_3 = \psi^t_2 J \psi_3 = -1$. The first part of the lemma follows by Corollary~\ref{cor:triangle} and the fact that $\norm{J}_{\infty} = 1$. For the second part, note that the joint distribution on $(\phi_2,\phi_3)$ is invariant under the transformation $(\phi_2,-\phi_3)$ when $\mathcal{C}$ is a Haar-random unitary (resp. orthogonal) transformation, because conditioned on $\phi_2$, $\phi_3$ is a Haar-random in the subspace in $\co^{2^n}$ (resp. $\R^{2^n}$) orthogonal to $\phi_2$. So in either case, \begin{equation}
        \E*{\phi^t_2 J \phi_3} = \frac{1}{2}\E*{\phi^t_2 J \phi_3} + \E*{\phi^t_2 J (-\phi_3)} = 0.
    \end{equation}
    Note that the function $F: O\mapsto \psi^t_2 O^t J O \psi_3$ is $2$-Lipschitz:
    \begin{equation}
        \abs{F(O_1) - F(O_2)} \le \abs*{\psi^t_2 (O_1 - O_2)^t J O_1 \psi_3} + \abs*{\psi^t_2 O_2^t J (O_1 - O_2) \psi_3} \le 2\norm{O_1 - O_2}_F. \label{eq:lipschitz}
    \end{equation}
    So by Lemma~\ref{lem:levy}, with probability at least $9/10$ over the randomness of the channel, we have that $\abs{\phi^t_2 J \phi_3} \le \mathcal{O}(1/2^{n/2})$. The second part of the lemma then follows from Corollary~\ref{cor:triangle}.
\end{proof}

\begin{lemma}\label{lem:realnorm}
    Condition on the outcome of Lemma~\ref{lem:tomo}. If $\mathcal{C}$ is a Haar-random orthogonal matrix channel, then $\abs{\wh{\phi}^t_1 \wh{\phi}_1} = 1$. If $\mathcal{C}$ is a Haar-random unitary matrix channel, then $\abs{\wh{\phi}^t_1 \wh{\phi}_1} < 1/2$ with probability at least $14/15$ over the randomness of the channel.
\end{lemma}

\begin{proof}
    Because $\phi_1$ is a unit vector with only real entries, $\wh{\phi}^t_1 \wh{\phi}_1 = 1$, so the first part of the claim follows by Corollary~\ref{cor:triangle}. For the second part, if the channel is Haar-random unitary, then $\phi_1$ is a Haar-random complex unit vector, so $\E{\wh{\phi}^t_1 \wh{\phi}_1} = 0$. The function $F: U\mapsto \psi^t_1 U^t U \psi_1$ is $2$-Lipschitz by a calculation completely analogous to \eqref{eq:lipschitz}. So by Lemma~\ref{lem:levy}, with probability at least $9/10$ over the randomness of the channel, we have that $\abs{\wh{\phi}^t_1\wh{\phi}_1} \le \mathcal{O}(1/2^{n/2})$. The second part of the lemma then follows from Corollary~\ref{cor:triangle}.
\end{proof}

We are now ready to prove Theorem~\ref{thm:symmetry_tomo}.

\begin{proof}[Proof of Theorem~\ref{thm:symmetry_tomo}]
    The algorithm will be to apply the state tomography algorithm of Lemma~\ref{lem:tomo} to the outputs of $\brc{\ket{\psi_i}}$ under the channel, yielding pure states $\brc{\ket{\wh{\phi}_i}}$. By a union bound, with probability $2/3$ we have that the state tomography algorithm succeeds for all $i = 1,2,3$, and Lemmas~\ref{lem:z2Jz3} and \ref{lem:realnorm} hold. We form the quantity $\abs{\wh{\phi}_2^t J \wh{\phi}_3}$ and check whether it exceeds $1/2$. If so, we conclude by Lemma~\ref{lem:z2Jz3} that $\mathcal{C}$ is symplectic. Otherwise, we form the quantity $\abs{\wh{\phi}^t_1 \wh{\phi}_1}$ and check whether it exceeds $1/2$. If so, we conclude by Lemma~\ref{lem:realnorm} that $\mathcal{C}$ is orthogonal, otherwise it is unitary.

\end{proof}

\bibliographystyle{alpha}
\bibliography{biblio}

\end{document}